\documentclass[a4paper,twocolumn,11pt,accepted=2017-05-09]{quantumarticle}
\pdfoutput=1
\usepackage[utf8]{inputenc}
\usepackage[english]{babel}
\usepackage[T1]{fontenc}
\usepackage{amsmath}
\usepackage{hyperref}
\usepackage{tikz}
\usepackage{lipsum}
\usepackage[numbers]{natbib}

\begin{document}
\title{Fractional shot noise of an SU(N) Kondo system}
\author{Damian Krychowski}
\email{krychowski@ifmpan.poznan.pl}
\affiliation{Department of Theory of Nanostructures and Quantum Materials, Institute of Molecular Physics, Polish Academy of Sciences, M. Smoluchowskiego 17, 60-179 Pozna\'{n}, Poland}
\orcid{0000-0002-1890-3334}
\author{Stanis{\l}aw Lipi\'{n}ski}
\email{lipinski@ifmpan.poznan.pl}
\orcid{0000-0001-9142-5554}
\maketitle

\begin{abstract}
  We consider transport through a multi-level interacting quantum dot (N-QD) in Kondo regime. Using the Kotliar-Ruckentein slave boson approach (SBMFA) for N-level Anderson model we define effectively noninteracting quasiparticles of the SU(N) Kondo system ($N=2,3,4,5,6$). Kondo resonance transmission coefficients determine linear noise describing quasiparticle partitioning. To discuss nonlinear conductance, susceptibilities and shot noise in the strong coupling regime, we apply Fermi liquid theory with parameters expressed by susceptibilities of pseudofermions determined within SBMFA. Nonlinear shot noise is dominated by two-quasiparticle scattering. However, we demonstrate that for occupation regions distant from the electron-hole symmetry point, the role of three-body correlations must be revealed.
\end{abstract}

\section{Introduction}

Quantum dot (QD)  structures are being intensively investigated  both for fundamental physics and for potential quantum information applications, detection and sensing \cite{Petta2005,Elzerman2004,Loss1998,Kouwenhoven2001}. To achieve these goals, the increasing ability to manipulate quantum states is crucial. As electrons are confined in fewer dimensions and as the size of the dot decreases the charging energy of a single excess charge on the dot increases.  Strong dynamic correlations start to play the dominant role when Coulomb interaction  exceeds electron kinetic energy.  For dots weakly coupled to the leads many-body resonances build up at low temperatures opening new paths for coherent transport. Due to the tunability of QDs by external fields, voltages and strains, strong correlations can be tested also in the regimes not accessible in solid-state physics. Paramagnetic SU(2) Kondo effect have been observed in semiconducting-based QDs \cite{Goldhaber1998,Cronenwett1998}, in carbon nanotubes (CNT) \cite{Makarovski2007,Lim2006,Laird2015,Krychowski2018,Nygard2000}, and in molecular structures \cite{Park2002,Liang2002}. In this phenomenon, an entangled state is formed as a result of screening of the localized spin by conduction electrons \cite{Kondo1964,Hewson1997,Yoo2018}. The growing interest in Kondo effect in nanoscopic systems is motivated primarily by cognitive purposes, as this phenomenon is a basis for understanding a large variety of intricate many-body problems. Potential applications are also relevant. Let us just mention a few. Kondo effect can be used e.g. as conductance control mechanism \cite{Goldhaber1998,Xing2022}, in probing of magnetic interactions \cite{Garnier2020}, or when polarized electrodes are connected, also for generation of spin polarized currents \cite{Sergueev2002,Bulka2003,Martinek2003,Choi2004,Lipinski2014}. A continuing present goal is to test fundamental correlations between different degrees of freedom  and to examine their role in quantum transport. The principal aim is to understand, how dot or multi-dot structures with internal spin, orbital or charge isospin may lead to variants of high symmetry Kondo effect involving different degrees of freedom \cite{Keller2014,Shiau2007,Mantelli2016,Krychowski2016,Choi2005,Herrero2005,Sasaki2004,Schmid2015,Cleuziou2013}. Spin and orbital degeneracies can occur simultaneously leading to Kondo ground state of SU(4) symmetry with orbital-spin entanglement. The simultaneous screening  of charge or orbital pseudospin and  the real spin has been reported in vertical QDs \cite{Sasaki2004} in capacitively coupled dots \cite{Keller2014} and in CNTs \cite{Herrero2005}. Suggestions for realization of SU(3) Kondo physics can be found e.g. in  \cite{Lopez2013,Lindner2018,Moca2012}.  One of the interesting directions of the study of Kondo effect is to incorporate higher rank SU(N) symmetries ($N>2$). The first SU(N) generalizations of the Anderson model appeared  in the literature on heavy fermion systems \cite{Bickers1987}, where large N expansion proved to be a powerful approximation for describing magnetic atoms with orbital degeneracy. From the perspective of potential applications it is important that the Anderson SU(N) model can be realized in a controlled way in various nanoscopic structures \cite{Carmi2011,Krychowski2020} and in correlated cold atomic gases \cite{Nishida2013,Kuzmenko2016}. Proposal of SU(6) Kondo effect for QD structure can be found in \cite{Nishida2013} and for cold atoms in  \cite{Kuzmenko2016}. There are also reports on the Kondo effects for $N>6$, e.g. SU(12) Kondo effect in CNT is analyzed  in \cite{Kuzmenko2014}. In the SU(N) case, the system has N flavors instead of spin up and spin down and the three Pauli matrices are generalized to  $(N^{2}-1)$ generators for an N-dimensional space of Lie group \cite{Haber2021,Gilmore2008,Gilmore20081}. Recently rapidly rises an  interest in quantum simulators - engineered quantum many-body systems that can controllably simulate complex quantum phenomena.  Multilevel dots or dot systems provide a versatile platform for such  simulations addressing the questions across various domains of condensed matter and field theory \cite{Hofstetter2018,Barthelemy2013,Brown1966,Hartanto2005}. Using spin or charge Kondo building blocks could eventually lead to quantum simulations of exotic lattice models, including those almost unrealistic in solid-state systems.
The motivation for studies of high symmetry Kondo effects in nanostructures is the fact, that Kondo temperatures dramatically increase with degeneracy and take on values that are experimentally accessible \cite{Hewson1997}. At the same time Kondo resonance peak remains narrower than Coulomb resonances. This makes SU(N) Kondo systems attractive for transport control, because in Kondo regime transport characteristics remain narrow with respect to gate voltage.

In this work we limit ourselves to a discussion of thermodynamic and transport properties of the basic unitary symmetries SU(N) N$=2,3,4,5,6$ that one encounters in nanoscopic systems. We address our calculations to multilevel 2DEG quantum dots \cite{Hong2018} and to single-walled and multi-walled carbon nanotube QDs \cite{Jarillo2004,Kontos2004}. It is worth mentioning that these symmetries are of a fundamental nature and concern systems from various, sometimes very distant fields of physics, including chromodynamics or Grand Unified Theories \cite{Anda2020}. Conductance, the most fundamental transport property provides information  on the time averaged electron transport.  In a rough picture this quantity can be understood by Landauer-B\"{u}ttiker type form, where conductance  is expressed by transmission, for the case of interacting electrons appropriately renormalized \cite{Meir1992}.  For Kondo correlated  systems, it is the many-body Kondo resonance  forming at the Fermi level that enables perfect electron transmission. In the  Kondo regime conductance and many other quantities  exhibit universal scaling, the Kondo temperature $k_{B}T_{K}$  being the only energy scale that governs low energy properties  \cite{Scott2009}. Kondo temperature can be extracted for example from the temperature dependence of conductance or from susceptibility \cite{Katanin2021,Turco2024,Kouwenhoven2001}. For bulk systems also spectroscopic measurements of temperature evolution of Kondo peak provide this information \cite{Madhavan1998}. Although some information about correlations is hidden in the renormalization of transmission, a deeper insight into the correlations of electronic wave functions   can be obtained from the shot noise and therefore we especially focus in this paper  on the analysis of this quantity. Despite high theoretical interest \cite{Tamir2022,Kobayashi2021}, relatively little experimental studies of shot noise are available \cite{Sikula2004,Gustavsson2007,Delattre2009,Basset2012,Onac2006,Jompol2015}. The noise  measurements are technically more  difficult than the  conventional conductance examinations and one of the reason is a need of separating  shot noise from background $1/f$ - noise caused by fluctuations in the physical environment \cite{Kobayashi2016}. The shot noise is a purely nonequilibrium property that results from the fact that current is not a continuous flow, but a sum of discrete pulses in time. Shot noise is a zero-frequency current noise out of equilibrium. Current and  shot noise, corresponding to the average and variance of the number of electrons passing through a dot per unit time, respectively, provide different information on a transport phenomenon. Equilibrium (thermal) fluctuations  can be related via the fluctuation-dissipation theorem to the linear conductance and thus equilibrium  noise does not carry extra information other than this from conductance \cite{Nazarov2009}.  Coulomb repulsion between electrons and their fermionic nature can regulate their motion,  effect which cannot be deduced from the time-averaged DC measurements, but  may be detected in the shot noise reduction or enhancement in respect to Poissonian value for  noninteracting carriers \cite{Ferrier2016,Kobayashi2016,Beenakker2003,Ferrier2020,Ferrier2017}. The increasing  interest in shot noise is also dictated by the possibility of  extracting information about the specific,  most fundamental form of correlations -  entanglement \cite{Saraga2003,Horodecki2009,Belzig2014,Terhal2025}. There are several theoretical proposal to form  and detect the entanglement in solid state devices by means of the shot noise measurement \cite{Krischek2011}. Shot noise studies are crucial for quantum computing. Shot noise, characterized by uncertainty in particle arrival times, can disrupt superposition of states and entanglement, leading to information loss and decoherence.

The considerations carried out in this work are in the spirit of the effective  Fermi liquid theory (FL) \cite{Landau1965,Nozieres1974,Luttinger1962,Hewson1993,Mora2015,Oguri2022,Teratani2020,Oguri2018,Hata2021,Schoeller2010}. One can analyze the low energy behavior of N-QD  in terms of quasiparticles and their weak residual interactions. The Kondo many-body singlet is described by us  in the extended slave boson approach using Kotliar and Ruckenstein representation \cite{Kotliar1986,Dong2001}.  In the mean field approximation (SBMFA)  the effective non-interacting quasiparticles  scatter elastically on the  Kondo resonance. SBMFA correctly  describes  spin or pseudospin fluctuations in the unitary  regime.  In the large infinite N limit this description of the SU(N) Kondo effect  is exact. For finite N, however, apart from elastic scattering of quasiparticles described by energy-dependent phase shift, one must also deal with two particle scattering off the singlet. At finite degeneracy, fluctuations of slave bosons  about their mean field coherent  states determine the size of current fluctuations.  Motions of the quasiparticles generate fluctuations which create interactions between quasiparticles \cite{Coleman1987}. The  Fermi liquid parameters can be expressed in terms of linear and nonlinear  susceptibilities \cite{Oguri2022}. We compute the general susceptibilities and three-body correlation functions in terms of quasiparticle Green's functions determined with respect to the SBMFA equilibrium ground state and using them we specify the effective interactions between quasiparticles, which in turn allows us to calculate the higher order Fermi liquid corrections to conductance, susceptibilities and shot noise.  Linear shot noise  is essential around equilibrium  and  it reflects the partition of the scattered particles. The meaning of this contribution to noise is consistent with Landauer-B\"{u}ttiker theory. Out of equilibrium interaction between quasiparticles shows up and noise is nonlinear and strongly enhanced. This is caused by two-particle and three-particle scattering events and the accompanying backscattering processes \cite{Hata2021}.

In the present paper we show, that SBMFA, which in strong coupling limit satisfactory describes strong correlations, can be used for a discussion  of  linear conductance and partition noise. This approach, however, as the mean field approximation uses the  picture of independent, dressed quasiparticles, and thus neglects fluctuations. This method therefore, cannot be used for analysis of  nonlinear noise and non-diagonal susceptibilities. One ways to solve this problem  is to perform tedious calculations of  fluctuations of boson fields  around the saddle point \cite{Lavagna1990,Coleman1987}. The role of interaction beyond MFA can be also included by a simple, intuitive  analysis based on Fermi liquid theory. Following the earlier reports the needed FL coefficients can be found perturbatively \cite{Mora2009}, and they are expressed by susceptibilities. In our proposition the diagonal susceptibilities are found directly within SBMFA formalism and for calculation of non-diagonal correlators occurring for interacting  systems, we use a simple extrapolation of the Wilson coefficient from the case of an isolated system to a system weakly coupled  with electrodes.This allows us to determine FL free energy and hence  the thermodynamics and nonlinear transport properties of the system.

Let us now give an overview of the most important topics of this article. The paper comprehensively analyzes the linear and nonlinear transport and thermodynamic properties of Kondo systems with SU(N) symmetries up to $N = 6$ with emphasis on the discussions of the shot noise. Particularly innovative are the results for systems with SU(5) and SU(6) symmetries, which have not been discussed in detail so far. Apart from numerical results for Kondo temperatures,  we also derived simple approximate analytical SBMFA formula for $T_{K}$. Characterizing the scheme used, the presented study is an analysis based on Fermi liquid theory with quasiparticles defined within SBMFA.  To extend the calculations for high voltages we went beyond MFA description of slave bosons and we determined the auxiliary bosons in self-consistent procedure including FL corrections. We illustrated  screening of generalized SU(N) spins by presenting temperature dependencies of susceptibilities and entropies. Discussing nonlinear shot noise, we have shown, for which occupancies the two-particle correlators, and for which three-particle are dominant and how it is related to the symmetry of the correlation functions.

The remainder of the paper is organized as follows. In the next chapter, we first introduce the model and  outline the many-body formalism we use - slave boson technique.  Next we  present the expressions for the susceptibilities, linear  conductance and shot noise, and we briefly explain the microscopic basis of the Fermi liquid theory with quasiparticles defined in the framework of SBMFA. This  allows us in the next step to present and justify  the formulas  for the nonlinear conductance and noise.  The next chapter is devoted to the numerical results and their analysis. We first discuss conductance, Wilson's ratios, shot noise, Fano factors and Kondo temperatures of SU(N) systems. We also show susceptibilities and entropies of N-QD linked with electrodes and compare them with the case of isolated multiorbital dot. Finally, we give conclusions and some final remarks.  We complete the text with two appendices containing an approximate analytical  SBMFA expression for the Kondo temperature in SU(N) systems and  formulas for the susceptibilities and Wilson coefficients of isolated N-QD.

\section{Formulation}
\subsection{Model}
We consider multilevel quantum dot (N-QD) coupled to electrodes, described by N-level Anderson Hamiltonian:
\begin{eqnarray}
&&{\mathcal{H}}=\sum_{\nu}E_{\nu}n_{\nu}+\sum_{\nu\nu'}Un_{\nu}n_{\nu'}+
\nonumber\\&&\sum_{k\alpha \nu}E_{k\alpha \nu}n_{k\alpha \nu}+\sum_{k\alpha \nu}t(c^{\dagger}_{k\alpha \nu}d_{\nu}+h.c).\label{1}
\end{eqnarray}
The first term represents orbital energies $E_{\nu}=E_{d}$ ($\nu=1,2,..N$), the second parameterized by $U$ describes Coulomb interactions and the last two terms describe electrons in the electrodes and their tunneling to the dot ($t$). Hamiltonian (\ref{1}) describes also capacitively coupled N-dot structure with dots connected to the separate leads.
The occupation number operators of the spin-orbital and of the Fermi sea in the left (right) leads are given by $n_{\nu} = d^{\dagger}_{\nu}d_{\nu}$ and $n_{\nu} = c^{\dagger}_{k\alpha\nu}c_{k\alpha\nu}$, respectively. We assume the coupling strength to the electrodes with the rectangular density of states 1/2W ($W$ denoting  the electron bandwidth of electrodes in the wide band approximation, where the leads are represented by flat density of states). $\Gamma=\pi t^{2}/2W$ is the coupling strength to leads. In the calculations, we use the natural units settings $\hbar=k_{B}=e=g=|\nu_{B}|=1$. We also take $W/50$ as the energy unit.
Although numerical calculations presented in this article only concern fully symmetric cases (degenerate levels), the presented introductory formulas are more general and refer also to nondegenerate. Our next publication will be devoted to changes of the shot noise along symmetry crossover \cite{Lipinski2025}. In the following we analyze strongly correlated states of N-QD characterized by SU(N) symmetry, N$=2,3,4,5,6$. In the absence of magnetic field, for even value of N, degeneracy applies to both orbital ($i$) and spin ($\sigma$) degrees of freedom and labeling $\nu$ can be understood as $\nu=i\sigma$.

To analyze correlation effects, we use finite U slave boson mean field approach (SBMFA) of Kotliar and Ruckenstein \cite{Kotliar1986}. In this approximation the effect of Coulomb interactions is effectively replaced by interaction of  quasiparticles with auxiliary bosons, which project the state space onto subspaces of different occupation numbers.  Mean field approach is correct for describing  spin and orbital fluctuations in the unitary regime and it leads to a local Fermi-liquid behavior at zero temperature. As an example we show MFA Hamiltonian (2) describing noninteracting quasiparticles in boson fields for the case of the highest of the described symmetries - SU(6):
\begin{eqnarray}
&&\nonumber \widetilde{{\mathcal{H}}}=\sum_{i\sigma}E_{i\sigma}n^{(f)}_{i \sigma}+\sum_{k\alpha\sigma}E_{ki\alpha}c^{\dagger}_{ki\alpha\sigma}c_{ki\alpha\sigma}
+\nonumber\\&&\sum_{k\alpha\sigma}t(c^{\dagger}_{ki\alpha\sigma}z_{i\sigma}f_{i\sigma}+h.c.)+U\sum_{i}d^{\dagger}_{i}d_{i}
+\nonumber\\&&U\sum_{ij\sigma\sigma',i<j}d^{\dagger}_{ij\sigma\sigma'}d_{ij\sigma\sigma'}+
3U\sum_{ij\sigma,i\neq j}t^{\dagger}_{i,j\sigma}t_{i,j\sigma}+ \nonumber\\&&3U\sum_{\sigma\sigma'\sigma''}t^{\dagger}_{\sigma\sigma'\sigma''}t_{\sigma\sigma'\sigma''}+6U\sum_{i}f^{\dagger}_{i}f_{i}+
\nonumber\\&&6U\sum_{ij\sigma\sigma',i<j}f^{\dagger}_{ij\sigma\sigma'}f_{ij\sigma\sigma'}+10U\sum_{i\sigma}q^{\dagger}_{i\sigma}q_{i\sigma}
+\nonumber\\&&15Us^{\dagger}s+\lambda({\cal{I}}-1)+\sum_{i\sigma}\lambda_{i\sigma}(n^{(f)}_{i \sigma}-Q_{i\sigma}),\label{2}
\end{eqnarray}
where $n^{(f)}_{i \sigma}=f^{\dag}_{i \sigma}f_{i \sigma}$  is pseudofermion occupation operator $\{e, p, d, t, f, q, s\}$ denote double, triple, quadruple and quint  boson fields respectively  and  $\lambda, \lambda_{i\sigma}$  are Lagrange multipliers introduced to eliminate unphysical states (details of the method are presented in Appendix A). The pole of retarded Green's function $G^{R}_{\nu,\nu}(E)=\langle\langle f_{\nu};f^{\dagger}_{\nu}\rangle\rangle^{R(A)}=1/(E-\widetilde{E}_{\nu}+ \textbf{i}\widetilde{\Gamma}_{\nu})$
in channel $\nu$, determine position $\widetilde{E}_{\nu=i\sigma}=E_{\nu}+\lambda_{\nu}$ and the width of quasiparticle resonance $\widetilde{\Gamma}_{\nu}=\Gamma z_{\nu}^{2}$. The corresponding characteristic resonance temperature is $T_{\nu}=\sqrt{\widetilde{E}^{2}_{\nu}+\widetilde{\Gamma}^{2}_{\nu}}$.
For fully symmetric systems there is a single resonance line and $T_{\nu}$ for all orbitals are equal and $T_{K}=T_{\nu}$ is the Kondo temperature. In addition to our numerical SBMFA estimations of the Kondo temperature $T_{\nu}$, we also use simple, derived by us, SBMFA approximation of $T_{K}$ presented below. The derivation of Equation (\ref{3}) is given in the Appendix A. Formula (\ref{3}) well reproduces the numerically calculated Kondo temperatures and has an advantage that it is expressed by the bare parameters of the Anderson model (insets on Figures \ref{fig1} and \ref{fig3}). For this reason it gives more intuitive insight into factors determining Kondo temperature:
\begin{eqnarray}
&&T_{K}(N,n)=
\nonumber\\&&We^{-\frac{|E_{d}+(n-1)U||E_{d}+nU|}{\frac{\Gamma}{\pi}(\Lambda^{n+1}_{n}|E_{d}+(n-1)U|+
\Lambda^{n}_{n-1}|E_{d}+nU|)}},\label{3}
\end{eqnarray}
$n$ denotes the total filling of the dot and coefficients $\Lambda^{n+1}_{n}(N)$ and $\Lambda^{n}_{n-1}(N)$ are given in Appendix A.

\subsection{Thermodynamic and transport properties. SBMFA and Fermi liquid approach}
In the following we present, based on the FL concept, formulas for basic thermodynamic and transport quantities of N-QD. FL theory is used to describe the behavior of interacting fermions at low temperatures. The starting point is a non-interacting Fermi gas. We keep the considerations at a microscopic level and  assume that the noninteracting quasiparticles  are defined within SBMFA formalism.  Elastic scattering off the many-body  Kondo singlet and weak interaction of quasiparticles induced by polarization of the spin singlet determine the higher order corrections to physical quantities. In the first step let us  write formulas for thermodynamic and transport quantities in the non-interacting quasiparticle picture.

The zero temperature current $I$ and the shot noise can be expressed by transmission ${\mathcal{T}}_{\nu}$ in the following form:
\begin{eqnarray}
&&I=\frac{e}{h}\int^{|V|/2}_{-|V|/2}\sum_{\nu}{\mathcal{T}}_{\nu}(E)dE\label{4}
\\&&\nonumber S=2\frac{e^{2}}{h}\int^{|V|/2}_{-|V|/2}\sum_{\nu}{\mathcal{T}}_{\nu}(E)[1-{\mathcal{T}}_{\nu}(E)]dE
\end{eqnarray}
 According to Landauer-B\"{u}tikker approach \cite{Meir1992,Kobayashi2016,Nazarov2009} the low bias current $I_{0}=G_{0}V$ and shot noise $S_{0}=A_{0}V$ can be expressed in the limit of zero temperature in terms of transmission (reflection) probabilities at Fermi energy alone.  The factors $(1-{\mathcal{T}}_{\nu})$ describe reduction of noise with respect to Poisson noise. Transmission ${\mathcal{T}}_{\nu}$ is determined by phase shift (${\mathcal{T}}_{\nu}(E)=\widetilde{\Gamma}^{2}_{\nu}/[(E-\widetilde{E}_{\nu})^{2}+\widetilde{\Gamma}^{2}_{\nu}]
=\widetilde{\Gamma}^{2}_{\nu}/[(E-\widetilde{\Gamma}_{\nu}\cot[\delta_{\nu}])^{2}+\widetilde{\Gamma}^{2}_{\nu}]$)  and  consequently  the linear conductance and linear noise ratio are given by:
\begin{eqnarray}
&&\nonumber G_{0}=(e^{2}/h)\sum_{\nu}\sin[\delta_{\nu}]^{2}=\\&&\frac{e^{2}}{h}\sum_{\nu}
\frac{\widetilde{\Gamma}^{2}_{\nu}}{\widetilde{E}^{2}_{\nu}+\widetilde{\Gamma}^{2}_{\nu}}\label{5}
\\&&\nonumber A_{0}=(e^{2}/h)\sum_{\nu}\frac{\sin[2\delta_{\nu}]^{2}}{4}
=\\&&\nonumber \frac{e^{2}}{h}\sum_{\nu}\frac{\widetilde{E}^{2}_{\nu}\widetilde{\Gamma}^{2}_{\nu}}{(\widetilde{E}^{2}_{\nu}+\widetilde{\Gamma}^{2}_{\nu})^2}
\end{eqnarray}
The linear zero temperature Fano factor $F_{0}$ reads:
\begin{eqnarray}
&&F_{0}=\lim_{V\mapsto0}\frac{S}{2|e|I}=\frac{\sum_{\nu}{\mathcal{T}}_{\nu}(0)[1-{\mathcal{T}}_{\nu}(0)]}{\sum_{\nu}
{\mathcal{T}}_{\nu}(0)}\label{6}
\end{eqnarray}
and for fully degenerate case, where all single channel transmissions are equal it simplifies to $F_{0}=1-{\mathcal{T}}_{\nu}(0)=A_{0}/G_{0}=\frac{\widetilde{E}^{2}_{\nu}}{T^{2}_{K}}$.
The above contribution to the noise is called partition noise, because it reflects fluctuations related to partition of the scattered particles. There are no fluctuations in the number of carriers at zero temperature, shot noise reflects character of transmission described by probabilities. Formally, the shot noise formula (\ref{4}) is analogous to the free electron formula, but one has to keep in mind, that in some way, part of the electron correlations is already included in this simple approach by renormalizing the SBMFA parameters and it properly captures the limit $V\mapsto0$. For example it correctly gives complete suppression of the shot noise for SU(2) Kondo systems and $S_{0}=2(e^{2}/h)$ for SU(4) \cite{Delattre2009}. Linear noise is a direct signature of symmetry class, because for each symmetry a different number of channels are involved in  transport. In the general case (finite temperature and voltage) it is necessary to take into account noise corrections  beyond MFA. One needs either consider fluctuations of  slave bosons and pseudofermion operators or to find higher order corrections to current and shot noise  in the framework of the Fermi liquid theory analyzing the elastic and inelastic scattering of quasiparticles. This second approach is the path we  follow in this article. FL theory applies to the low energy properties of quasiparticles. The great advantage of this theory  is that it can be also applied to nonequilibrium processes.  In the interested us Kondo problem quasiparticles are scattered elastically by the singlet. They also interact through polarization of the singlet. Fermi liquid energy functional which is biquadratic with respect to quasiparticle creation/annihilation operators (bilinear with respect to quasiparticle fluctuations) is parameterized by four coefficients: the two postulated by Nozi$\grave{e}$res \cite{Nozieres1974}, they modify the excitation energy in the first order and parameters of second order necessary for description of  such  nonequilibrium quantities, like nonlinear current and noise \cite{Sela2006,Mora2009}.  Their gate, field or temperature dependencies are mostly quadratic. Many recent works  on the Kondo systems \cite{LeHur2009} emphasize the importance of residual interactions in current fluctuations.  As shown e.g. in \cite{Oguri2022,Krychowski2025}, all four FL parameters can be expressed using Friedel sum rule by zero-temperature susceptibilities and their derivatives with respect to the local level positions.
Static two-body susceptibilities $\widetilde{\chi}_{\nu_{1}\nu_{2}}=\int^{1/k_{B}T}_{0}d\tau\langle\delta n_{\nu_{2}}(\tau)\delta n_{\nu_{1}}(0)\rangle^{<}$ and three-body correlation functions $\widetilde{\chi}^{[3]}_{\nu_{1}\nu_{2}\nu_{3}}=-\int^{1/k_{B}T}_{0}d\tau_{3}\int^{1/k_{B}T}_{0}d\tau_{2}\langle T_{[\tau]}\delta n_{\nu_{3}}(\tau_{3})\delta n_{\nu_{2}}(\tau_{2})\delta n_{\nu_{1}}$
$(0)\rangle^{<}$ are expressed through derivatives of the free energy
with respect to site energies. $\delta n_{\nu}$ denote deviations from the ground state distribution  $\delta n_{\nu}\equiv n_{\nu}-\langle n_{\nu}(0)\rangle$.In the mean field approximation the free energy corresponding to Hamiltonian (2) is a sum of slave boson free energy $\widetilde{F}_{b}$, and fermionic contribution $\widetilde{F}_{f}$ ($\widetilde{F}=\widetilde{F}_{f}+\widetilde{F}_{b}$).
The diagonal fermionic susceptibilities at low temperatures $\widetilde{\chi}_{\nu\nu}=-\left\langle\frac{\partial^{2} \widetilde{F}_{f}}{\partial \widetilde{E}^{2}_{\nu}}\right\rangle$, $\widetilde{\chi}^{[3]}_{\nu\nu\nu}=-\left\langle\frac{\partial^{3} \widetilde{F}_{f}}{\partial \widetilde{E}^{3}_{\nu}}\right\rangle$ are given by Equations (\ref{A13}-\ref{A14}).

Commonly used quantities serving as a measure of  relative strength of correlations are Wilson ratios $W_{\nu\nu'}$, defined by the proportions of the off-diagonal component of susceptibility to the diagonal component \cite{Hewson1993,Oguri2022}:
\begin{eqnarray}
W_{\nu\nu'}-1\equiv-\frac{\widetilde{\chi}_{\nu\nu'}}{\sqrt{\widetilde{\chi}_{\nu\nu}\widetilde{\chi}_{\nu'\nu'}}}\label{7}.
\end{eqnarray}
Most frequently cited in the  literature spin Wilson ratio \cite{Hewson1993} is defined as the  proportion of the spin susceptibility $\chi_{(s)}$ and linear specific heat coefficient $\gamma_{(0)}$ and can be written as:
\begin{eqnarray}
&&W_{s}=\frac{4\pi^{2} \chi_{(s)}}{3 \gamma_{(0)}}=1-\frac{\widetilde{\chi}_{\nu\nu'}}{\widetilde{\chi}_{\nu\nu}}\label{8}
\end{eqnarray}
where $\gamma_{(0)}=(\pi^2/3)N\widetilde{\chi}_{\nu\nu}$ is the linear specific heat coefficient \cite{Hewson1997}.
For the fully symmetric case SU(N) the linear susceptibility has only two independent, generally different, components: diagonal and off-diagonal.
All diagonal elements $\widetilde{\chi}_{\nu\nu}=\widetilde{\chi}_{d}$ are equal and similarly all off-diagonal elements $\widetilde{\chi}_{\nu\nu'}=\widetilde{\chi}_{nd}$ ($\nu\neq\nu'$).
Consistently Wilson ratio is a single number in this case.
Spin $\chi_{(s)}$ and charge $\chi_{(c)}$ susceptibilities can be written as:
\begin{eqnarray}
&& \nonumber\chi_{(s)}=\frac{1}{4}\sum_{i}\int^{1/k_{B}T}_{0}d\tau\langle\delta S^{Z}_{i}(\tau)\delta S^{Z}_{i}(0)\rangle^{<}=
\\&&\frac{1}{4}\sum_{i\sigma}\sigma\overline{\sigma}\int^{1/k_{B}T}_{0}d\tau\langle\delta n_{i\sigma}(\tau)\delta n_{i\sigma}(0)\rangle^{<}=\label{9}
\\&&\nonumber\frac{1}{4}\sum_{i\sigma\sigma'}\sigma\sigma'\widetilde{\chi}_{i\sigma i\sigma'}
\\&&\chi_{(c)}=\int^{1/k_{B}T}_{0}d\tau\langle\delta Q(\tau)\delta Q(0)\rangle^{<}=
\\&&\nonumber\sum_{i\sigma i'\sigma'}\int^{1/k_{B}T}_{0}d\tau\langle\delta n_{i\sigma}(\tau)\delta n_{i'\sigma'}(0)\rangle^{<}=
\sum_{\nu\nu'}\widetilde{\chi}_{\nu\nu'}\label{10}
\end{eqnarray}
Assuming that the Wilson ratio is known  and using correlator equations (16,18), we can write the spin and charge susceptibilities for SU(N) system in the form:
\begin{eqnarray}
&& \chi_{(s)}=\frac{1}{4}\left(N\widetilde{\chi}_{d}-N\widetilde{\chi}_{nd}\right)=
\\&&\nonumber\frac{1}{4}\left[N\widetilde{\chi}_{d}+N(W_{\nu\nu'}-1)\widetilde{\chi}_{d}\right]=
\frac{NW_{\nu\nu'}\widetilde{\chi}_{d}}{4}\label{11}
\\&& \nonumber\chi_{(c)}=N\widetilde{\chi}_{d}+(N^{2}-N)\widetilde{\chi}_{nd}=
N\widetilde{\chi}_{d}-\\&&\nonumber(N^{2}-N)(W_{\nu\nu'}-1)\widetilde{\chi}_{d}=
\\&&[N^{2}+W_{\nu\nu'}(N-N^{2})]\widetilde{\chi}_{d}\label{12}
\end{eqnarray}
Typically, due to residual interactions Wilson ratio increases, but in general it depends on the type and strength of the interactions and the resulting many-body phenomena \cite{Hewson1997}. One of the possible reasons for the weakening of Wilson ratio may be strong charge fluctuations. In the case discussed here (SU(N) Kondo states) values of $W_{\nu\nu'}$ are larger than one and hence spin susceptibilities are enhanced by residual interactions, whereas charge susceptibility is suppressed. The characteristic temperature $T^{\star}$ in which Kondo correlations become visible in susceptibilities is defined as $T^{\star}=1/(4\sqrt{\widetilde{\chi}_{\nu\nu}\widetilde{\chi}_{\nu'\nu'}})$, and can be considered as Kondo temperature determined from susceptibility.
For the special case of SU(2) symmetry Kondo susceptibility expressed by characteristic temperature $T^{\star}$ is
$\chi_{(s)}=\frac{1}{4T^{\star}}$ \cite{Nishikawa2012}.
The off-diagonal linear  susceptibilities $\chi_{\nu\nu'}$ and nonlinear susceptibilities $\chi^{[3]}_{\nu\nu'\nu'}$ can be expressed using only Wilson coefficients and diagonal elements as:
\begin{eqnarray}
&& \widetilde{\chi}_{\nu\nu'}=-(W_{\nu\nu'}-1)
\sqrt{\widetilde{\chi}_{\nu\nu}\widetilde{\chi}_{\nu'\nu'}}\label{13}\\
&& \nonumber\widetilde{\chi}^{[3]}_{\nu\nu'\nu'}=\frac{\partial \widetilde{\chi}_{\nu\nu'}}{\partial \widetilde{E}_{\nu'}}=-(W_{\nu\nu'}-1)\frac{\partial \sqrt{\widetilde{\chi}_{\nu\nu}\widetilde{\chi}_{\nu'\nu'}}}{\partial \widetilde{E}_{\nu'}}-
\\&& \nonumber\frac{\partial W_{\nu\nu'}}{\partial \widetilde{E}_{\nu'}}\sqrt{\widetilde{\chi}_{\nu\nu}\widetilde{\chi}_{\nu'\nu'}}=\frac{\chi_{\nu\nu'}}{\chi_{\nu'\nu'}}\chi^{[3]}_{\nu'\nu'\nu'}
\end{eqnarray}

Another thermodynamic quantity revealing quenching of local moment is the entropy of SU(N) Kondo dot, which we can generally express in the form:
\begin{eqnarray}
&& S_{N}=-\frac{\partial\widetilde{F}}{\partial T}\label{14}
\end{eqnarray}
where $\widetilde{F}=\widetilde{F}_{f}+\widetilde{F}_{b}$. For $T\ll T_{K}$, $S_{N}\approx0$, which indicates transition to the SU(N) Kondo singlet. An increase in temperature leads to a saturation of the entropy at a value, characteristic of the local moment, where the screening effect is removed and the entropy reaches a value of $S_{N}=k_{B}\ln[N!/(n!(N-n)!)]$. The essence of  Kondo effect is a formation of a singlet of  localized electron with conduction electrons. In the case of SU(2) symmetry the screened  quantity is spin or orbital (charge) pseudospin characterized by two-dimensional Pauli matrices.  For SU(3) system  three possible one-electron states are labeled by flavor: up (u), down (d) and strange (s), terminology borrowed from quark theory \cite{Lindner2018,Brown1966}. Eight dimensional SU(3) spin corresponds to the eight hermitian, traceless $3\times3$ generators  of SU(3) Lie group - Gell-Mann matrices (G-M). In the language of information theory  Pauli matrices (N$=2$) act on qubits, Gell-Mann matrices (N$=3$)  operate on qutrits, and for generic N-level system the generalized  G-M matrices act on qudits \cite{Balantekin2024}.   Among the generalized SU(3) spin components three have vector character and five are quadrupoles.  Generalizing the i-th component of SU(N) spin can be defined as follows \cite{Mantelli2016,Carmi2011}: $S^{i}=\sum_{\nu\nu'}d^{\dagger}_{\nu}T^{i}_{\nu\nu'}d_{\nu'}$, where $T^{i}$ are $N\times N$ generators of SU(N) group, $i = 1,..N^{2}-1$ and  $\nu$, $\nu'$   range over N channels.
Operators $T^{i}$ span the full space of local physical observables.
$T^{i}=(1/2)\Lambda_{i}$, where $i$ is a generalized G-M matrix. The above choice of generators called fundamental is not the only one \cite{Haber2021,Gilmore20081,Omolo2025}, but to treat all symmetries in a unified way we will work with this set. In the following we will use the terms SU(N) spin and generalized spin interchangeably. Any SU(N) group element can be written as:
\begin{eqnarray}
&&U_{N,\vec{\alpha}}=e^{\textbf{i}\sum_{i=1}^{N^2-1}\alpha_{i}T^{i}}\label{15}
\end{eqnarray}
where $\vec{\alpha}$ is $N^{2}-1$ dimensional  vector of real parameters.
The contribution of conduction electrons to the total spin has a similar form
$s^{i}=\sum_{k\alpha\nu\nu'}c^{\dagger}_{k\alpha\nu}T^{i}_{\nu\nu'}c_{k\alpha\nu'}$ and the components of the total SU(N) spin $S_{tot}$ are
$S^{i}_{tot}=S^{i}+s^{i}$.
$S{tot}$ commutes with SU(N) Hamiltonian (1).  In SU(N) symmetry,  there are $N-1$ commuting generators that play the role of $S_{Z}$ (represented by diagonal matrices) and $N(N-1)/2$ symmetric and $N(N-1)/2$ antisymmetric matrices corresponding to transverse components of $S$. The squared sum of the generators gives the quadratic Casimir operator $C_{2}=\sum_{i=1}^{N^{2}-1}(S^{i}_{tot})^{2}$, a quantity that plays a central role in the theory of Lie groups. This operator is a group invariant and commutes with every generator of the group. The eigenvalues of Casimir elements can be used to classify irreducible representations of the group. As it is seen $C_{2}$ can be interpreted as the squared SU(N) spin operator.
We can write the generalized SU(N) spin susceptibility as:
\begin{eqnarray}
&&\chi_{N}(T)=\frac{\langle S^{2}_{tot}\rangle-\langle S_{tot}\rangle^{2}}{T}\label{16}
\end{eqnarray}
The SU(N) susceptibility can be expressed as:
\begin{eqnarray}
&&T\chi_{N}= TC_{2}(N)=
\\&&\nonumber T\frac{(N+1)}{4}(\sum_{\nu}\widetilde{\chi}_{\nu\nu}-\sum_{\nu\neq\nu'}\widetilde{\chi}_{\nu\nu'})\label{17}
\end{eqnarray}
and for $t=0$ it is reduced to:
\begin{eqnarray}
&&\nonumber T\chi_{N}=a_{N}(\sum_{i=\nu}Q_{i}-b_{N}\sum_{i<k}Q_{i}Q_{k})=\\&&TC^{(0)}_{2}(N)\label{18}
\end{eqnarray}
where $a_{N}=\{3/4,4/3,15/8,12/5,35/12...\}$ and $b_{N}=\{2,1,2/3,1/2,14/5...\}$. $C^{(0)}_{2}$ is the quadratic Casimir operator in the
free pseudospin momentum limit ($t=0$).
We see, that for $Q=1e$, the inter-state charge transfer correlator is $Q_{ik}=Q_{i}Q_{k}=0$.
Correspondingly, successive values of $N$ reach the numbers $T\chi_{N}\approx a_{N}$. In the Figure \ref{fig6}, we presented $(T_{K}+T)\chi_{N}$. It reaches
in the low temperature limit $T\mapsto0$ characteristic value:
\begin{eqnarray}
&& \nonumber T_{K}\chi_{N}=T_{K}\frac{N(N+1)\widetilde{\chi}_{\nu\nu}}{2}=\frac{N(N+1)\widetilde{\Gamma}_{\nu}}{2\pi T_{K}}=\\&&\frac{N(N+1)\sin[\delta_{\nu}]}{2\pi}\label{19}.
\end{eqnarray}
$T_{K}\chi_{N}$ depends on the degree of degeneration $N$ and the charge (via the phase shift of $\delta_{\nu}$), where $(T_{K}+T)\chi_{N}=(T_{K}+T)\frac{N+1}{4}(N+(W_{\nu\nu'}-1)(N^{2}-N))\widetilde{\chi}_{\nu\nu}(T)\stackrel{T\mapsto0}
=T_{K}\frac{N(N+1)}{2}\widetilde{\chi}_{\nu\nu}(0)$. This measurable quantity provides information about the residual interaction of quasiparticles with pseudospin. At high temperatures $(T_{K}+T)\chi_{N}\mapsto T\chi_{N}$ and describes the square of the generalized spin.

\subsection{Non-linear current and non-linear noise}
As we have mentioned in the previous subsection, linear noise ($eV\ll T_{K}$) is completely described by non-interacting quasiparticles. At higher voltages, some nonlinearity in the noise also appears due to nonlinear conductance, being a consequence of strong energy dependence of transmission. However, the decisive role in enhancement  of current fluctuations is played by residual interactions. Scattering of quasiparticles is characterized by an effective charge e$^{\star}$ different from the electron charge e. This is a consequence of the simultaneous backscattering of one and two quasiparticles \cite{Sela2006}. The noise refereing to fractional effective charge is called fractional shot noise. The enhancement value is closely related to the Wilson ratio and is universal for the Fermi liquid in the Kondo regime as it depends only on the symmetry group of the system \cite{Ferrier2016}. To analyze higher order corrections to the shot noise generated by interactions one has supplement the mean field slave boson  free energy (\ref{A9}) by at least bilinear terms with respect to population fluctuations.
Nozi\'{e}res  first formulated microscopic Fermi liquid  theory for the  Kondo model \cite{Oguri2018}.  He derived FL coefficients ($\alpha_{1,\nu}, \varphi_{1,\nu\nu'}$) analyzing  off - singlet quasiparticle  scattering and  expanding the corresponding  phase shifts   ($\delta_{\nu}(E,n_{\nu})$) in leading order in energy and deviations of the quasiparticle distribution function from its ground state.  To describe  nonequilibrium  properties away from particle hole symmetry points apart from Nozi\'{e}res coefficients, which are first order in the excitation energy, also additional second order coefficients are required.  A clear presentation of microscopic scheme for determining FL coefficients and nonlinear noise is presented e.g.in \cite{Mora2015,Oguri2022,Oguri2018}.
The current-current correlation function which is the main subject of this paper, $S=(e^{2}/h)\int^{+\infty}_{-\infty}dt\langle\delta\hat{I}_{\nu}(t)\delta\hat{I}_{\nu'}(0)
+\delta\hat{I}_{\nu'}(0)\delta\hat{I}_{\nu}(t)\rangle$, $\delta\hat{I}_{\nu}(t)=\hat{I}_{\nu}(t)-\langle\hat{I}_{\nu}(0)\rangle$,
depends on the collision term of two quasiparticles, which is described by the Keldysh vertex corrections. Oguri et al \cite{Oguri2022} calculated the  vertex functions up to the linear - order with respect to bias voltage $V$, temperature $T$ and the energy $E$. The Keldysh Green's functions were expanded up to terms of order $V^{2}$, $T^{2}$ and  $E^{2}$. Fermi liquid coefficients were deduced from derivatives of self-energy and vertex functions  using the Ward identities \cite{Zawadowski1978,Oguri2022}, which relate these quantities. It was shown that free energy coefficients can only be expressed by static linear $\widetilde{\chi}_{\nu\nu'}$ and nonlinear $\widetilde{\chi}^{[3]}_{\nu\nu'\nu''}$ correlation functions. The derived formula for the free energy is \cite{Mora2015}:
\begin{eqnarray}
&&\nonumber \Delta\widetilde{F}=-\frac{1}{\pi T_{K}}\sum_{\nu,E}\left(\alpha_{1,\nu}E+\frac{\alpha_{2,\nu}E^{2}}{T_{K}}\right)\delta n_{\nu}
+\\&&\nonumber \frac{1}{\pi T_{K}}\sum_{\nu<\nu',EE'}\left(\varphi_{1,\nu\nu'}+\frac{\varphi_{2,\nu\nu'\nu'}(E+E')}{2T_{K}}\right)\cdot\\&&\cdot\delta n_{\nu}\delta n_{\nu'}+\\&&\nonumber-\frac{1}{\pi T_{K}}\sum_{\nu<\nu'<\nu'',EE'E''}\frac{\varphi_{2,\nu\nu'\nu''}}{(N-2)T_{K}}\delta n_{\nu}\delta n_{\nu'}\delta n_{\nu''}\label{20}
\end{eqnarray}
where $\alpha_{1,\nu}/\pi=\widetilde{\chi}_{\nu\nu}$ ($\alpha_{2,\nu}/\pi=-(1/2)\widetilde{\chi}^{[3]}_{\nu\nu\nu}$), $\varphi_{1,\nu\nu'}/\pi=-\widetilde{\chi}_{\nu\nu'}$ ($\varphi_{2,\nu\nu'\nu''}/\pi=2\widetilde{\chi}^{[3]}_{\nu\nu'\nu''}$) are the FL coefficients \cite{Oguri2018}.
Formula (32) includes all first and second-order terms in the low energy coupling strength $1/T^{2}_{K}$.
The first term of (\ref{20}) describes elastic scattering off Kondo singlet and is reflected in energy (spin) dependence of the phase shift. Second, interaction terms is related to the inelastic scattering and quasiparticle interaction. The last term of the expansion (\ref{20}) can be formally interpreted as effective three-body interaction. The  three-body contribution is essential outside electron-hole symmetry point ($n = N/2$) and for systems with broken symmetry. Single particle energies of quasiparticles are measured with respect to the Fermi energy, $\delta n_{\nu}$ denote deviations from the ground state distribution and FL coefficients are expressed by static, in our considerations SBMFA  susceptibilities. Part of the bilinear correction, if included in the Hartree approximation, only modifies the elastic scattering contribution, while the rest describes inelastic collisions \cite{Oguri2018}. In general for specifying FL parameters both diagonal and off-diagonal elements of susceptibility are necessary. Off-diagonal susceptibility  appears only if residual interactions are present in the system.
Our calculation methodology is the  following. We directly use formula (\ref{20}) and determine the FL coefficients using diagonal correlators calculated in the SBMFA scheme. Off-diagonal correlators, which are the result of residual interactions are also required. They do not appear in the mean-field formalism, but  can be determined using diagonal correlators and  Wilson coefficient. We introduce a simple extrapolation of the Wilson coefficient from the solvable case of an isolated system to a weakly coupled system with electrodes. When the Wilson coefficients are known, the off-diagonal correlators can be directly expressed in terms of the diagonal terms Equations (\ref{7}) and (\ref{13}). All susceptibility elements  are easily found for interacting dots disconnected from electrodes ($t = 0$). The spectrum of isolated N-QD  can be calculated exactly, and the corresponding exact free energy can be found. Susceptibilities are then calculated in the standard way differentiating the free energy over the appropriate energies. Consequently one finds also Wilson ratios $W^{(0)}_{\nu\nu'}$ for isolated N-QD. They can be can be expressed by quasiparticle occupations as follows (Appendix C):
\begin{eqnarray}
&&W^{(0)}_{\nu\nu'}-1=\frac{n_{\nu}n_{\nu}-n_{\nu\nu'}}{\sqrt{\delta n^{2}_{\nu}\delta n^{2}_{\nu'}}}=\\&&\nonumber
\frac{Q_{\nu}Q_{\nu'}-Q_{\nu\nu'}}{\sqrt{Q_{\nu}(I-Q_{\nu})Q_{\nu'}(I-Q_{\nu'})}}\label{21}
\end{eqnarray}
For simplicity of calculation we introduce the following  approximation, which will help us easily find approximate values of Wilson ratios and consequently also non-diagonal susceptibility elements for system  linked with the electrodes. Considering the low value of the quasiparticle renormalization weight $z_{\nu}$ in the K-R SBMFA approach, it is reasonable to assume that extrapolation of the formula (\ref{21})  to the range of small values of hopping parameter $t$ should be a good approximation for Wilson ratio and static correlators.
Using Equation (\ref{7}) for N-QD connected to electrodes, the assumed quasiparticle occupation should correspond to this case ($t\neq0$).
This simplification allows to calculate all the components of susceptibility avoiding  at this stage the use of the tedious advanced many-body approaches. One can easily improve this approximation by introducing a self-consistency procedure for susceptibility and Wilson coefficients, however, this seems unnecessary because already this simple proposed above approximation gives satisfactory results. It well reproduces NRG calculations of two- and three-body quantities \cite{Teratani2020}.

Before commenting on the contribution to the  noise from interactions it is worth mentioning that for higher voltages part of nonlinearities also appear without interactions due to non-linear conductance. To analyze  higher order noise corrections generated by interactions  one has consider, free energy interaction term $\Delta \widetilde{F}$. Part of it (terms $\alpha_{1(2),\nu}$) describe elastic scattering off Kondo singlet and is reflected  in  energy (spin) dependence of the phase shift. Interaction terms ($\varphi_{1(2),\nu\nu'}$) are the source of inelastic scattering, but as we mentioned earlier Hartree terms stemming from this contribution also formally contribute  to elastic scattering and they  give mean-field energy shift of the Kondo resonance. A full account of the consequences of $\Delta \widetilde{F}$ on the conductance and shot noise requires the use of Keldysh formalism.
Collision terms of two quasiparticles are microscopically described by the Keldysh vertex corrections.
The Keldysh Green's functions are expanded up to the terms of order $(eV)^{3}$ to include multiple collision processes contributing to the noise,  the  vertex function has to be determined up to linear order. Leaving all the details, we write down after \cite{Oguri2022} the final expressions Equations 34-35 for the current and shot noise given with accuracy of the order of $(eV)^{2}$ and  $(eV)^{3}$ respectively and expressed by equilibrium  susceptibilities. To maintain the accuracy of $(1/T_{K})^{2}$, both two- and three-body correlations contributing to nonlinear current and noise  have to be taken into account \cite{Oguri2022}. The details of derivation  can be found in \cite{Oguri2022}:
\begin{eqnarray}
&& I=I_{0}+I_{K}=G_{0}V+\sum_{\nu}c_{V,\nu}V^{3}\label{22}\\
&& S=S_{0}+S_{K}=A_{0}V+\sum_{\nu}c_{S,\nu}V^{3}\label{23}
\end{eqnarray}
Current and shot noise refer to nonequilibrium, where the time is broken and therefore in formulas (\ref{22}) and (\ref{23}) occur only odd terms of $V$.
The formulas for linear  current ($I_{0}$) and partition noise ($S_{0}$) have been given earlier and the nonlinear coefficients $c_{V,\nu}$ and $c_{S,\nu}$  derived in \cite{Oguri2022} are presented in Appendix D. The generalized Fano factor $F_{K}$, which describes the relation between the nonlinear current and current noise is \cite{Oguri2022}:
\begin{eqnarray}
&&\nonumber F_{K}=\frac{d^{2}S/dV^{2}}{2|e|d^{2}I/dV^{2}}=\frac{S-S_{0}}{2|e|(I-I_{0})}
=\\&&\frac{S_{K}}{2|e|I_{K}}=\frac{\sum_{\nu}c_{S,\nu}}{\sum_{\nu}c_{V,\nu}}\label{24}
\end{eqnarray}

\section{Results}
We present numerical results for the  N-degenerate quantum dot described by N level Anderson model. Our study examines the impact of correlations on transport and thermodynamic properties of SU(N) Kondo dot with special focus on the shot noise. As mentioned in the Introduction, the following considerations are not limited to the multilevel dot, but  they can  as well be addressed to multiple dots and various other strongly correlated systems described by Hamiltonian (\ref{1}) and characterized by special unitary symmetries. To discuss the correlations we use extended Kotliar-Ruckenstein slave boson formalism in the mean field approximation(SBMFA)  \cite{Krychowski2018,Kotliar1986,Dong2001}.
The calculations were carried assuming Coulomb parameter $U = 3$ and coupling to the leads $\Gamma = 0.025$.
Before we get to the results let us mention that, an assumption of the weak electrode coupling ($U/\Gamma=120$), ensures achieving Kondo's unitary limits in calculations for all the analyzed symmetries,
which manifests in the occurrence of the flat plateaus of conductance with unitary values and correspondingly also flat lines of Fano coefficients in Kondo regimes and convergence of Wilson ratios to the values $1/(N-1)$. The assumed ratio of coupling strength and Coulomb interaction parameter are within the range of experimental data. The value of $U$ can be inferred from the size of Coulomb diamonds. For semiconducting
carbon nanotubes, the charging energy is of order of tens meV \cite{Herrero2005,Makarovski2007} and for QDs in 2DEG $U=1-4$ meV \cite{Schmid1998,vanderWiel2000,Cronenwett1998}. $\Gamma$ informs us a quality of contacts and is of order of meV or even its fractions \cite{Makarovski2007,Schmid1998}. On the following figures we present   Kondo temperatures, conductances, Wilson ratios, Fano factors, susceptibilities and entropies drawn as the functions of current or gate voltage.

\subsection{Even symmetries of Kondo states}
\begin{figure}
\includegraphics[width=0.8\linewidth]{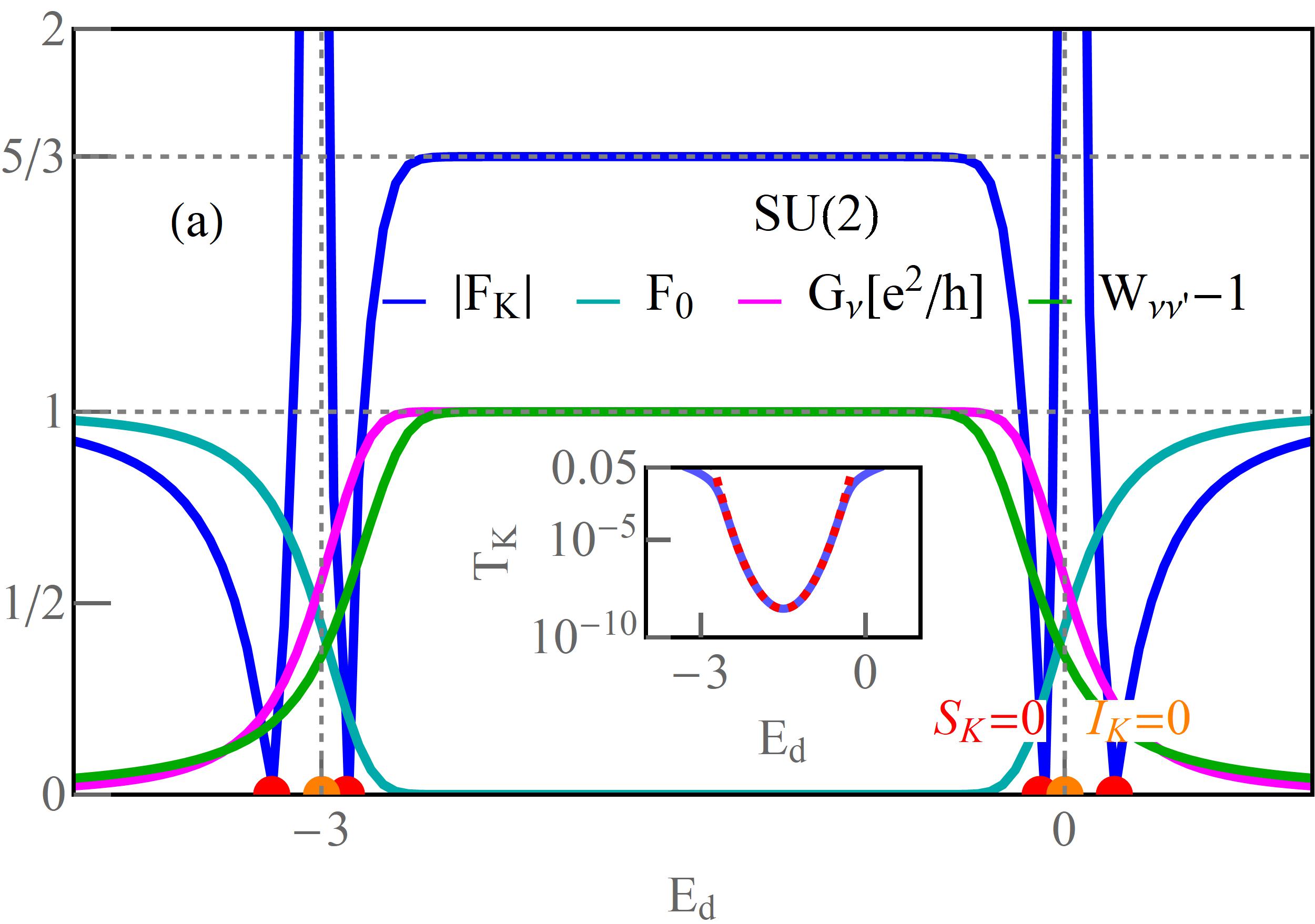}
\includegraphics[width=0.8\linewidth]{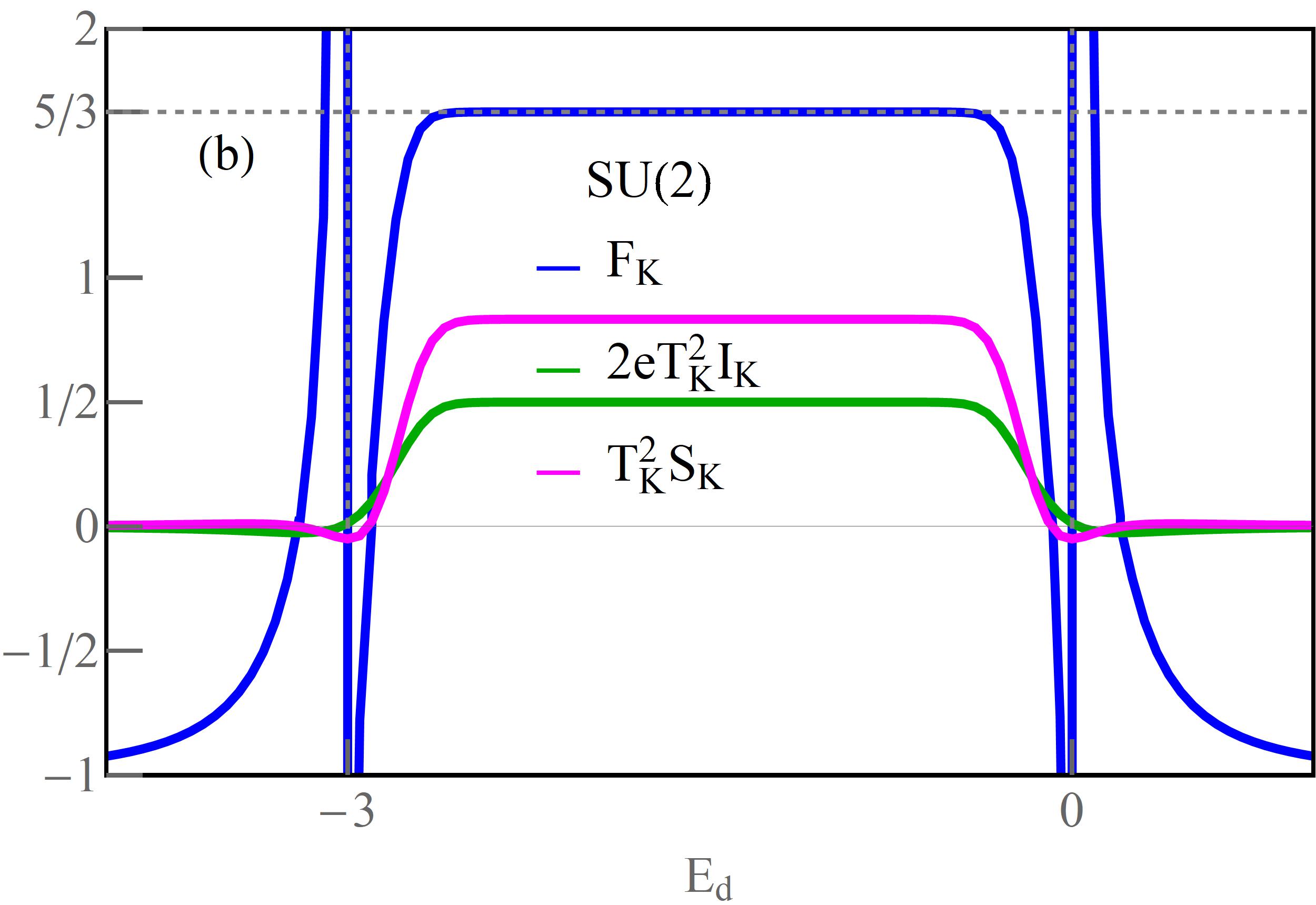}\\
\includegraphics[width=0.8\linewidth]{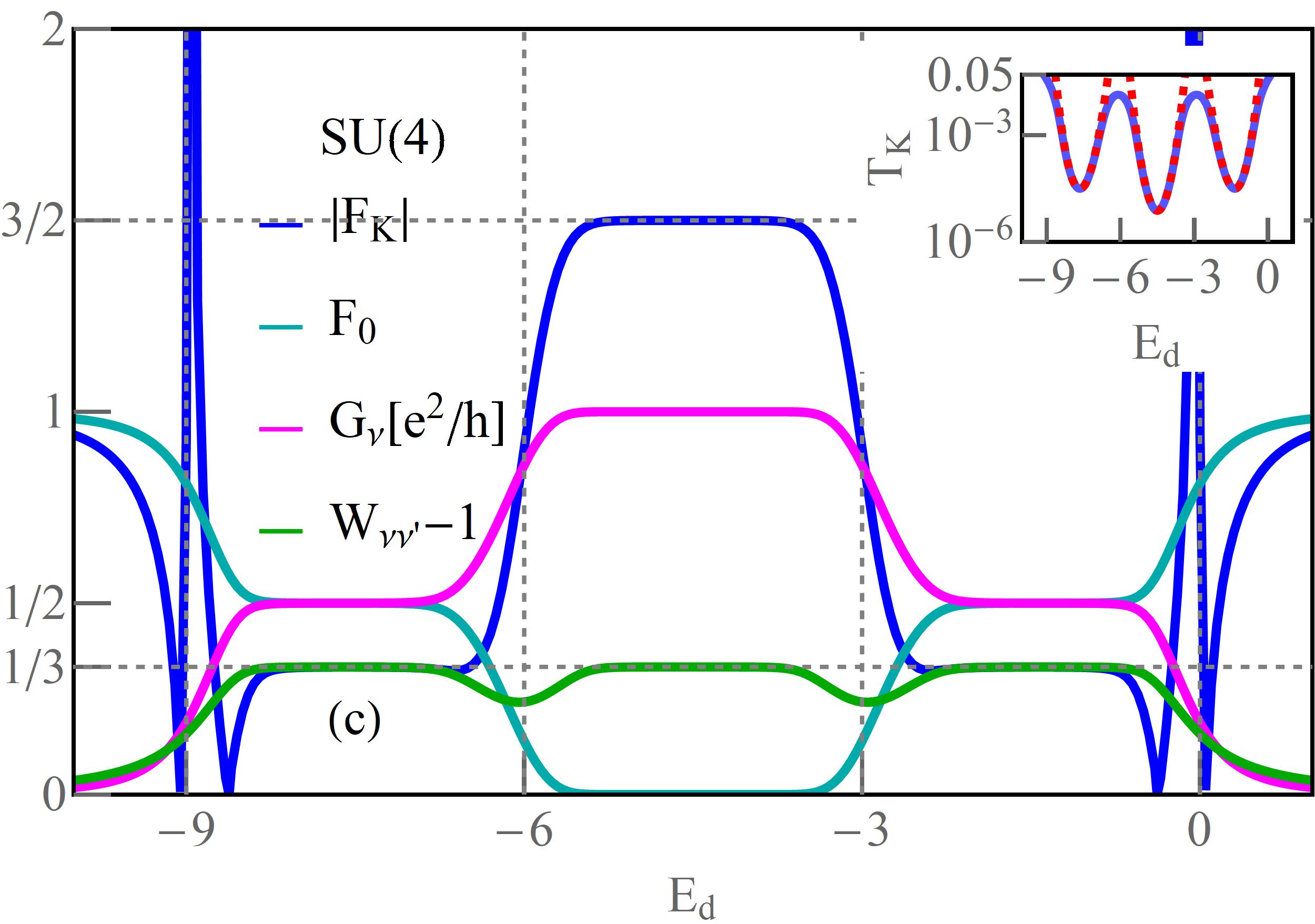}
\includegraphics[width=0.8\linewidth]{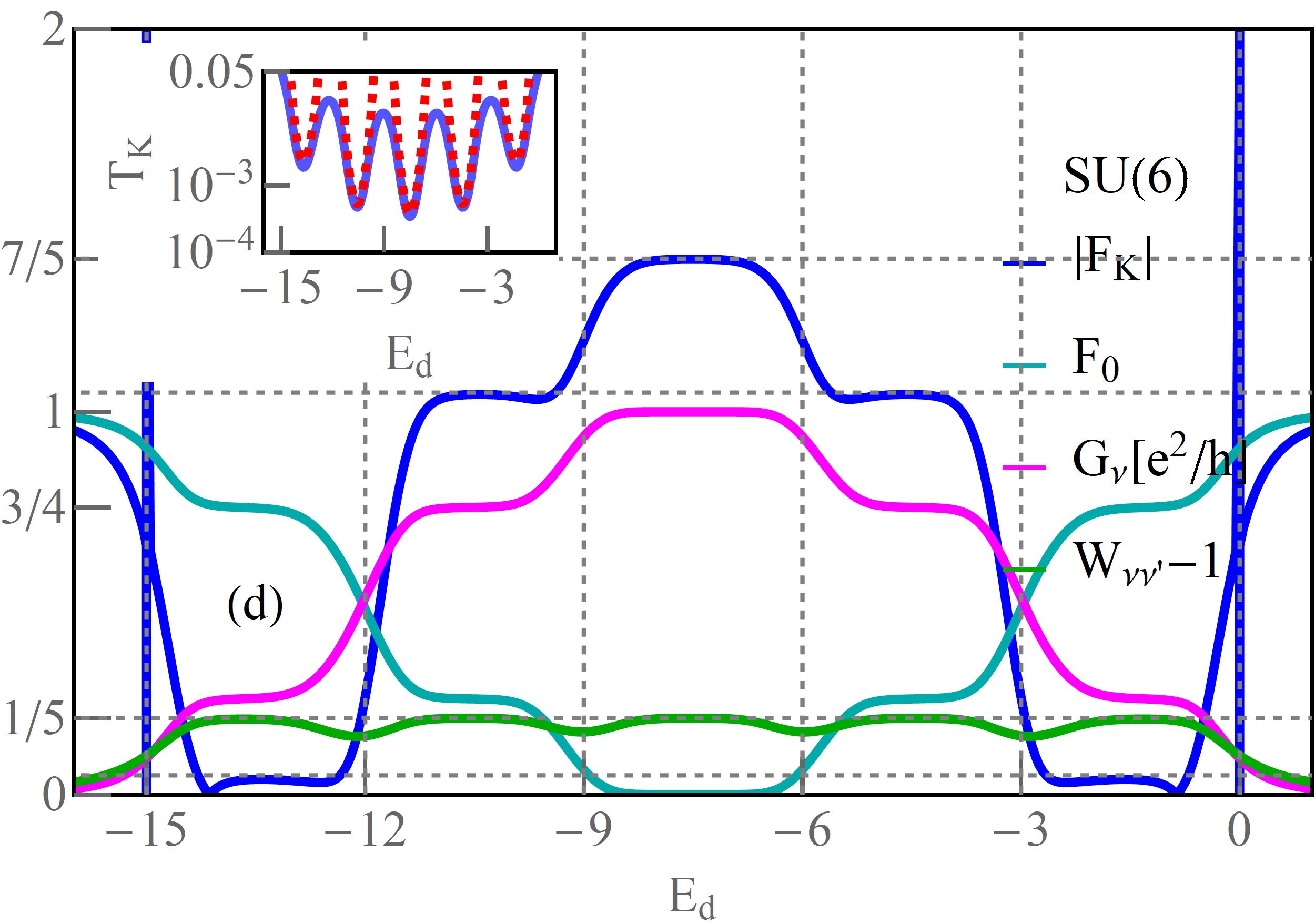}\caption{\label{fig1} Kondo effect with even symmetry SU(N=2,4,6): (a,c,d) linear and nonlinear Fano factor $F_{0(K)}$, single-channel quantum conductance $G_{\nu}$
and Wilson ratio $W_{\nu\nu'}-1$ as a function of dot energy $E_{d}$ for $N=2, 4, 6$. (b) Gate-dependent $F_{K}$ compared with nonlinear current and shot-noise, rescaled by the square of Kondo temperature. Inset shows a numerical and analytical approximation of $T_{K}$ (blue and red dashed lines) ($U=3, \Gamma=0.025, T=0$, energies are given in the units $W/50$). Dark cyan curve in the inset of (c) presents $T_{K}$ for $\Gamma=0.05$.}
\end{figure}

Figure \ref{fig1} concerns symmetries of even rank SU(N$=2,4,6$). Some aspects regarding the first two symmetries have been discussed earlier e.g. in \cite{Makarovski2007,Lim2006,Keller2014,Choi2005,Herrero2005,Delattre2009,LeHur2009,Teratani20201}, and here we present these results  mainly for the sake of completeness of the discussion and for comparative purposes. The chosen ratio $U/\Gamma=120$ ensures the strong correlation limit of the analyzed systems. As earlier mentioned for $\Gamma=1-5$ meV the value of Coulomb parameter $U$ ($U=5-30$ meV) corresponds to carbon nanotubes \cite{Makarovski2007,Herrero2005} and for $U=1.5-4$ meV refers to semiconducting quantum dot systems with $\Gamma=0.1-0.4$ meV \cite{Goldhaber1998,Cronenwett1998}. The calculated  SU(2) conductance per single channel visible in Figure 1(a) for $n = 1$  approaches unitary Kondo limit $G_{\nu}  = e^{2}/h$, which agrees with the predicted value based on the Friedel sum rule linking conductance  with the phase shift, or equivalently with the charge on the dot. The phase shift   $\delta_{\nu}=\pi/2$  determines the total conductance $G =(e^{2}/h)\sum_{\nu} \sin[\delta_{\nu}]^{2} = 2(e^{2}/h)$. The Fano factor in the linear range reduces to zero ($A_{0} =(e^{2}/h)\sum_{\nu} \sin[2\delta_{\nu}]^{2}/2 = 0$).  SU(2) Kondo resonance is pinned at the Fermi level ($\widetilde{E}_{\nu}\approx0$) and  consequently transmission ${\mathcal{T}}_{\nu}\mapsto1$, and thus an absence of partition  noise is observed. As can be seen in Figure \ref{fig1}(a), the general dependence of the Fano factor $F_{0}$ on the gate potential resembles an upside-down curve of the conductance. Although partition noise  for  SU(2) Kondo device  is equal to zero ($F_{0}=0$),  for finite bias voltage transport in this state  is  noisy due to quasiparticle interactions  (Figures \ref{fig1}(a), \ref{fig4}(a)). For $N = 4$  (Figure \ref{fig1}(c)) three plateaus of conductance  are observed, lower $G_{\nu} = (1/2)(e^{2}/h)$   for odd Kondo effects ($n =1,3$)  ($\delta_{\nu} =\pi/4, 3\pi/4$)  and higher for even Kondo effect $G_{\nu}  = e^{2}/h$  ($n = 2 , \delta_{\nu} =  \pi/2$).
For half-filling,  six  local two-particle states participate in the formation  of  Kondo  resonance. In general for a given degeneracy $N$ and occupancy  $n$ the number of  $n$-particle local  states involved in Kondo fluctuations is $N!/(n! (N -n)!)$.  At quarter filling  Kondo resonance is shifted from the Fermi energy and transmission ${\mathcal{T}}_{\nu} = 1/2$, which creates a strong partition noise. For $n = 2$ perfect conductance appears and consequent absence of partition noise. Two characteristic values of  $\delta_{\nu}$  for SU(4) symmetry   reflect in two values of plateaus in the  linear Fano Factors  $F_{0}$  presented on Figure \ref{fig1}(c). In the regions of odd occupancies $F_{0} = 1/2$, whereas  for even occupancies $F_{0}$  vanishes (noiseless,  ballistic transport for ultra-low voltages).  The noise measurements carried in the single wall carbon nanotube based quantum dots (SU(4))  by  Delattre et al. \cite{Delattre2009} confirm this result and prove, that  $F_{0}$ calculated by  SBMFA, describing shot noise of renormalized independent particles, accurately estimates the noise in the low temperature, low-voltage regime. The earlier calculations of partition noise for SU(4) Kondo systems can be found in \cite{LeHur2009,Oguri2022,Teratani2020,Oguri2018,Hata2021,Krychowski2025}. For SU(6) symmetry (Figure \ref{fig1}(d)) Kondo effects occur for regions characterized by occupancies $n = 1,2,3,4,5$.  At half-filling, the resonance is again centered at the Fermi level and twenty local three-particle states participate in  Kondo fluctuations.  Similarly to the earlier discussed symmetries, for $n = 3$, fully transmitting, noiseless channels occur at low voltages  (${\mathcal{T}}_{\nu}\mapsto1$ , $A_{0}\mapsto0$) (Equation (\ref{6})). Conductances  of  SU(6) Kondo states per single quantum channel for different gate voltages are  $G_{\nu} =(e^{2}/h)\sin[\delta_{\nu}]^{2}=1/4(e^{2}/h)$, $\delta_{\nu}=\pi/6$ for $n = 1,5$, $G_{\nu} = 3/4(e^{2}/h)$, $\delta_{\nu}=\pi/3$ for $n = 2,4$, and $G_{\nu} = e^{2}/h$, $\delta_{\nu}=\pi/2$ for $n = 3$.   Outside half filling region partition noise of SU(6) Kondo states  is finite. Insets of Figure \ref{fig1}(a,c,d) present Kondo temperatures  $T_{K}=\sqrt{\widetilde{E}^{2}_{\nu}+\widetilde{\Gamma}^{2}_{\nu}}$. For each degeneracy $N$ the lowest value of Kondo temperature occur at half-filling. It is related to the fact that for $n = N/2$  the electrode channels link  with the  largest number of local states $M$  ($M > N$). This indicates that the number of states participating in fluctuations or the number of relevant bosonic fields decisively influence on the values of coefficients  $\Lambda^{n}_{n-1}$ (\ref{B2}) and  $\Lambda^{n+1}_{n}$ (\ref{B4}) and for half filling it is maximal.   The red dashed lines shown in the insets correspond to the analytical slave boson formula (Equation (\ref{3})) expressing Kondo temperature through the bare parameters of Anderson model. The solid blue lines present numerical calculations based on the formula $T_{K}=\sqrt{\widetilde{E}^{2}_{\nu}+\widetilde{\Gamma}^{2}_{\nu}}$, which is similar to (\ref{B10}).
The agreement of analytical solution with the numerical calculations is satisfactory. One should exclude in comparison the boundary regions between Coulomb valleys, because they are poorly described within SBMFA. While the SBMFA picture of the free quasiparticles is correct near the equilibrium state, for description of  the enhancement of current fluctuations beyond the equilibrium it is necessary to take into account the residual interactions between quasiparticles responsible for the two-particle scattering processes. The importance of residual interactions is manifested in the increase of Wilson ratio in the Kondo range, for SU(2) symmetry it reaches value   $W_{\nu\nu'} = 2$ (Figure \ref{fig1}(a)). Similarly, Wilson ratios for Kondo states in SU(4) and SU(6) symmetries are $W_{\nu\nu'} = 4/3$ (Figure \ref{fig1}(c)) \cite{Nishikawa2012} and  $W_{\nu\nu'} = 6/5$ (Figure \ref{fig1}(d)) respectively.  For the four fold degeneracy, both spin and orbital pseudospin Wilson coefficients  are equal. Wilson ratios decrease with increasing $N$ signaling the weakening of correlations. It is worth noting, that approaching the regions of occupancies $n = 0$ or $n = N$, $W_{\nu\nu'}$ tends to one, which means that, the picture  of non-interacting particles gradually begins to apply there.  At the borders of Coulomb valleys (mixed valence range) Wilson ratios still take significant values. This indicates, that significant correlations persist also in these areas due to the proximity of strongly correlated regions. The key source of nonlinear noise is backscattering.  Both, the flow of single quasiparticle and pairs can be viewed as a current carried by two different charges e and 2e and they are  backscattered with different probabilities \cite{Sela2006,Mora2008}. Simultaneous presence of one- and two-particle scattering can be interpreted in the language of the universal effective average charge $e^{\star}$ \cite{Sela2006}. A direct measure of an average charge, is the ratio of the shot noise and the backscattered current (nonlinear Fano factor) $F_{K}=S_{K}/2eI_{K}$, $F_{K} = e^{\star}/e$  (fractional shot noise). At half filling $e^{\star}/e = 5/3$ for SU(2) symmetry, $e^{\star}/e = 3/2$  for SU(4)  and $e^{\star}/e = 7/4$ for SU(6) respectively.  The mentioned values has been confirmed in the experiments for SU(2) systems in QDs \cite{Yamauchi2011,Ferrier2017,Ferrier2020} and in CTNQD for SU(4) \cite{Delattre2009,Basset2012}, which confirmed the necessity of taking into account in the discussion of nonlinear noise the effects of interactions. They were neglected in the earlier theoretical discussion \cite{Delattre2009,Lipinski2010,Meir2002}. The backscattering events reflect in the increase of mixed two- and three-body correlation functions (Equations (\ref{13})) and in consequent increase of coefficients  $c_{V,\nu}$  and $c_{S,\nu}$ (Equations (\ref{D1}) and (\ref{D2})), which determine nonlinear current $I_{K}$ and nonlinear noise $S_{K}$. With the increase of bias voltage shot noise terms of power $V^{3}$ begin to dominate over the linear contribution.
The significance of three-body correlations has been confirmed experimentally in \cite{Hata2021}.
\begin{figure}
\includegraphics[width=0.8\linewidth]{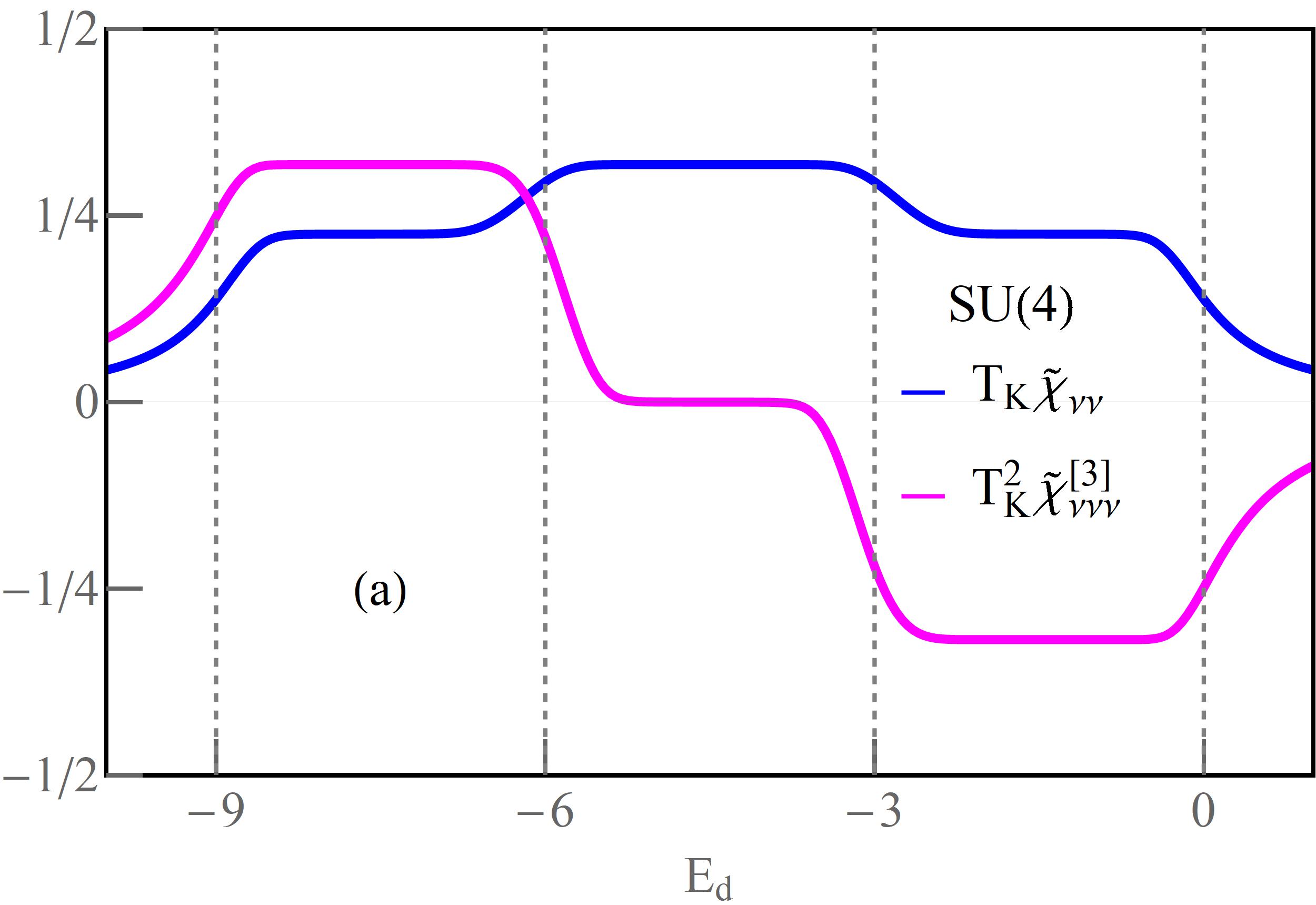}
\includegraphics[width=0.8\linewidth]{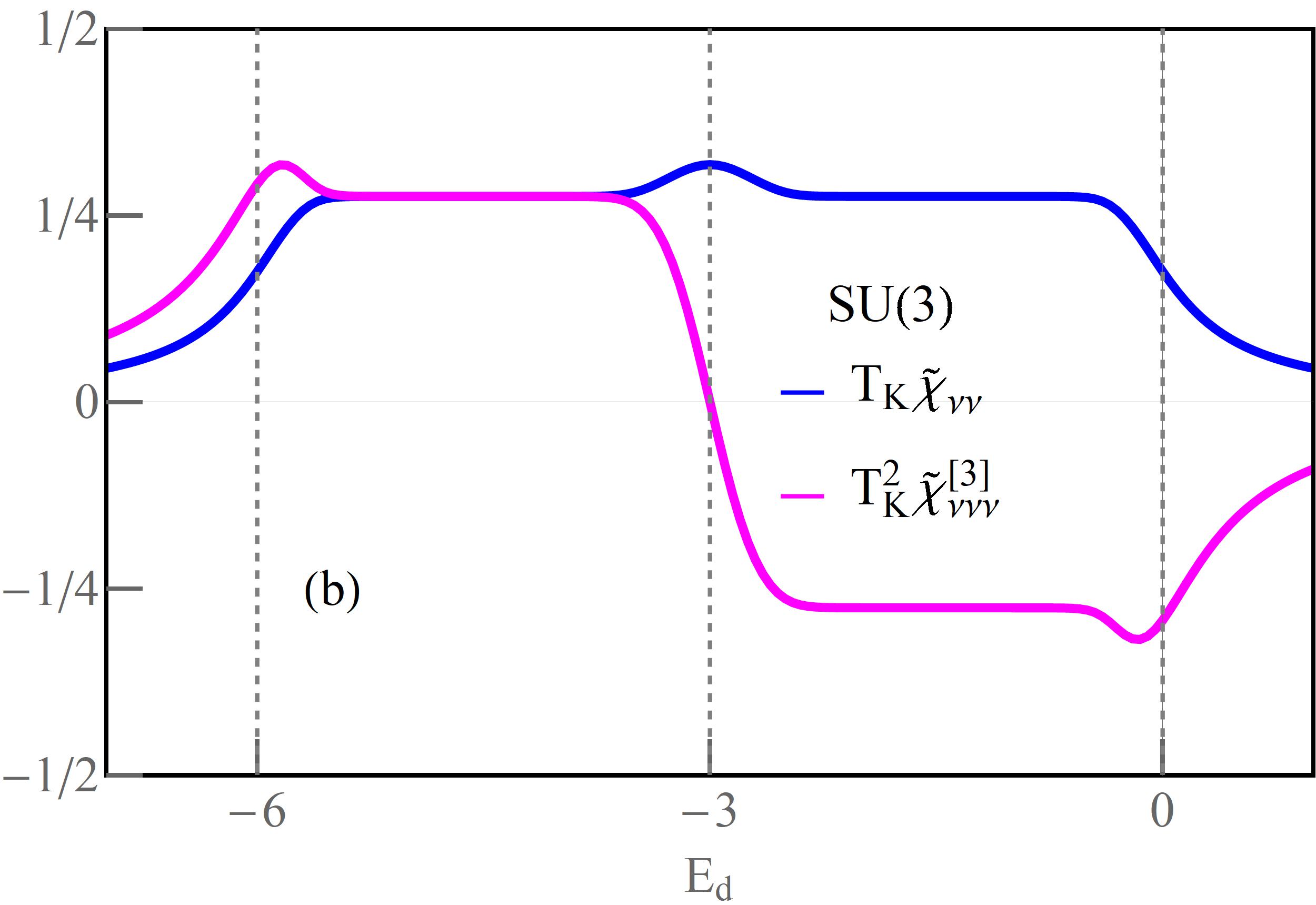}\caption{\label{fig2} Two and three body correlators rescaled by $T_{K}$ for SU(4) and SU(3) Kondo state (blue and magenta lines). The parameters used in calculations are the same as in Figure 1.}
\end{figure}

Figure \ref{fig1}(b) shows the variation of the nonlinear Fano factor $F_{K}$ as a function of the atomic level of the quantum dot $E_{d}$ for the Kondo state with SU(2) symmetry. In the region of the empty state $0e$ and the fully occupied state $2e$, the Fano factor reaches the value $-1$, which is related to the dominant influence of the two body correlation functions $\widetilde{\chi}_{\nu\nu'}$ in $c_{V,\nu}$ (see Equation (\ref{D1})). In the region with one electron $Q=1e$, $F_{K}$ is $5/3$, which is in agreement  with literature reports \cite{Sela2006,Mora2008,Yamauchi2011,Ferrier2016,Oguri2018}.
In this area, $\widetilde{\chi}^{[3]}_{\nu\nu\nu}$ and $\widetilde{\chi}^{[3]}_{\nu\nu'\nu'}$ are close to zero. The three-body correlators, as the odd functions of the gate voltage change the sign in the region with single electron. The green and magenta lines on Figure \ref{fig1}(b) show the values of the nonlinear current normalized by the squared  Kondo temperature $T^{2}_{K}I_{K}=1/2$ and the shot noise $T^{2}_{K}S_{K}=5/6$. This gives the value of nonlinear Fano factor $F_{K} =e^{\star}/e=5/3$. Formally we can write the nonlinear current $I_{K}$ and its fluctuating part $S_{K}$ (first moment of the current correlator) as a sum of two body and three body contributions  $T^{2}_{K}I_{K}=T^{2}_{K}I^{[2]}_{K}+T^{2}_{K}I^{[3]}_{K}$ and similarly $T^{2}_{K}S_{K}=T^{2}_{K}S^{[2]}_{K}+T^{2}_{K}S^{[3]}_{K}$. Near the triple degeneracy points (equality of SB amplitudes $p^{2}_{\sigma}=e$ and $p^{2}_{\sigma}=d$) current $I_{K}$ changes its sign and for $I_{K}=0$, nonlinear Fano factor diverges $F_{K}\mapsto\pm \infty$. This is characteristic for the transition between the Kondo state and Coulomb blockade regime. Around this transition region we observe tiny negative nonlinear shot-noise $S_{K}<0$.  $S_{K}$ is the nonlinear contribution to the shot noise and it might be negative, but the total value $S$ remains positive. In the empty region and for double occupancy $n=2$, nonlinear noise is positive, but small, $S_{K}>0$. There are two pairs  of points below and above  $E_{d} = 0$ and $E_{d} = -U$, for which $F_{K}$ is zeroing  (Figures \ref{fig1}(a,b)).  Similar behavior is also observed for higher symmetries (Figures \ref{fig1}(c,d)) (close to $E_{d}=0$ and $E_{d}=-3U$ or $E_{d}=-5U$). For SU(2) dot near $E_{d} = 0$ and $E_{d} = -U$ also points of  vanishing of current  are  observed ($E^{0}_{d}$) (orange dots on Figure \ref{fig1}(a)). Slightly moving from points $E^{0}_{d}\approx 0,-U$ current $I_{K}$ changes sign in a narrow ranges of gate voltage due to the  three-body correlations (three-body contributions) to coefficients $c_{V,\nu}$  (Figure \ref{fig1}(b)). Around the $S_{K}$ zeroing points there is also a change in the sign of the nonlinear contribution to the noise.  The  giant values of  $|S_{K}|$ observed  in Figures \ref{fig1}(a,c,d) can be interpreted according to the accepted terminology as hyper-Poissonian noise  \cite{Oguri2018,Krychowski2025}.  Careful insight into Figure \ref{fig1} reveals that this strong enhancements of $F_{K}$  is due to the occurrence of very small current in these regions. For even symmetries SU($N =2,4,6$)  at half-fillings nonlinear noise factor  $F_{K}$  takes the value $F_{K}=(8+N)/(4+N)$. The equivalent expression can be found in \cite{Teratani2020}.\newline
As earlier mentioned \cite{Sela2006,Mora2008,Ferrier2016,Oguri2018}, three body correlations vanish at e-h symmetric point and around it they are small and change signs.  In this region it is the two-body correlation $\widetilde{\chi}_{\nu\nu'}$ that modify the current $I_{K}$  and noise $S_{K}$. For $n = 1, 3$  in SU(4) system $F_{K} = 1/3$, the noise is reduced and there, of particular importance are three body correlation functions  $\widetilde{\chi}^{[3]}_{\nu\nu\nu}$   and $\widetilde{\chi}^{[3]}_{\nu\nu'\nu'}$ (Figure \ref{fig2}(a)). In the $Q=1e,5e$ charge regions of SU(6) system, the linear Fano factor is $F_{0}=3/4$, and for high voltages the Fano factor is reduced to $F_{K}=1/20$ (sub-Poissonian noise). Again responsible for this behavior are the three-particle correlations $\widetilde{\chi}^{[3]}_{\nu\nu'\nu'}$ dominating in these sectors. In the charge regions $Q=2e, 3e$ there is a reversed behavior, increase of $F$, $F_{0}=1/3,0$, and $F_{K}=21/20, 7/5$ respectively, indicating that the residual quasiparticle interactions in the Kondo system cause the super-Poissonian shot noise. Since the interplay of two- and three-body correlations functions decides about value of nonlinear Fano factor (Equations (\ref{D1}-\ref{D3})) we present on Figures \ref{fig2}(a,b) the gate dependencies of these correlations for exemplary chosen even - SU(4) and odd - SU(3) symmetries. The quantities have been rescaled by the Kondo temperature and its square, respectively (see Equations (\ref{B8}) and (\ref{B9})). $T_{K}\widetilde{\chi}_{\nu\nu}$ is an even function of the gate voltage, whereas $T^{2}_{K}\widetilde{\chi}^{[3]}_{\nu\nu\nu}$ is an odd function with a negligible contribution around the e-h symmetry point. The small peaks visible on the correlators for $SU(3)$ symmetry occur in the region of Coulomb blockade boundaries, similar peaks appear also for other odd symmetries (not presented). As can be seen in the following Figures, this feature is reflected in the gate dependencies of other physical quantities for odd symmetries.

\subsection{Odd symmetries of Kondo states}
Figure 3 shows results for Kondo effects with odd degeneracies ($N =3, 5$). The e-h symmetry points in these cases correspond to non-integer occupancies ($n = 3/2$, $n = 5/2$) and high conductances  at these points are associated with the charge resonances centered at the Fermi level.  The dependence of conductance on gate potential at half-filling is characterized by the occurrence of a peak instead of a plateau as in the case of Kondo resonances.  In general, however, at e-h symmetric points linear conductance  for any value of N, regardless of the type of resonance at the Fermi level, takes the unitary limit. For fillings ($n =1,..N-1$) conductance plateaus are observed, which  manifest Kondo effects for these occupations. For SU(3) symmetry the value of conductance for $n = 1$ is $(3/4)(e^{2}/h)$  and correspondingly total conductance $G=(9/4)(e^{2}/h)$, and what's interesting it is the highest total value in single charge sector among SU(N) Kondo systems. The curves of linear Fano factors around  half fillings form the  deeps with $F_{0} = 0$ for e-h symmetric points and in Kondo regions $F_{0}$ take the values according to the known relationship $F_{0}=1-{\mathcal{T}}_{\nu}=\frac{\widetilde{E}^{2}_{\nu}}{T^{2}_{K}}$.  Wilsons ratios decrease with the increase of N (weaker correlations) and are reduced in the areas of charge  fluctuations, but what is worth noting, they still have significant values there.  Apart from  singularities of $F_{K}$ occurring for $I_{K} = 0$,  points located close to empty ($n = 0$) or full occupied regions ($n = N$), an enhancement  of $F_{K}$  near the e-h symmetric point is observed $F>1$ (the super-Poissonian noise). For $N = 3$, $n\approx3/2$  three  single electron states become degenerate with three two-electron states. The probability of occurrence of pairs of quasiparticles is growing in this region, which is associated with an increase in the probability of  backscattering of pairs. For $N=5$ similarly as for other cases, the numerically calculated plateaus of Kondo conductance for given charge sectors agree with  formula  $G_{\nu}  = (e^{2}/h)\sin^{2}[Q_{\nu}]$.

For integer values of charge, Wilson ratio $W_{\nu\nu'}-1 = 1/4$ and at the boundaries of charge sectors $W_{\nu\nu'}-1 = 1/5$. For occupancies  $n = 1,4$ i.e. for ranges close to the  empty or fully occupied  regions, probability of occurrence of pairs is low and the backscattering of pairs weakens so much, that the nonlinear Fano coefficient $F_{K} = 1/5$  becomes smaller than linear noise factor $F_{0} = 2/3$. For higher occupancies $n = 2,3$ pairs of quasiparticles are starting to play a significant role and, as we have checked numerically, both  pair correlation functions  $\widetilde{\chi}_{\nu\nu'}$  and nonlinear correlations $\widetilde{\chi}^{[3]}_{\nu\nu'\nu'}$ significantly contribute to the nonlinear noise and  $F_{K}$ in this range reaches value  $F_{K} = 5/4$  and dominates the linear noise ($F_{0} = 1/8$).
\begin{figure}
\includegraphics[width=0.8\linewidth]{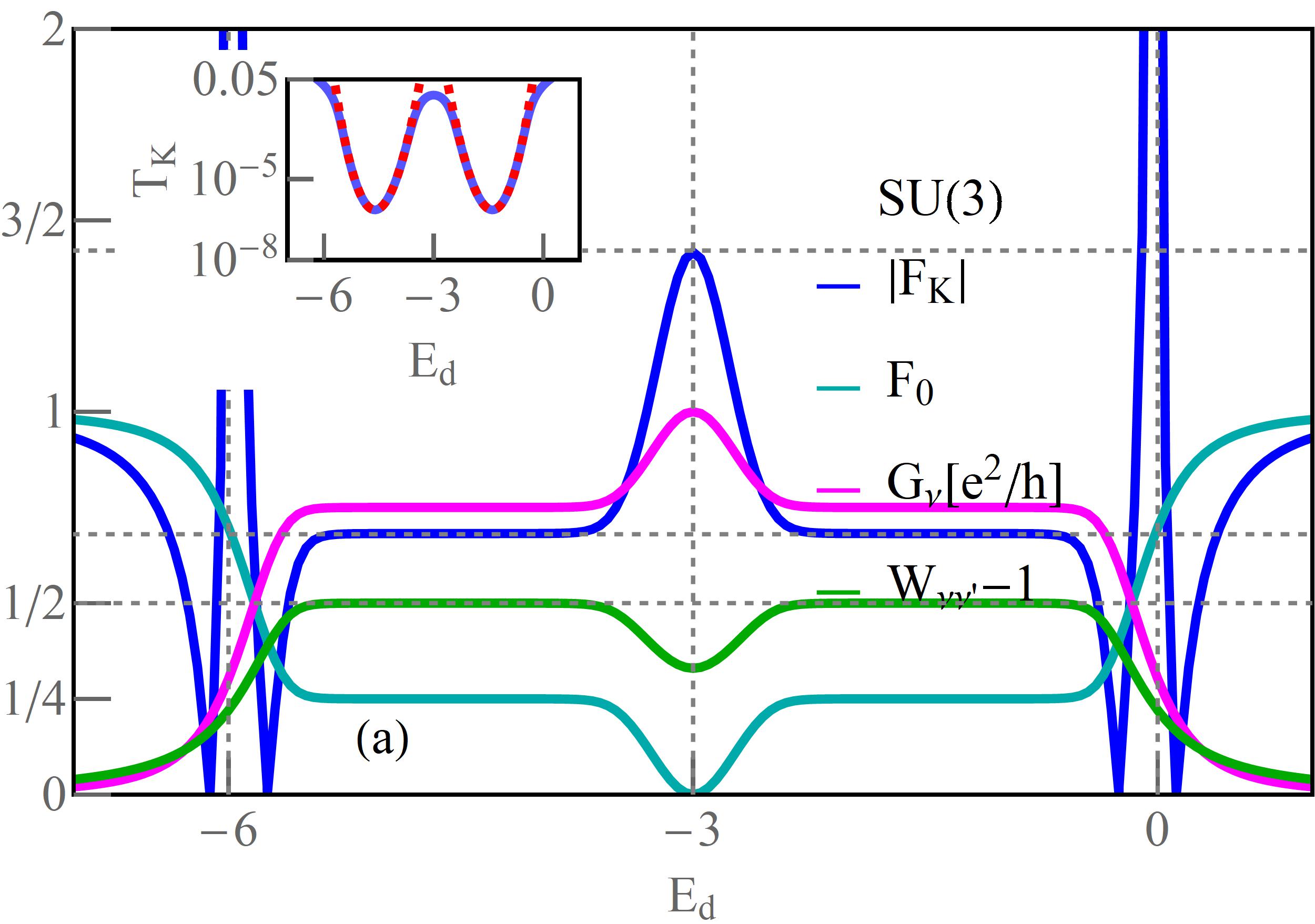}
\includegraphics[width=0.8\linewidth]{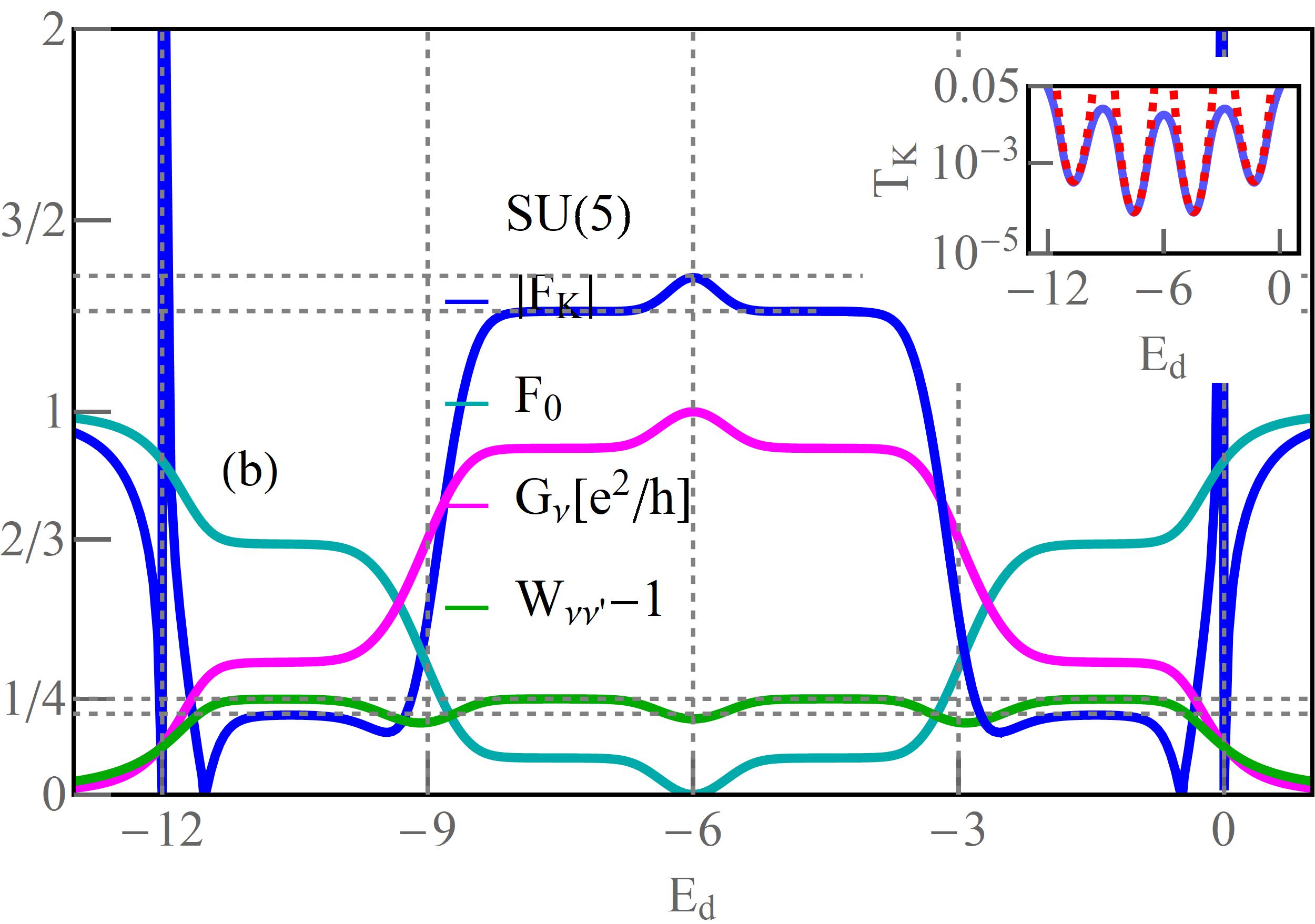}\caption{\label{fig3} Odd SU(3) and SU(5) Kondo effect: (a,b) Gate-dependent $F_{0(K)}$, $G_{\nu}$ and $W_{\nu\nu'}-1$. Insets show a numerical and analytical approximation of $T_{K}$ (blue and red dashed lines) ($U=3, \Gamma=0.025, T=0$).}
\end{figure}

\subsection{Equilibrium versus non-equilibrium noise}
\begin{figure}
\includegraphics[width=0.8\linewidth]{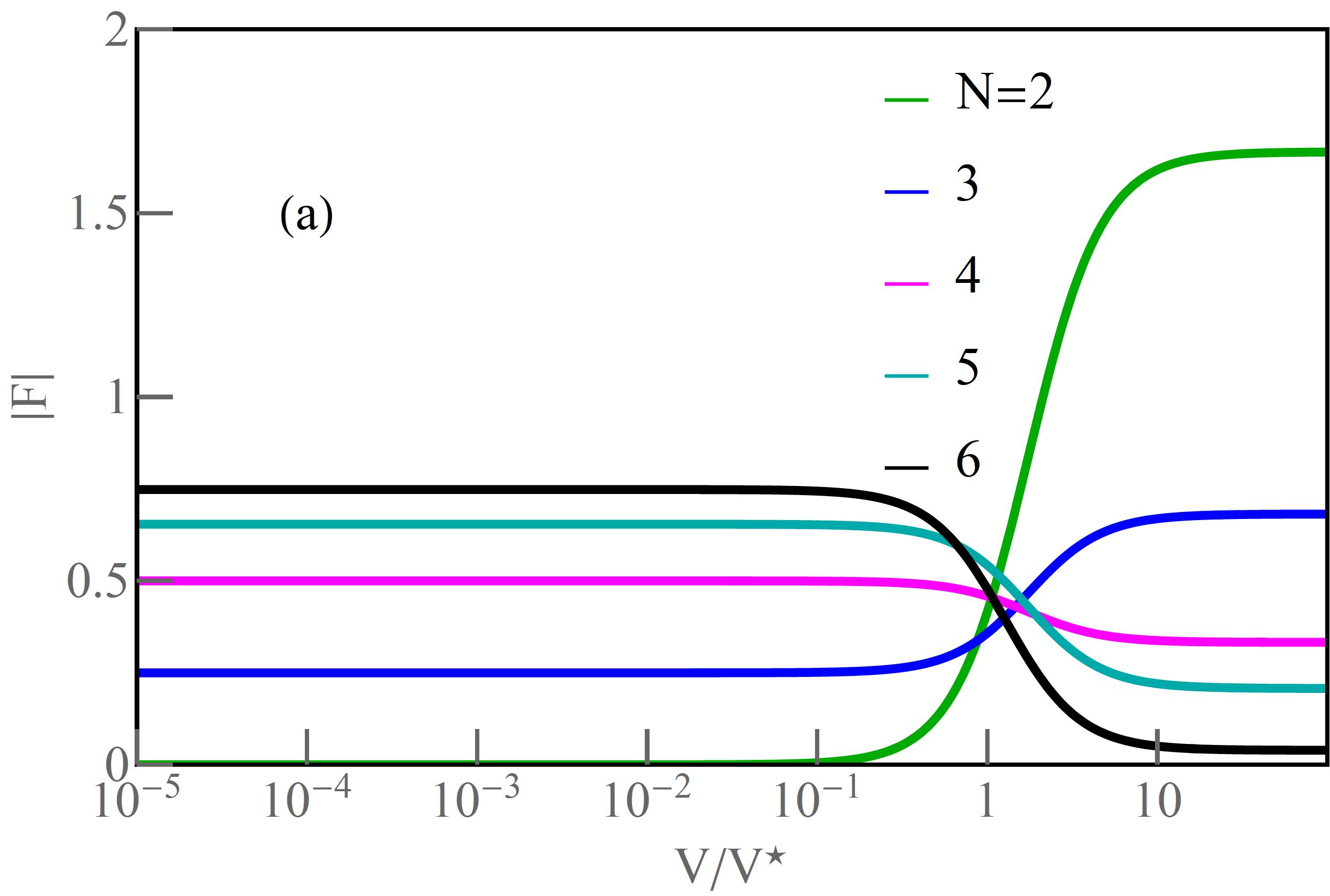}
\includegraphics[width=0.8\linewidth]{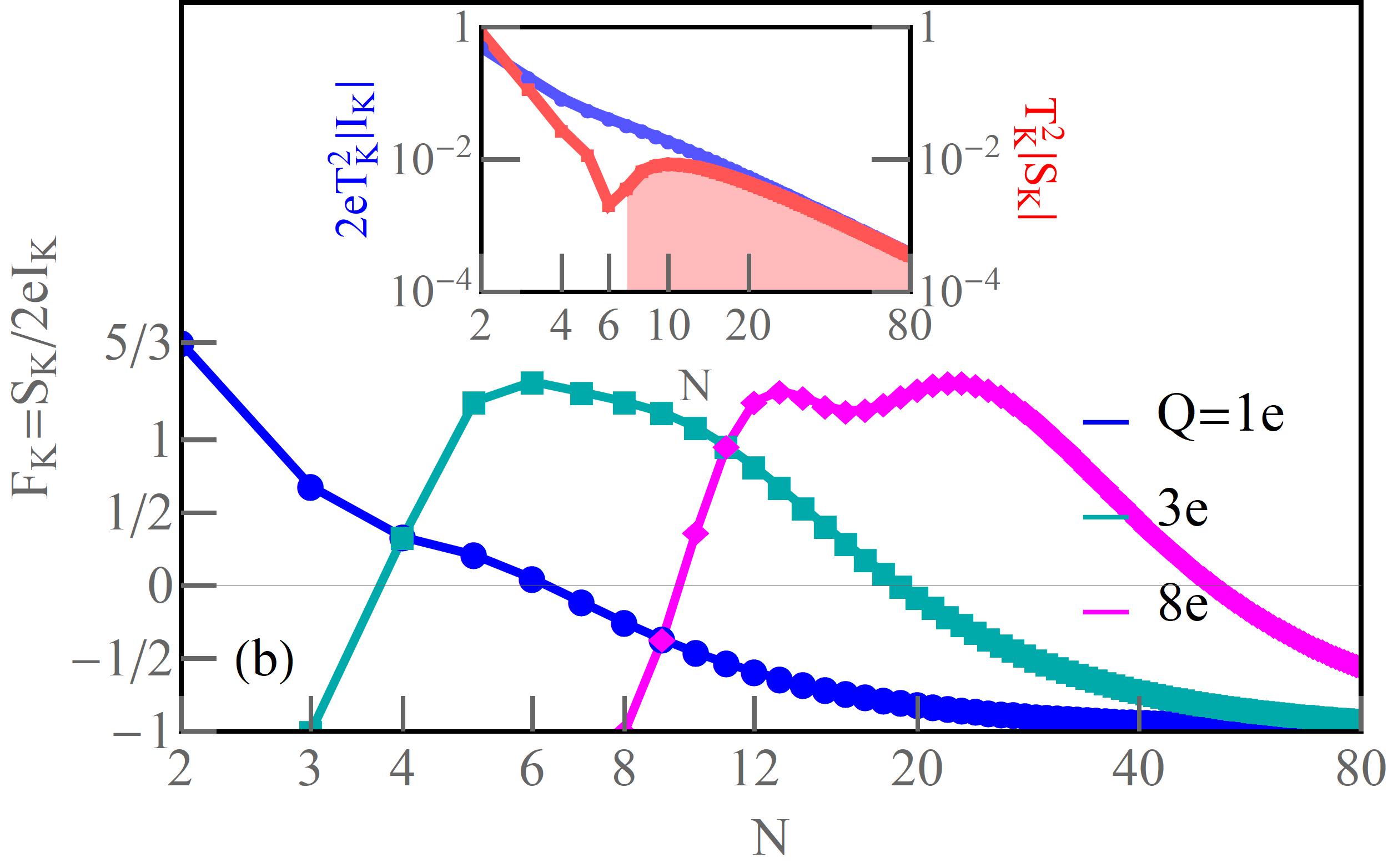}\\
\includegraphics[width=0.8\linewidth]{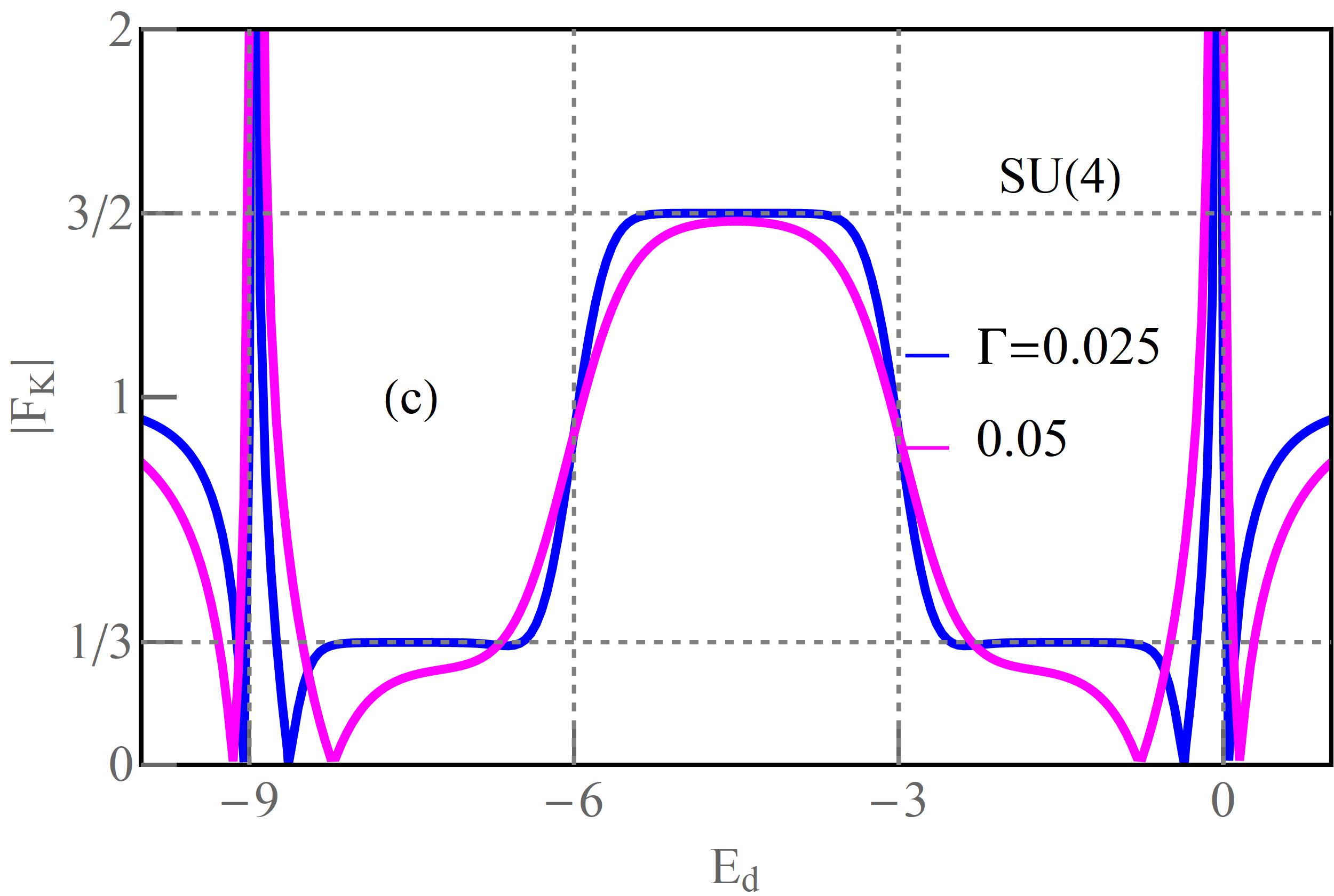}
\includegraphics[width=0.8\linewidth]{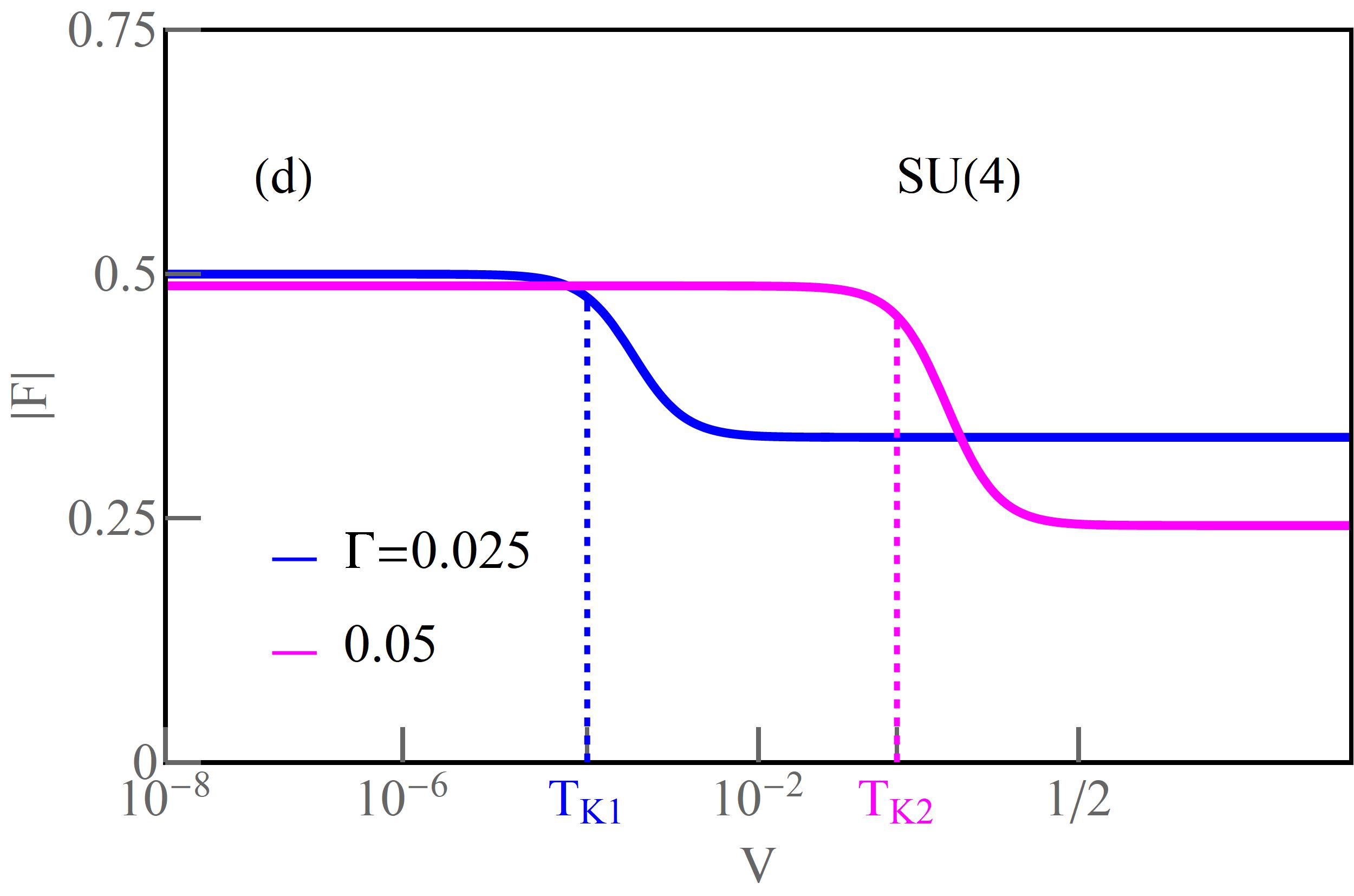}\caption{\label{fig4} (a) Total Fano factor $|F|$ versus $V/V^{\star}$ for SU(N) Kondo states ($Q=1e$). (b) Nonlinear Fano factor $F_{K}$ as a function of degeneracy number N for $Q=1e, 3e, 8e$ ($\Gamma=0.025$, $V/V^{\star}=10$). Inset compares $2eT^{2}_{K}I_{K}$ and $T^{2}_{K}|S_{K}|$ with an increase of N (blue and red lines, $Q=1e$). The area filled in red indicates $S_{K}<0$. (c,d) $F_{K}$ as a function of $E_{d}$ and bias-dependent $|F|$ for SU(4) Kondo state ($U=3$, $T=0$). $T_{K1}$ and $T_{K2}$ in Figure (d) represent two Kondo temperatures for $V = 0$.}
\end{figure}

Figure \ref{fig4}(a) illustrates voltage dependencies of the total Fano factor $F(V/V^{\star})$ for exemplary chosen filling $n =1$ (characteristic value $V^{\star}$ is defined in Appendix A). In the linear regime Fano factors $F_{0}$ depend only on symmetry and they  increase with the increase of degeneracy.
This tendency reflects the increasing shift of the Kondo  resonance from the Fermi level with the growth of $N$.
For $N = 2,3$, where residual interactions are stronger than for higher degeneracies, one observes the increase of $F$ with voltage and its  saturation  for high values of  $V$,  $F = 5/3,15/22$ respectively.
For $N\geq4$ , $F$ at high voltages  tends  to  very small values  $F\mapsto1/3,0.316,1/20$ for $N = 4,5,6$. These limits are solely determined by symmetry, which directly shows Equation (D4). In this equation, $|F|$ depends on degeneracy $N$ and the Wilson ratio which is directly related to symmetry $W_{\nu\nu'}=1+1/(N-1)$.
\begin{figure}
\includegraphics[width=0.8\linewidth]{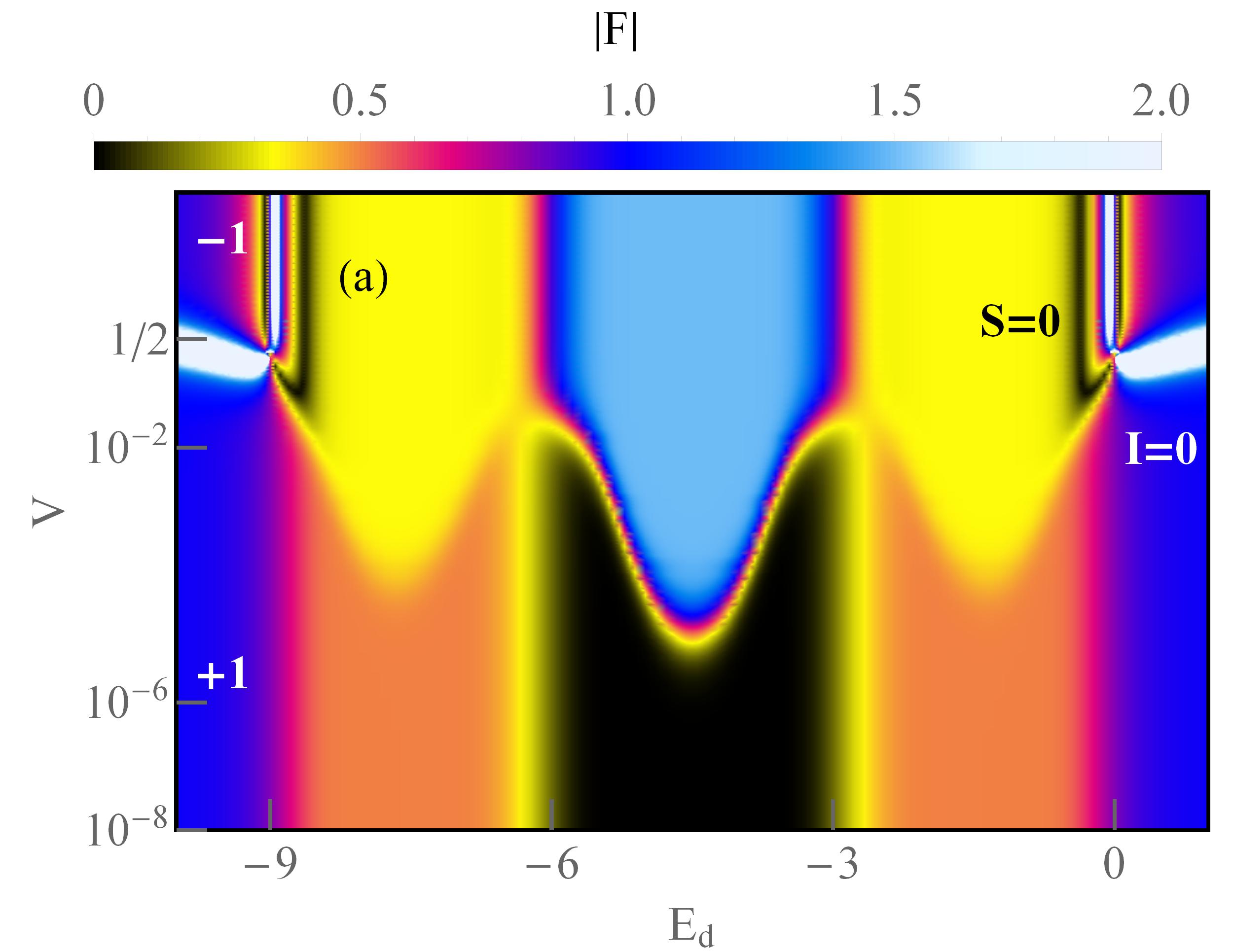}
\includegraphics[width=0.8\linewidth]{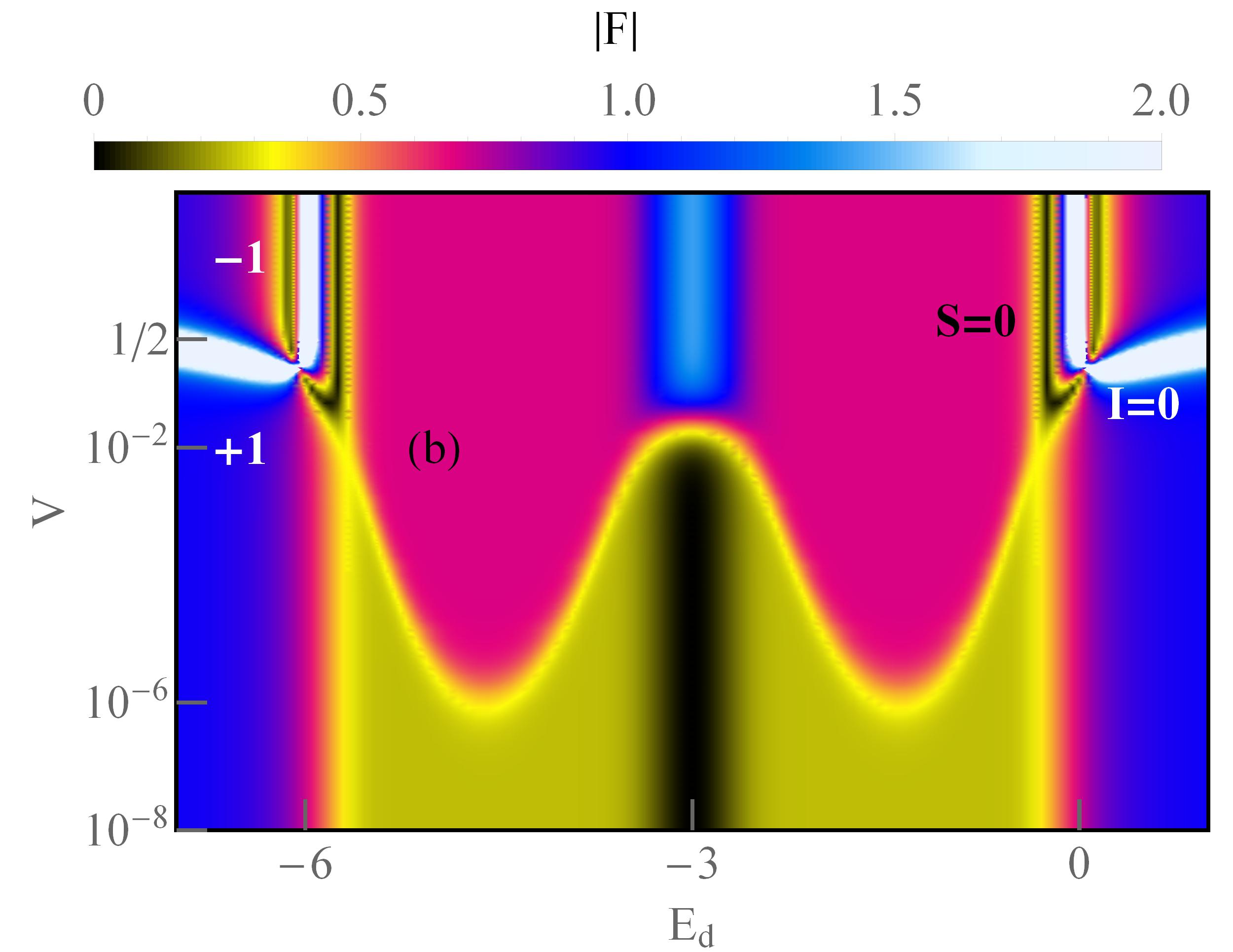}
\caption{\label{fig5} (a,b) Density plots of gate-dependent total Fano factor |F| as function of bias voltage $V$ for SU(4) and SU(3) Kondo states  ($U=3, \Gamma=0.025, T=0$). $\pm1$ indicates the sign of the Fano factor.}
\end{figure}

Figure \ref{fig4}(b) compares the dependencies of nonlinear Fano factor on the rank of SU(N) group for the selected occupancies.
$F_{K}>1$ indicates the dominance of bunching processes. This occurs for $Q=N/2$. In the systems with a larger number of electrons, the ranges in which quasiparticle grouping processes dominate ($F_{K}>1$) shift to higher degenerations (N) (Figure \ref{fig4}(b)).
For $N=3$, the noise lowers $|S_{K}|<2e|I_{K}|$.
For $N=7$, the influence of three-particle correlation becomes dominant $\widetilde{\chi}^{[3]}_{\nu\nu\nu}>\widetilde{\chi}_{\nu\nu}$, which consequently leads to negative noise ($S_{K}<0$, inset of Figure \ref{fig4}(b)).
For $Q=3$e and 8e, and degenerations $N=3$ and $N=8$, the nonlinear Fano factor takes the value $F_{K}=-1$, because two-particle effects in the current dominate there and $I_{K}<0$.
It is interesting to note the formation of a maximum for $N=6$ ($Q=3e$) and two maxima for $Q=8e$ ($N=13,23$). These effects have no fundamental cause, they occur only due to the trigonometric functional dependencies of the phase shifts: $2\delta_{\nu}$ and $4\delta_{\nu}$ in the nonlinear noise coefficient (see Formula \ref{D2}). The nonlinear contribution to noise (D4) leads to $F_{K}=-1$ in the limit $N\mapsto\infty$ (Figure \ref{fig4}(b)).

The inset in Figure \ref{fig4}(b) shows the nonlinear current and noise as a function of N, scaled by $T_{K}$ for $Q=1e$.
The calculations were performed for a voltage value of $V/V^{\star}=10$.
In this range, $T_{K}^{2}I_{K}\approx-T_{K}^{2}S_{K}$, the negative value of the noise is close to the non-linear value of the current and therefore $F_{K}=-1$ (area shaded in red for $N>=7$, inset \ref{fig4} (b)).
For $N=2$, super-Poisson processes are dominant, consequently $T^{2}_{K}S_{K}>2eT^{2}_{K}I_{K}$ (inset 4(b)), when $N>2$ there is a reversal of the trend $T^{2}_ {K}S_{K}<2eT^{2}_{K}I_{K}$. This is due to the increase of the number of three-particle correlators, which grows with the number of quantum channels. In the case of $Q=1e$ it happens for $N>=7$.

Figures \ref{fig4}(c,d) show the effect of the strength of coupling $\Gamma$ on $|F|$.
An increase of $\Gamma$ increases Kondo temperature (vertical dashed lines in Figure \ref{fig4}(d)).
It causes also a departure from the strictly unitary limits of SU(4) Kondo effect, i.e., from $F_{0}=1/2$ and $F_{K}=1/3$. The strongest modification occurs for $Q=1e,3e$, where the charge boundaries $1e/0e$ and $3e/4e$ are closest. The region is broadened
due to the three-particle correlations. For half-filling, we observe a slight deviation from the value $F_{K}=3/2$, due to slight drop of the Wilson ratio $W_{\nu\nu'}=4/3$.

The density maps in Figure \ref{fig5} show the Fano factor $F$ as a function $V$ and $E_{d}$ for SU(4) and SU(3) symmetries.
We observe the oscillating line, around $V\approx V^{\star}$, separating the linear region $F_{0}$ from the non-linear Fano factor $F_{K}$.
This line corresponds to characteristic voltage $V^{\star}=\sqrt{\frac{G_{0}}{3c_{V,\nu}}}\approx T_{K}$, where the slope saturates up to $F_{K}$. For $V>V^{\star}$ the picture of non-interacting Kondo particles falls down and nonlinear current and shot-noise are significant (see Appendix A and D).

For SU(4) Kondo state, regions 1e and 3e show the change from $F_{0}=1/2$ (orange sector) to $|F_{K}|=1/3$ (yellow sector). In the case of half-filling $Q=N/2=2e$, there is a transition from $|F|=F_{0}=0$ up to $|F|=|F_{K}|=3/2$. For SU(3) balance of two-body ($\widetilde{\chi}_{\nu\nu}$) and three-body correlations ($\widetilde{\chi}^{[3]}_{\nu\nu\nu}$) in shot-noise reverses the trend, in this case $F_{0}=1/4$ (yellow sector) and $F_{K}\approx0.68$ (magenta sector). This opposite behavior in the case of these two symmetries is explained by Figure \ref{fig2}, where we see that for SU(4) $\widetilde{\chi}^{[3]}_{\nu\nu\nu} <\widetilde{\chi}_{\nu\nu}$, whereas for the SU(3) symmetry $\widetilde{\chi}^{[3]}_{\nu\nu\nu}=\widetilde{\chi}_{\nu\nu}$. The second important result is the transition from $F=+1$ to $F=-1$ for the empty and fully occupied states, where the total current is suppressed at the boundary ($I=0$, bright white horizontal line). In the region, where $F_{K}$ dominates, we observe a characteristic vertical white line, along this line nonlinear current vanishes $I_{K}=0$. This white line is surrounded by zeroing contour of nonlinear noise $S_{K}=0$ (black envelope at the boundaries $1e/0e$ and $(N-1)e/Ne$).

\subsection{Susceptibility and entropy}
\begin{figure}[h!]
\includegraphics[width=0.8\linewidth]{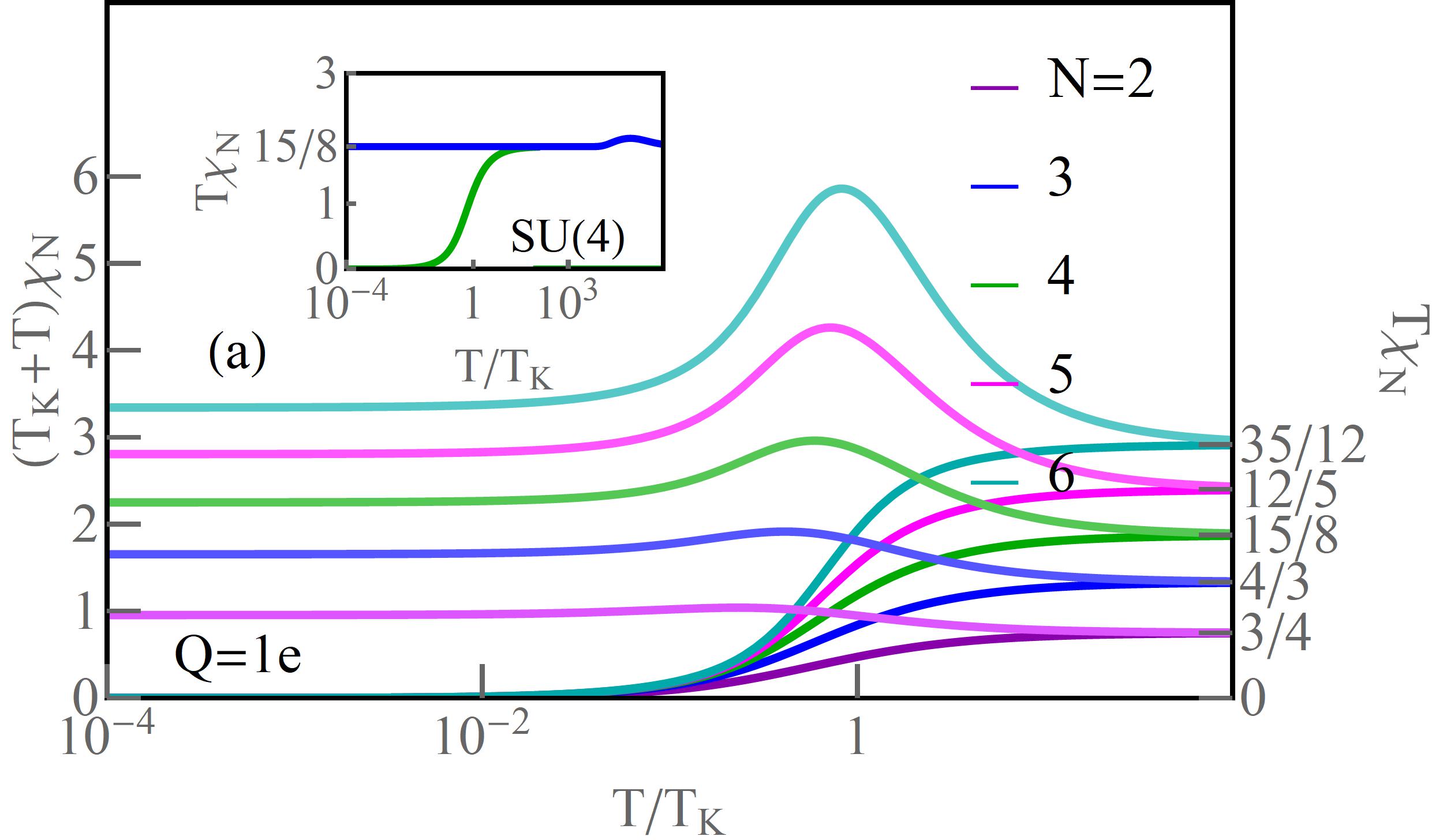}
\includegraphics[width=0.8\linewidth]{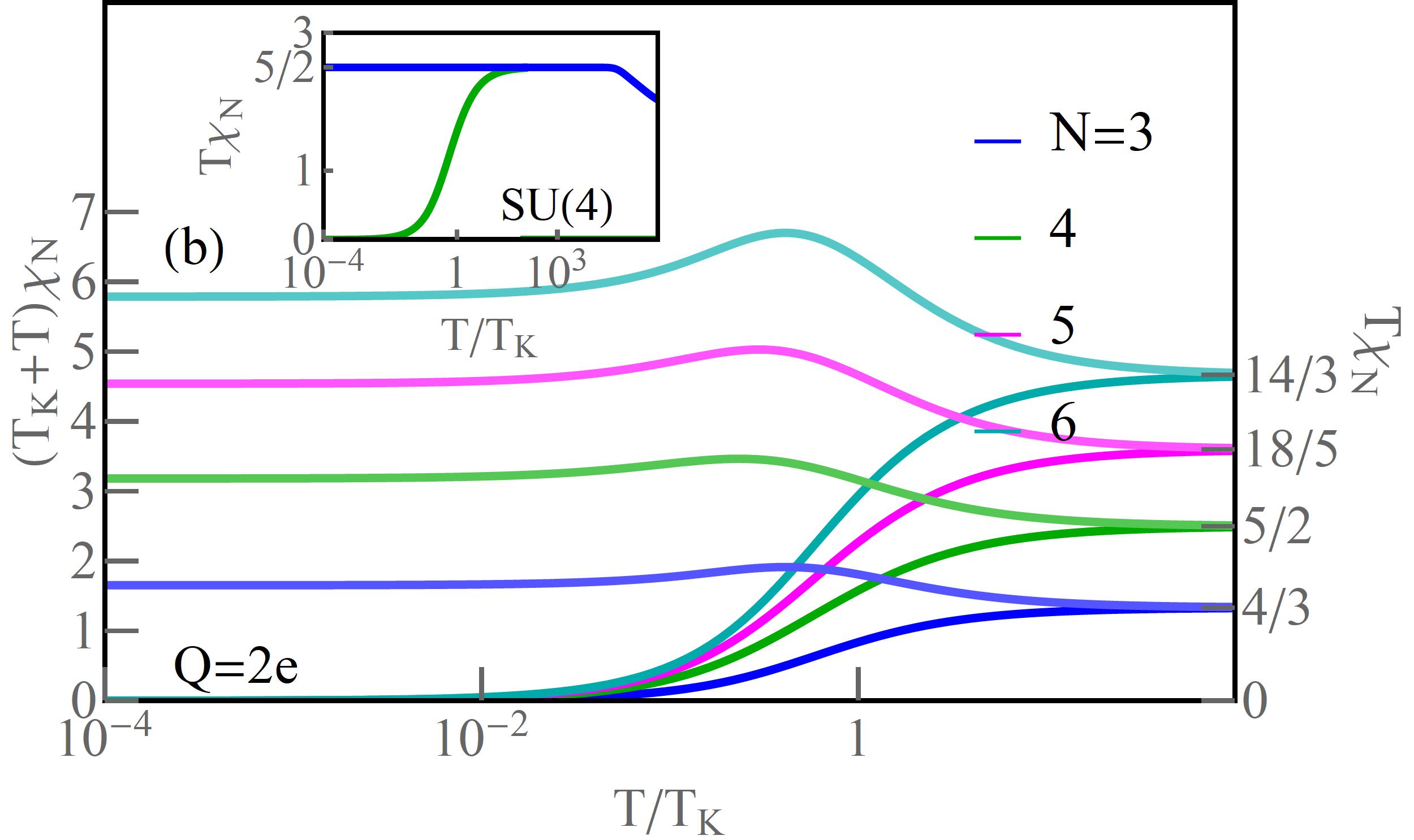}\\
\includegraphics[width=0.8\linewidth]{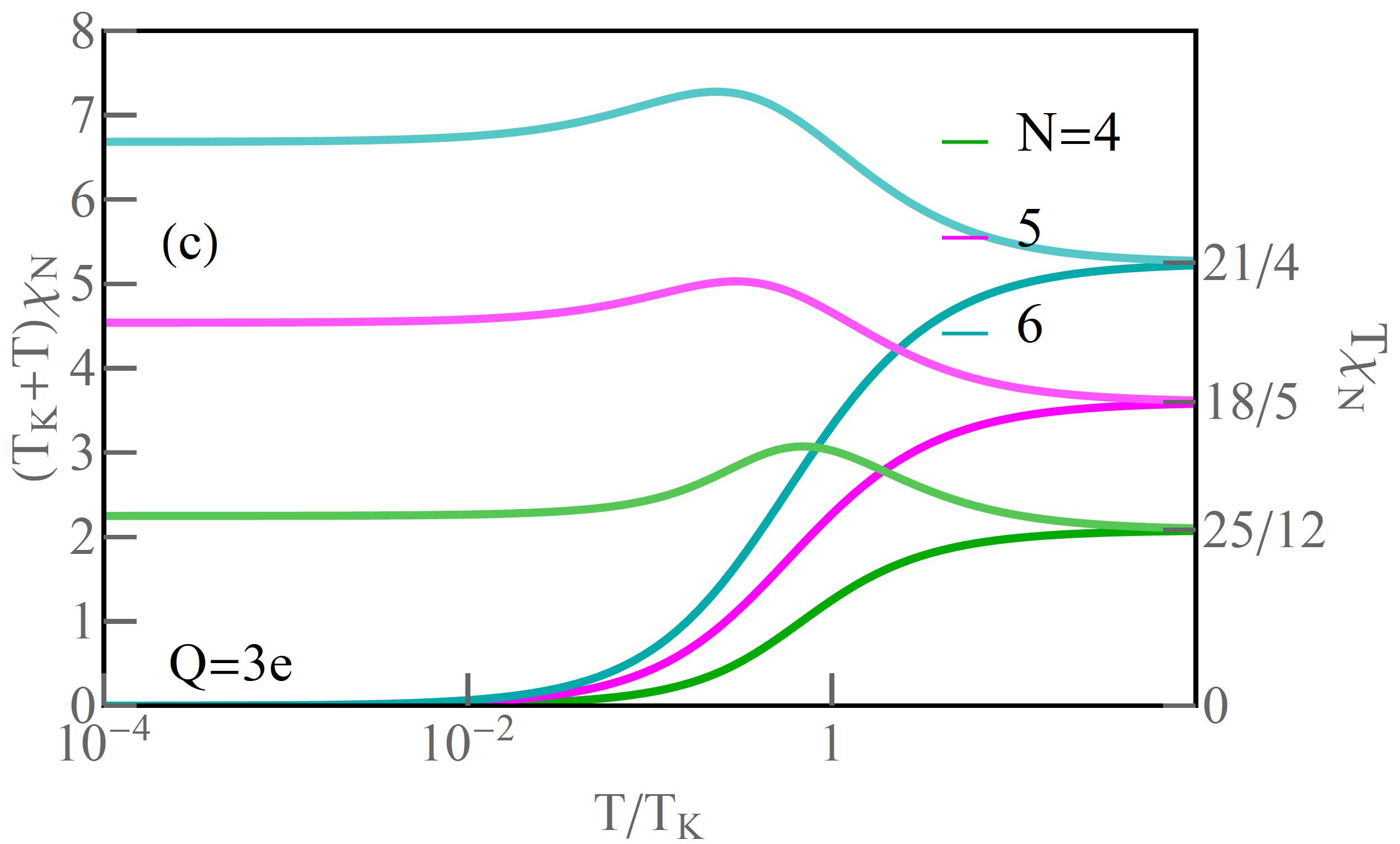}\\\caption{\label{fig6} (a-c) Generalized spin $T\chi_{N}$ and $(T_{K}+T)\chi_{N}$ versus $T/T_{K}$ (dark and light lines). The insets illustrate temperature dependencies of isolated and screened SU(4) spin for $Q=1e$ and $Q=2e$ respectively (blue and green lines) ($U=3, \Gamma=0.025, V=0$).}
\end{figure}
\begin{figure}[h!]
\includegraphics[width=0.8\linewidth]{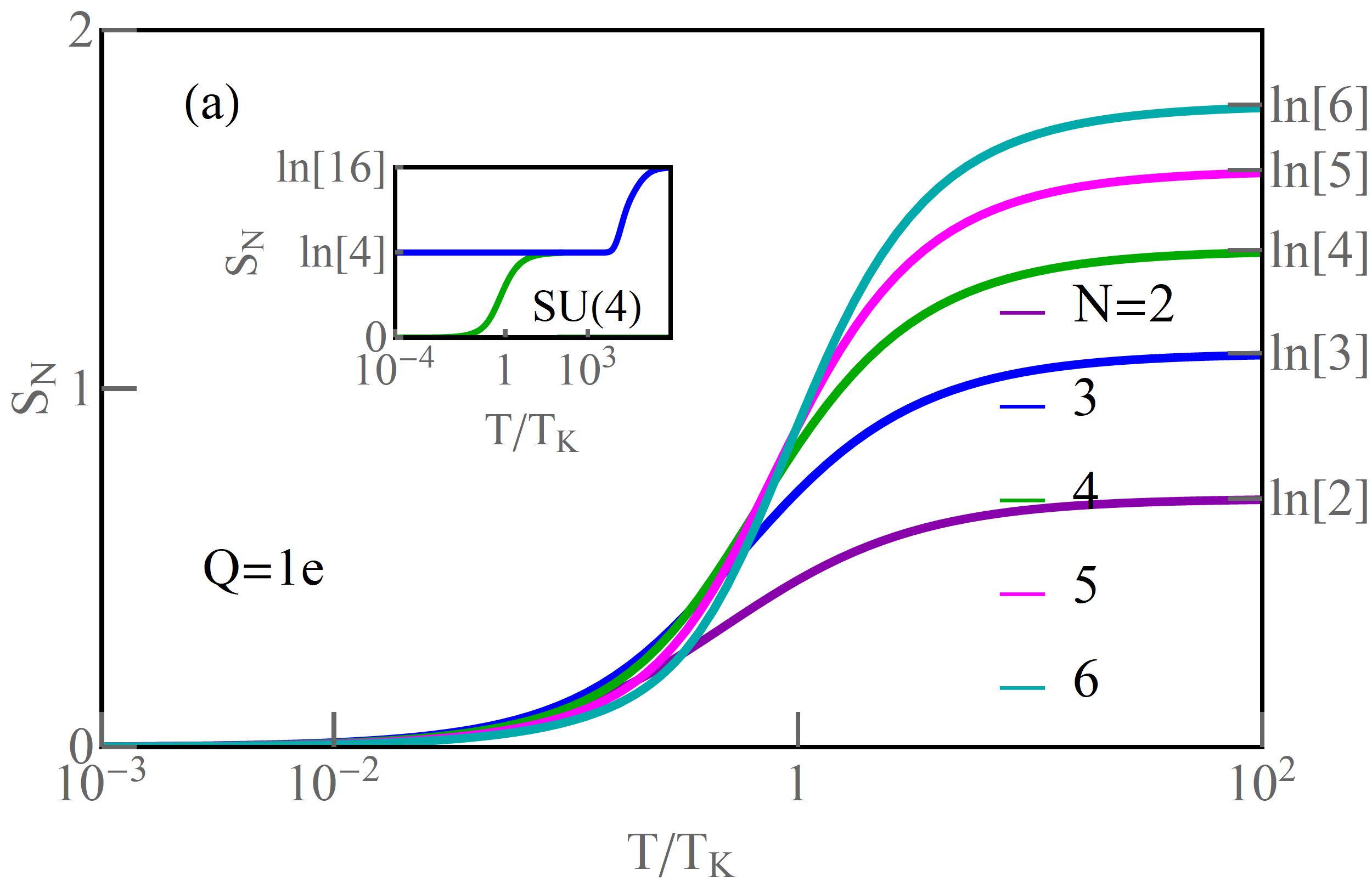}
\includegraphics[width=0.8\linewidth]{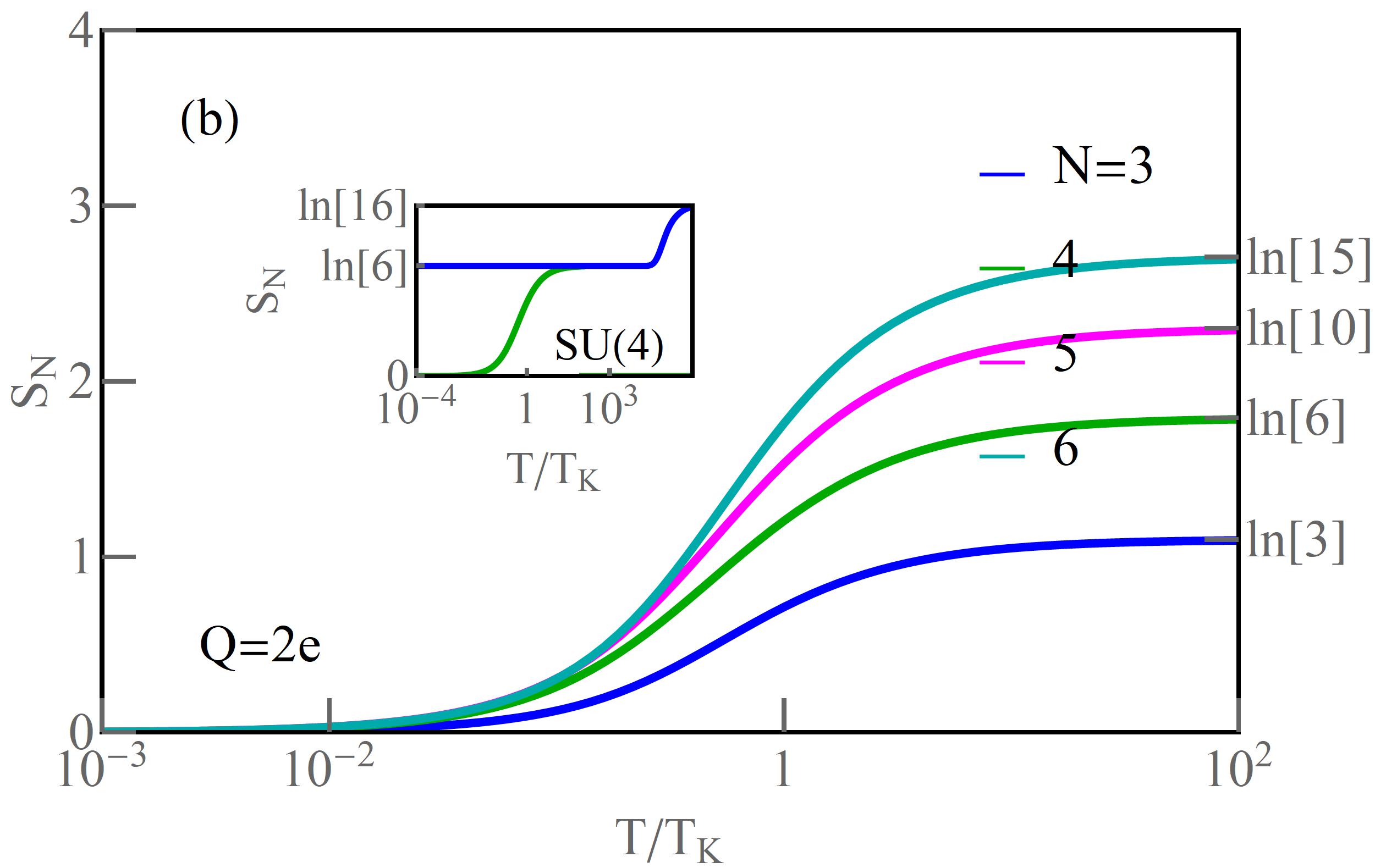}\\
\includegraphics[width=0.8\linewidth]{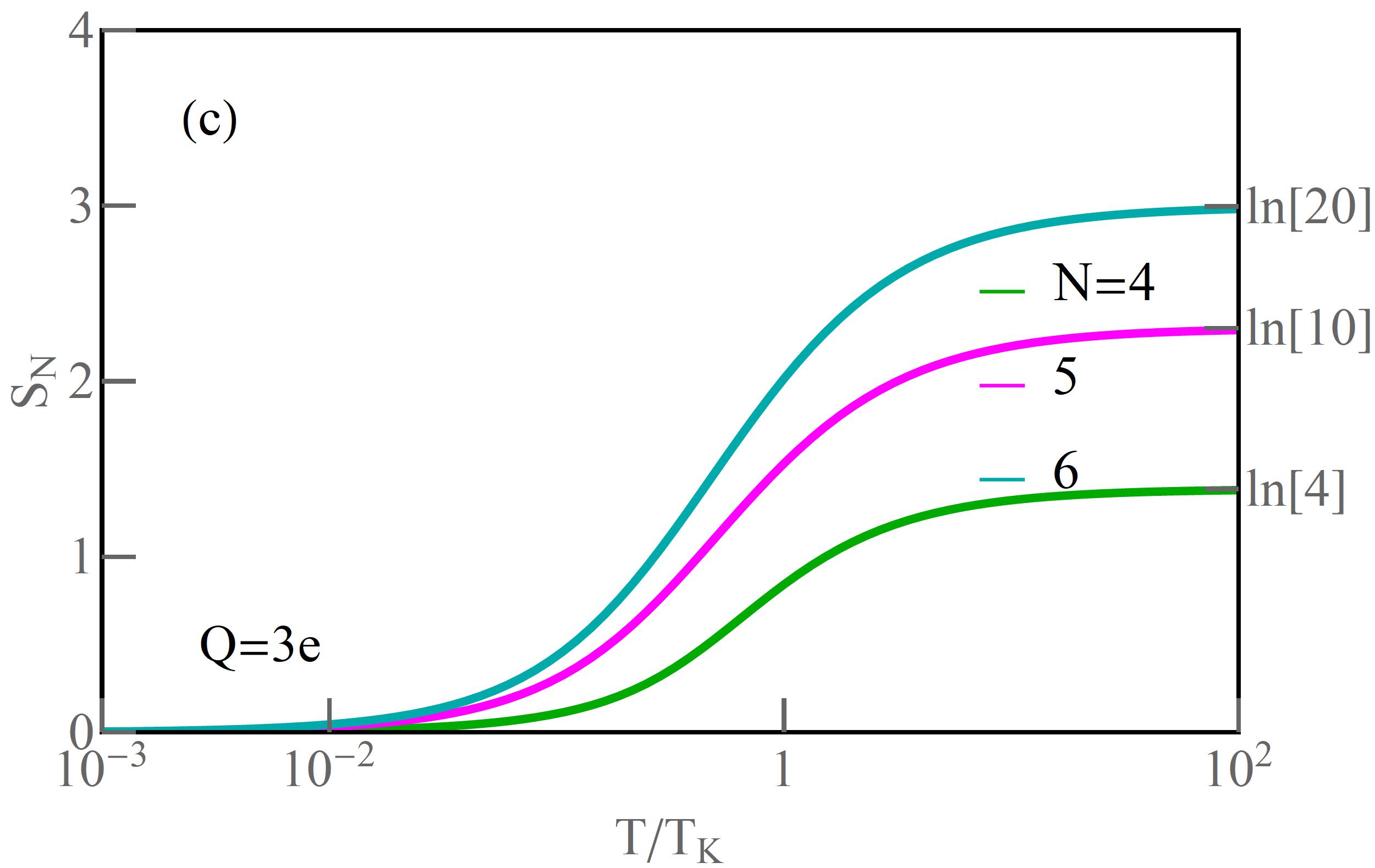}\\\caption{\label{fig7} (a-c) The impurity entropies $S_{N}$ versus $T/T_{K}$ for selected charge regions $Q=1e, 2e, 3e$. The insets present temperature dependent $S_{N=4}$ of isolated and screened SU(4) impurity (blue and green lines) ($U=3, \Gamma=0.025, V=0$).}
\end{figure}

Generalized spin susceptibility and entropy provide direct, thermodynamic evidence for the formation of Kondo singlet.
Figures \ref{fig6} and \ref{fig7} present the temperature dependencies of  susceptibilities, generalized moments and entropies for different SU(N) systems and various occupancies.  $T\chi_{N}$ (dark curves in Figure \ref{fig6}), where $\widetilde{\chi}_{N}$ denoting generalized spin susceptibility, illustrate the fluctuation with increase of the temperature of the squared generalized SU(N) spins defined by Equation (\ref{16}).
Local Coulomb repulsion favors single occupancies of spin-orbitals and local SU(N) moments form at high temperatures. This fact is illustrated by the saturation of the continuous lines $T\chi_{N}$ in Figures \ref{fig6}(a,b,c) to the values of squared generalized local SU(N) spins $T\chi_{N} = (N+1)NQ_{\nu}(1-Q_{\nu})/2$. The saturation values are described by Equation (\ref{19}). For  SU(2) in $n = 1$ range the corresponding value is $3/4$,   for SU(3)  $4/3$, for SU(4) $15/8$, for SU(5)  $12/5$   and $35/12$ for SU(6).
Similarly, for $n=2$ and $n=3$ the saturation values can be read from Figures \ref{fig6}(b,c).

The light lines on Figure \ref{fig6} show temperature dependencies of SU(N) susceptibilities. As it is seen they evolve from Curie type behavior for high temperatures to Pauli like, temperature independent at low temperatures \cite{Piquard2023}. For example, we show in the insets of Figures \ref{fig6}(a,b)  the exemplary comparison in a wide temperature range of the moments for an isolated SU(4) system (blue lines) with the  moments of SU(4) QD coupled to electrodes (green lines). Around $T\sim\sqrt{\Gamma U}$ there is a transition from the range of local moment on the dot to the free orbital
regime, where probability of occupation of all spin-orbitals is equal \cite{Zitko2006,Andrei1986,Wiegmann1983}.  At low temperatures, below $T_{K}$, the role of Kondo fluctuations becomes dominant and for $T\mapsto0$ moment of the dot is dynamically screened by electrode electrons and a many-body SU(N) Kondo singlet is formed. Apart from observations in transport, quite recently appeared also direct measurement of the progressive screening of the spin of single Anderson impurity (SU(2)) \cite{Piquard2023}.

Examples of the temperature dependence of the moments for SU(4) symmetry  are illustrated by  solid (green)  lines on Figures \ref{fig6}(a,b). They tend to zero for $T\mapsto0$. For high temperatures moments saturate to a value  $T\chi_{N=4} = 15/8$.  In this specific case quenched is also spin $\sigma$ and orbital moment $\tau$ and in addition also multipole $\sigma\tau$. In general, however, the screening of SU(N) moment is not necessarily accompanied by quenching of conventional spin or orbital moments. This is the case for example for discussed  SU(3) or SU(5) symmetries. Thermodynamic behavior of SU(N) moments coupled to electrode electrons  reflects also in temperature dependencies of  entropy $S_{N}$. Kondo ground state is a singlet and therefore, as it is seen on Figure \ref{fig7}, for  $T \mapsto0$  $S_{N}\mapsto0$. The rise of temperature causes an increase of entropy.   For $T > T_{K}$, the local moment becomes unscreened, free and for $n =1$  the entropy is determined by N lowest, degenerate single electron states and $S_{N} = k_{B}\ln[N]$.  For $n = 2$ the low temperature entropy is determined by the lowest degenerate two-particle states and for $n = 3$ by the degenerate three-particle states. Entropy expressed by the slave boson amplitudes is $S_{N} =-k_{B}\sum_{n}b^{2}_{n}\ln[b^{2}_{n}]=k_{B}\ln[N!/(n!(N-n)!)]$.  For still higher temperatures,  more states become excited, allowing thermal energy to be more widely spread. For $T \gg \sqrt{\Gamma U}$   all $N^{2}$ states of SU(N) system  are thermally accessible and entropy reaches value $2k_{B}\ln [N]$. Examples of entropies in a wide temperature range are presented in the insets of Figures \ref{fig7}(a,b).

\section{Conclusions}
We have investigated some transport and thermodynamic properties of multilevel degenerate QD symmetrically coupled to the leads.
In the analysis of correlations, we innovationely combine the description of the equilibrium Kondo state based on the generalized slave boson techniques (SBMFA) with microsopic Fermil liquid approach for high voltages, with parameters derived from slave boson fields.
Our study, which is the first systematic discussion of all SU(N) Kondo states to high order up to $N=6$.
We put special attention on the analysis of the shot noise. Compared to the conductance, shot noise is more difficult investigate experimentally, but its examination is essential, because it  provides extended information about dynamics of electron transfer and correlations. The systems under investigations are characterized by SU(N) group. For $N>2$ spin is replaced by generalized spin, which refers to the representation of SU(N) Lie group. The square of  SU(N) spin $S^{2}$, and its longitudinal component $S_{Z}$ commute with Hamiltonian and hence the invariant subspaces can by classified by quantum numbers $S$, and $S_{Z}$. For even $N$ many-body resonances appearing in the particle-hole symmetric point of SU(N)  structures represent Kondo states, while for odd N these are charge resonances. For half filling  both types of resonances  are centered at the Fermi level and therefore perfect transmission occurs  and  the partial conductance $G$ reaches the unitary limit at low temperatures ($G  = e^{2}/h$)  and the total conductance is $G = Ne^{2}/h$. The linear Fano factor, also  directly related to transmission, vanishes  at $E_{F}$ in this case. For this occupancy Kondo temperature reaches its lowest value. For even N, Kondo resonances hardly move with the change of gate voltage and hence plateaus are observed in the dependencies of conduction and Fano factors on gate voltage. For odd N, charge resonances broaden and shift already for small changes of gate voltage,  and   peaks of conductance or dips of linear Fano factor are seen. The Kondo resonances away from half filling are shifted relative to the Fermi level, hence the transmissions decrease and consequently the values of conduction plateaus decrease and correspondingly $F_{0}$ plateaus rise. Also decrease of $T_{K}$ is observed for $n \neq N/2$. Important information is the strength of correlations and therefore we also discussed Wilson ratio $W_{\nu\nu'}$, which quantifies it. If this ratio is different from $1$, it indicates that the correlations are significant. In Kondo systems, where localized electrons  interact strongly with conduction electrons, the Wilson ratio can be enhanced compared to the Fermi gas value. This enhancement is  associated with strong fluctuations of the generalized spin and the formation of a Kondo resonance. In  Kondo regime charge susceptibility in the limit   $U \mapsto \infty$ is zero or almost zero for strong interactions and this leads, that Wilson ratios take the universal values characteristic for a given $N$ and weakly depend on the occupation sector. Interestingly, $W_{\nu\nu'}$  decrease in charge fluctuation regions, but they still retain significant values indicating the importance of correlations also in these areas. The strength of correlations in a system tends to decrease as the degree of degeneracy increases, because more degenerate states  allow for compensatory changes to disturbances introduced by interactions. In the limit $N\mapsto\infty$, the quasiparticles become independent, $W_{\nu\nu'}\mapsto1$.  Approaching empty or fully occupied regions Wilson ratio also tend to unity for any $N$.

Screening  of magnetic moment by conduction electrons  is a key process of the spin Kondo effect. In this paper we have considered more general case, the effects of strong coupling of $N$ dimensional generalized spin with conduction electrons. To illustrate quenching of the generalized moments, we plotted generalized susceptibilities and entropies for each of the examined symmetries.  As the temperature is lowered, the susceptibility turns into a Pauli-like, temperature independent  and effective  moment tends to zero. In the Kondo states, the entropy associated with the localized moment is reduced due to the formation of a Kondo singlet resulting from  screening of the  SU(N) spin. Above $T > T_{K}$ entropy reaches $S_{N} = k_{B}\ln[N!/(n!(N-n)!)]$ and for $T = 0$ it vanishes. Shot noise measures out- of- equilibrium current fluctuations. As we mentioned earlier in the text, linear Fano factor corresponds to partition noise i.e it reflects fluctuations related to the partition of the scattered particles. Nonlinear shot noise arises from the fluctuations in the number of discrete charge carriers passing through a system. It results from interactions between quasiparticles. The important thing is, that nonlinear shot noise contains signature of 2e scattering, which is not seen in the conductance. Our study  confirms a conviction of previous  authors \cite{Oguri2022}, that taking into account the non-linear contribution can completely change the picture  of fluctuations, in particular whether they are super-Poissonian in nature or sub-Poissonian.  These characteristics  have been studied by us based on the Fermi liquid theory, assuming  dressed, non-interacting pseudofermions  defined within  SBMFA as quasiparticles. Fermi liquid  parameters describing  quasiparticle interaction were expressed by two- and three-body correlation functions. Competition of these correlations going beyond linear contribution to the shot noise can determine the nature of nonlinear noise. The dominance of particular  scattering processes changes as the energy or chemical potential moves away from the symmetry point.
We  illustrated on the example of SU(4) group, the general property of correlations the two-body correlations are symmetric, whereas  three-body correlations are antisymmetric with respect to electron-hole symmetry point. This behavior is a consequence of fermion nature of particles and  related antisymmetry of the wave functions. Consequently, the three-body residual interactions have a different impact on the behavior of electrons and holes. The three-body correlations vanish at e-h symmetric point,  are very weak in its neighborhood, and their absolute value is significant in the Kondo  regions beyond  half-filling. On the other hand, the two-body correlations are positive and  dominate  for  $n = N/2$ and they get weaker away from half filling, so that three-particle residual interactions dominate over two-particle interactions for regions  with distinct differences in number of electrons and holes. Our calculations point out, that non-linear Fano factor, tends to decrease with increase distance away from the e-h symmetry point. The current fluctuations of carriers with dominant population become more correlated with the total current, while those of the remaining carriers become less correlated. This leads to less noisy current flow. The rank of the unitary group has also a significant impact on nonlinear shot noise. Higher-symmetry Kondo states  tend to have increased shot noise compared to  lower-symmetry states due to enhanced many-body correlations involving more  components of generalized spin.

Summarizing, we systematically presented the main transport and thermodynamics characteristics of all SU(N) Kondo states to high order $N=6$. We discuss linear and nonlinear conductances, shot noise, temperature dependencies of susceptibilities, entropies and
generalized spins associated with the representations of Lie group characterizing the symmetry of these systems. In the analysis of correlations, we innovatively combined the description of the equilibrium Kondo state based on the extended slave boson technique (SBMFA) with microscopic Fermi liquid (FL) approach for high voltages, with parameters derived from slave boson medium. Using Wilson ratios for systems decoupled from the leads, we have proposed simple extrapolation scheme for calculation off-diagonal correlators necessary for determinations of FL parameters. We have defined generalized SU(N) spins and presented they screening by showing temperature dependencies of susceptibilities and entropies.  We have also indicated the occupancy ranges of the dominance in the noise of  the two-body correlations and three-body. Our study illustrated that, Kondo shot noise increases  with the  increase of  degeneration, which is related to the participation of a larger number of components of generalized spin. Based on our calculations it is seen, that  the nonlinear shot noise decreases as one moves away from the electron-hole symmetry point. Important observation is also the fact, that in the strong coupling limit the Wilson ratios take the universal values for a given degeneracy and occupation. We have presented that  Kondo temperature depends not only on the two-body susceptibility, but for its determination essential are also three-body correlators.

The conclusions drawn in this paper can be easily adopted for the case of other structures described by the same Hamiltonian e.g. to degenerate impurities, materials with multiple valleys, cold atom arrangements and others. Apart from enriching the knowledge about correlations, the discussed problems are of importance for quantum computing and other quantum technologies. The multidimensional generalized spin (qudit) offers greater information capacity and algorithmic efficiency compared to traditional spin. The Kondo effect provides a powerful, tunable platform to observe and control entanglement. Operating qudit involves carefully tuning the system (e.g. by magnetic or electric field) to be near or far from the Kondo regime, depending on the desired operation (e.g. initialization, entanglement, or readout). The main limitation of the possible applications of ordinary spin SU(2) Kondo effect in nanoobjects is its ultra low Kondo temperature. Increase of SU(N) symmetry removes this problem, because $T_{K}$ is growing exponentially with N and higher temperatures open up possibilities for practical use of these systems. Noise is a key area of interest. By studying changes in the shot noise, researchers can infer the level of decoherence occurring within the quantum system.  This knowledge is extremely important because decoherence is a critical factor that affects the stability and reliability of quantum information.
\vskip 0.1in
\textbf{Data availability statement}\\
All data that support the findings of this study are included within the article (and any supplementary
files).

\appendix
\section*{Appendix A. Slave boson formulation}
\renewcommand{\theequation}{A\arabic{equation}}
\setcounter{equation}{0}
For the discussion of correlation effects we use generalized finite-U slave boson mean field approximation (SBMFA) of Kotliar and Ruckenstein (K-R) \cite{Kotliar1986}. We present here only the basics of the slave boson formalism using a system with N$=6$ levels as an example. For lower multiciplicities (N$<6$), the formalism is analogous, but simplified, due to the smaller number of describes states, which allows the use of fewer bosons. This reduces the number of self-consistent minimization equations. For SU(6) we introduce a set of boson operators for each electronic configuration of the system. The auxiliary bosons $\{e, p, d, t, f, q, s\}$ project onto empty, single, double, triple, fourfold, fivefold and fully occupied states. The single occupation projectors $p_{i \sigma}$ ($q_{i \sigma}$) are labeled by orbital or site and spin numbers. Among double ($d_{i}$,$d_{ij\sigma\sigma'}$) and fourfold ($f_{i}$,$f_{ij\sigma}$)  occupancy  bosons, two classes can be distinguished, the first corresponds to occupancy of single orbital by two electrons ($d_{i}$) or two holes ($f_{i}$) and the second describes the double occupancy of electrons ($d_{ij\sigma\sigma'}$) or holes ($f_{ij\sigma\sigma'}$)  on different orbitals.  The three electron occupations are represented by twelve  bosons ($t_{i,j\sigma}$) corresponding to double occupation of one of the orbitals and single occupation of another and eight ($t_{\sigma\sigma'\sigma''}$) project onto states with single occupation on the orbital.
For SU(6) symmetry the use of seven independent SB operators is sufficient, in general for SU(N) their number is $(N+1)$. Apart from slave bosonic operators one also introduces in SB approach auxiliary fermionic operators, in terms of which the physical electron operators $d_{i\sigma}$  are expressed by $z_{i\sigma}f_{i\sigma}$, where $z_{i\sigma}$ is the transfer bosonic like-operator which modifies coupling to the leads:
\begin{eqnarray}
&&z_{i\sigma}=\widetilde{z}_{i\sigma}/\sqrt{\delta n^{2}_{i\sigma}}=(e^{\dagger}p_{i\sigma}+p^{\dagger}_{i\overline{\sigma}}d_{i}
+\\&&\nonumber\sum_{j\sigma'}p^{\dagger}_{j\sigma'}d_{ij\sigma\sigma'}
+\sum_{j}d^{\dagger}_{j}t_{j,i\sigma}+\sum_{j\sigma'}d^{\dagger}_{ij\overline{\sigma}\sigma'}t_{i,j\sigma'}+
\\&&\nonumber\sum_{i'<j,\sigma'\sigma''}d^{\dagger}_{i'j\sigma'\sigma''}t_{\sigma\sigma'\sigma''}
+\sum_{j}t^{\dagger}_{\overline{j},i\overline{\sigma}}f_{j}
+\\&&\nonumber\sum_{j\sigma'}t^{\dagger}_{i,j\sigma'}f_{ij\overline{\sigma}\sigma'}
+\sum_{i'<j,\sigma'\sigma''}t^{\dagger}_{\overline{\sigma}\overline{\sigma'}\overline{\sigma''}}
f_{i'j\sigma'\sigma''}+\\&&\nonumber f^{\dagger}_{i}q_{i\overline{\sigma}}
+\sum_{j\sigma'}f^{\dagger}_{ij\sigma\sigma'}q_{j\sigma'}+q^{\dagger}_{i\sigma}s)/\sqrt{\delta n^{2}_{i\sigma}}\label{A1}
\end{eqnarray}
and $f_{i\sigma}$ represents the pseudo-fermionic operators renormalizes orbital-lead hybridization, $\delta n^{2}_{i\sigma}=\langle Q_{i\sigma}(1-Q_{i\sigma})\rangle$ is the square of the orbital-spin fluctuations number and:
\begin{eqnarray}
&&Q_{i\sigma}=\widetilde{z}^{\dagger}_{i\sigma}\cdot\widetilde{z}_{i\sigma}=p^{\dagger}_{i\sigma}p_{i\sigma}+d^{\dagger}_{i}d_{i}+
\\&&\nonumber\sum_{j\sigma'}d^{\dagger}_{ij\sigma\sigma'}d_{ij\sigma\sigma'}
+\sum_{j\sigma'}t^{\dagger}_{i,j\sigma'}t_{i,j\sigma'}+\sum_{j}t^{\dagger}_{j,i\sigma}t_{j,i\sigma}
+\\&&\nonumber\sum_{\sigma\sigma'}t^{\dagger}_{\sigma\sigma'\sigma''}t_{\sigma\sigma'\sigma''}
+\sum_{j\sigma'}f^{\dagger}_{ij\overline{\sigma}\sigma'}f_{ij\overline{\sigma}\sigma'}
+\sum_{j}f^{\dagger}_{j}f_{j}+\\&&\nonumber\sum_{i'<j,\sigma'\sigma''}f^{\dagger}_{i'j\sigma'\sigma''}f_{i'j\sigma'\sigma''}+
f^{\dagger}_{i'j\sigma'\sigma''}f_{i'j\sigma'\sigma''}+\\&&\nonumber q^{\dagger}_{i\overline{\sigma}}q_{i\overline{\sigma}}
+\sum_{j\sigma'}q^{\dagger}_{j\sigma'}q_{j\sigma'}+s^{\dagger}s\label{A2}
\end{eqnarray}
denotes the charge operator. Tunneling (transfer) operator $\widetilde{z}_{i\sigma}$ is normalized by variance of orbital-spin occupancy number $\delta n^{2}_{i\sigma}$. The above form of effective resonant line narrowing factors $z_{i\sigma}$ are chosen in order to obtain the correct MFA limit in the uncorrelated case.
To eliminate additional unphysical states introduced by SB representation one supplements SB Hamiltonian by conditions of charge conservation $Q_{i\sigma}$ and completeness relation expressed by equating the sum of bosonic amplitudes with the unit operator ${\cal{I}}=e^{\dagger}e+\sum_{i\sigma}p^{\dagger}_{i\sigma}p_{i\sigma}+\sum_{i}d^{\dagger}_{i}d_{i}
+\sum_{ij\sigma\sigma',i<j}d^{\dagger}_{ij\sigma\sigma'}d_{ij\sigma\sigma'}+\sum_{ij\sigma,i\neq j}t^{\dagger}_{i,j\sigma}t_{i,j\sigma}
+\sum_{\sigma\sigma'\sigma''}t^{\dagger}_{\sigma\sigma'\sigma''}t_{\sigma\sigma'\sigma''}+\sum_{i}f^{\dagger}_{i}f_{i}
+\sum_{ij\sigma\sigma',i<j}f^{\dagger}_{ij\sigma\sigma'}f_{ij\sigma\sigma'}
+\sum_{i\sigma}q^{\dagger}_{i\sigma}q_{i\sigma}+s^{\dagger}s$.
These constraints are built into SB Hamiltonian by introducing Lagrange multipliers $\lambda$ and $\lambda_{i \sigma}$. The corresponding extended K-R Hamiltonian then reads:
\begin{eqnarray}
&&\nonumber \widetilde{{\mathcal{H}}}=\sum_{i\sigma}E_{i\sigma}n^{(f)}_{i \sigma}+\sum_{k\alpha\sigma}E_{ki\alpha}c^{\dagger}_{ki\alpha\sigma}c_{ki\alpha\sigma}
+\\&&\nonumber\sum_{k\alpha\sigma}t(c^{\dagger}_{ki\alpha\sigma}z_{i\sigma}f_{i\sigma}+h.c.)+U\sum_{i}d^{\dagger}_{i}d_{i}
+\\&&\nonumber U\sum_{ij\sigma\sigma',i<j}d^{\dagger}_{ij\sigma\sigma'}d_{ij\sigma\sigma'}+
3U\sum_{ij\sigma,i\neq j}t^{\dagger}_{i,j\sigma}t_{i,j\sigma}+\\&& 3U\sum_{\sigma\sigma'\sigma''}t^{\dagger}_{\sigma\sigma'\sigma''}t_{\sigma\sigma'\sigma''}+6U\sum_{i}f^{\dagger}_{i}f_{i}+\label{2}
\\&&\nonumber 6U\sum_{ij\sigma\sigma',i<j}f^{\dagger}_{ij\sigma\sigma'}f_{ij\sigma\sigma'}+10U\sum_{i\sigma}q^{\dagger}_{i\sigma}q_{i\sigma}
+\\&&\nonumber 15Us^{\dagger}s+\lambda({\cal{I}}-1)+\sum_{i\sigma}\lambda_{i\sigma}(n^{(f)}_{i \sigma}-Q_{i\sigma}),\label{A3}
\end{eqnarray}
where $n^{(f)}_{i \sigma}=f^{\dag}_{i \sigma}f_{i \sigma}$ is the pseudofermion occupation number operator \cite{Kotliar1986}.
The MFA ground state is found using the saddle-point approximation [K-R], in which all boson fields are replaced by their expectation values, found from the condition that $\langle\widetilde{{\mathcal{H}}}\rangle$  has an absolute minimum as a function of variables $\{b_{n=1..64}\}=\{e,p_{i\sigma},d_{i},d_{ij\sigma\sigma'},t_{i,j\sigma},t_{\sigma\sigma'\sigma''},f_{i},f_{ij\sigma\sigma'},q_{i\sigma},s\}$ and
Lagrange multipliers $\lambda$, $\lambda_{i\sigma}$. In this way the problem is formally reduced to the effective quasiparticle model with renormalized hopping integrals and renormalized dot energies. The resulting self-consistent minimalization equations read:
\begin{eqnarray}
&&\nonumber \left\langle\frac{\partial\widetilde{{\mathcal{H}}}}{\partial b_{n}^{\dagger}}\right\rangle=\left\langle\Delta \widetilde{{\mathcal{H}}}_{{n}}\right\rangle+\Delta \widetilde{E}_{{n}}\left\langle b_{n}\right\rangle=0\\
&&\left\langle\frac{\partial\widetilde{{\mathcal{H}}}}{\partial \lambda}\right\rangle={\cal{I}}-1=0\label{A4}\\
&&\nonumber\left\langle\frac{\partial\widetilde{{\mathcal{H}}}}{\partial \lambda_{i\sigma}}\right\rangle=\langle f^{\dagger}_{i\sigma}f_{i\sigma}\rangle^{<}-Q_{i\sigma}=0
\end{eqnarray}
where:
\begin{eqnarray}
&&\nonumber\left\langle\Delta \widetilde{{\mathcal{H}}}_{n}\right\rangle=
\\&&\sum_{k\alpha i\sigma}t\left\langle\frac{\partial z_{i\sigma}}{\partial b_{n}^{\dagger}}\right\rangle\langle c^{\dagger}_{k\alpha i\sigma}f_{i\sigma}\rangle^{<}+c.c.,\label{A5}
\end{eqnarray}
and $\Delta \widetilde{E}_{n}=\{(\lambda)_{n=1} ,(\lambda_{i\sigma}+\lambda)_{n=2..7},(U+\sum_{\sigma}\lambda_{i\sigma}+\lambda)_{n=8,..10},(U+\lambda_{i\sigma}+\lambda_{j\sigma'}+\lambda)_{n=11,..22}
,(3U+\lambda_{j\sigma}+\sum_{\sigma}\lambda_{i\sigma}+\lambda)_{n=23,..34},(3U+\lambda_{1\sigma}+\lambda_{2\sigma'}+\lambda_{3\sigma''}+\lambda)_{n=35,...42},
(6U+\sum_{j\sigma\neq i\sigma}\lambda_{j\sigma}+\lambda)_{n=43,..45},(6U+\sum_{i<j,\sigma\sigma'}(\lambda_{i\sigma}+\lambda_{j\sigma'})+\lambda)_{n=46,..57},(10U+\sum_{j\sigma\neq i\sigma}\lambda_{j\sigma'}+\lambda)_{n=58,..63},(15U+\sum_{i\sigma}\lambda_{i\sigma}+\lambda)_{n=64}\}$ correspond to effective corrections of the dot energies for different occupations. The correlators can be expressed by corresponding non-equilibrium Green's functions (NGF):
\begin{eqnarray}
&& \langle f^{\dagger}_{i\sigma}f_{i\sigma}\rangle^{<}
=\int^{+W}_{-W}\frac{dE G^{<}_{i\sigma,i\sigma}}{2\pi \textbf{i}}\\
&&\nonumber \sum_{k}\widetilde{t}\langle c^{\dagger}_{k\alpha i\sigma}f_{i\sigma}\rangle^{<}=\sum_{k}\int^{+W}_{-W}\frac{dE \widetilde{t}G^{<}_{k\alpha i\sigma,i\sigma}}{2\pi \textbf{i}}\label{A6}
\end{eqnarray}
$G^{<}_{i\sigma,i\sigma}=G^{R}_{i\sigma,i\sigma}\widetilde{\Sigma}^{<}_{i\sigma}G^{A}_{i\sigma,i\sigma}$ and $\sum_{k}\widetilde{t}G^{<}_{k\alpha i\sigma,i\sigma}=G^{R}_{i\sigma,i\sigma}\widetilde{\Sigma}^{<}_{\alpha i\sigma}+G^{<}_{i\sigma,i\sigma}\widetilde{\Sigma}^{A}_{\alpha i\sigma}$ are the non-equilibrium Green's functions that can be found by equations of motion method applied to  SB Hamiltonian (2).
$\widetilde{\Sigma}^{<}_{i\sigma}=\sum_{\alpha}f_{\alpha}\widetilde{\Sigma}^{A}_{\alpha i\sigma}$ and $\widetilde{\Sigma}^{A}_{\alpha i\sigma}=+\textbf{i}\frac{1}{2}\widetilde{\Gamma}_{i\sigma}$ are lesser and advanced self-energies.
The retarded and advanced Green's functions in channel $\nu$ are $G^{R(A)}_{\nu,\nu}(E)=\langle\langle f_{\nu};f^{\dagger}_{\nu}\rangle\rangle^{R(A)}=1/(E-\widetilde{E}_{\nu}\pm \textbf{i}\widetilde{\Gamma}_{\nu})$,
where the poles determine position $\widetilde{E}_{\nu=i\sigma}=E_{\nu}+\lambda_{\nu}$ and the width of quasiparticle resonance $\widetilde{\Gamma}_{\nu}=\Gamma z_{\nu}^{2}$.
The corresponding characteristic resonance temperature $T_{\nu}=\sqrt{\widetilde{E}^{2}_{\nu}+\widetilde{\Gamma}^{2}_{\nu}}$.
The renormalized level position and the width also specify charge located and thus also phase shift ($\delta_{\nu}$) at a given orbital $Q_{\nu}=\delta_{\nu}/\pi$.
The link between the complex pole of the Green's function, charge and characteristic temperature can be expressed as follows \cite{Coleman1987}:
\begin{eqnarray}
&&\ln[\widetilde{E}_{\nu}-\textbf{i}\widetilde{\Gamma}_{\nu}]=\ln[T_{\nu}]-\textbf{i}\pi Q_{\nu}\label{A7},
\end{eqnarray}
or equivalently:
\begin{eqnarray}
&&T_{\nu}=(\widetilde{E}_{\nu}-\textbf{i}\widetilde{\Gamma}_{\nu})e^{\textbf{i}\pi Q_{\nu}}\label{A8}.
\end{eqnarray}

In the mean field approximation the free energy corresponding to Hamiltonian (2) is a sum of slave boson free energy $\widetilde{F}_{b}$, and fermionic contribution $\widetilde{F}_{f}$ ($\widetilde{F}=\widetilde{F}_{f}+\widetilde{F}_{b}$). Using the Matsubara Green's function the fermionic and bosonic free energies can be written as:
\begin{eqnarray}
&& \widetilde{F}_{f}=-k_{B}T \textrm{Tr}\hat{G}^{-1}_{f}=
\\&&\nonumber -k_{B}T\sum_{\nu,\textbf{i} w_{n}}ln[\Lambda_{\nu}-\textbf{i} w_{n}]\label{A9}
\end{eqnarray}
where $\hat{G}_{f}$ is the full f-electron propagator in the presence of the Bose field (represented by the Green's function matrix in general), $iw_{n}$ are the Matsubara frequencies, and $\Lambda_{\nu}=\widetilde{E}_{\nu}+\textbf{i}\widetilde{\Gamma}_{\nu}=\widetilde{E}_{\nu}+\textbf{i}\Gamma z^{\dagger}_{\nu}z_{\nu}$ are the complex poles of the quasiparticle Kondo resonance. Performing integration of (12) on the complex plane one gets \cite{Coleman2015}:
\begin{eqnarray}
&&\widetilde{F}_{f}=\sum_{\nu}\int^{\Lambda_{\nu}}_{-\infty}dz\textrm{Im}\{X[z]\}\label{A10},
\end{eqnarray}
where $X[z]=(1/(2\pi))\sum_{\alpha={L,R}}\{\Psi_{0}[1/2+(z\pm V_{\alpha})/(2\pi \textbf{i} k_{B}T)]-ln[W/(2\pi \textbf{i} k_{B}T)]\}$ and $\Psi_{0}$ is the digamma function \cite{Coleman1987}. Slave boson contribution is:
\begin{eqnarray}
&&\nonumber\widetilde{F}_{b}=-\sum_{\nu}\lambda_{\nu}Q_{\nu}+U\sum_{i}d^{\dagger}_{i}d_{i}
+\\&&\nonumber U\sum_{ij\sigma\sigma',i<j}d^{\dagger}_{ij\sigma\sigma'}d_{ij\sigma\sigma'}+
3U\sum_{ij\sigma,i\neq j}t^{\dagger}_{i,j\sigma}t_{i,j\sigma}+\\&&3U\sum_{\sigma\sigma'\sigma''}t^{\dagger}_{\sigma\sigma'\sigma''}t_{\sigma\sigma'\sigma''}+\label{A11}
\\&&\nonumber+6U\sum_{i}f^{\dagger}_{i}f_{i}+6U\sum_{ij\sigma\sigma',i<j}f^{\dagger}_{ij\sigma\sigma'}f_{ij\sigma\sigma'}+
\\&&\nonumber 10U\sum_{i\sigma}q^{\dagger}_{i\sigma}q_{i\sigma}
+15Us^{\dagger}s+\lambda({\cal{I}}-1)
\end{eqnarray}

Phase shift $\delta_{\nu}$, which according to Friedel sum rules determines the elastic scattering, is linked with the corresponding partial  density of states at the Fermi level as follows $\widetilde{\varrho}_{\nu}(0)=\sin[\delta_{\nu}]^2/(\pi\widetilde{\Gamma}_{\nu})$.
At zero temperature $\delta_{\nu}$ is also related to the occupation number by:
\begin{eqnarray}
&&  n_{\nu}=Q_{\nu}=\langle f^{\dagger}_{ls}f_{ls}\rangle^{<}=\left\langle\frac{\partial \widetilde{F}_{f}}{\partial \widetilde{E}_{\nu}}\right\rangle=\label{A12}
\\&&\nonumber\sum_{\alpha}\textrm{Im} \left\{\frac{-ln\left[\frac{W}{2\pi \textbf{i} T}\right]}{2\pi}\right\}+
\\&&\nonumber\sum_{\alpha}\textrm{Im}\left\{\Psi_{0}\frac{\left[1/2+\frac{\widetilde{E}_{\nu}+\textbf{i}\widetilde{\Gamma}_{\nu}\pm V_{\alpha}}{2\pi \textbf{i} T}\right]}{2\pi}\right\}
\stackrel{V=T=0}=\frac{\delta_{\nu}}{\pi}.
\end{eqnarray}

Similarly static two-body susceptibilities $\widetilde{\chi}_{\nu_{1}\nu_{2}}=\int^{1/k_{B}T}_{0}d\tau\langle\delta n_{\nu_{2}}(\tau)\delta n_{\nu_{1}}(0)\rangle^{<}$ and three-body correlation functions $\widetilde{\chi}^{[3]}_{\nu_{1}\nu_{2}\nu_{3}}=-\int^{1/k_{B}T}_{0}d\tau_{3}\int^{1/k_{B}T}_{0}d\tau_{2}\langle T_{[\tau]}\delta n_{\nu_{3}}(\tau_{3})\delta n_{\nu_{2}}(\tau_{2})$
$\delta n_{\nu_{1}}(0)\rangle^{<}$ are expressed through derivatives of the free energy
with respect to site energies. $\delta n_{\nu}$ denote deviations from the ground state distribution  $\delta n_{\nu}\equiv n_{\nu}-\langle n_{\nu}(0)\rangle$.

The diagonal fermionic susceptibilities at low temperatures are given by \cite{Coleman1987}:
\begin{eqnarray}
&&\widetilde{\chi}_{\nu\nu}=-\left\langle\frac{\partial^{2} \widetilde{F}_{f}}{\partial \widetilde{E}^{2}_{\nu}}\right\rangle=\label{A13}\\
&&\nonumber \lim_{V,T\mapsto0} \sum_{\alpha}\textrm{Im} \left\{\frac{-\Psi_{1}\left[1/2+\frac{\widetilde{E}_{\nu}+\textbf{i}\widetilde{\Gamma}_{\nu}\pm V_{\alpha}}{2\pi \textbf{i} T}\right]}{4\pi^{2} \textbf{i} T}\right\}=\\&&\nonumber
\frac{\widetilde{\Gamma}_{\nu}}{\pi(\widetilde{E}_{\nu}^{2}+\widetilde{\Gamma}_{\nu}^{2})}
=\frac{\widetilde{\Gamma}_{\nu}}{\pi T^{2}_{K,\nu}}=\frac{\sin^{2}[\delta_{\nu}]}{\pi \widetilde{\Gamma}_{\nu}}
=\frac{\sin^{2}[\delta_{\nu}]\delta n_{\nu}}{\pi \Gamma \widetilde{z}^{2}_{\nu}}\\
&&\widetilde{\chi}^{[3]}_{\nu\nu\nu}=-\left\langle\frac{\partial^{3} \widetilde{F}_{f}}{\partial \widetilde{E}^{3}_{\nu}}\right\rangle=\label{A14}\\
&&\nonumber \lim_{V,T\mapsto0}\sum_{\alpha}\textrm{Im} \left\{\frac{\Psi_{2}\left[1/2+\frac{\widetilde{E}_{\nu}+\textbf{i}\widetilde{\Gamma}_{\nu}\pm V_{\alpha}}{2\pi \textbf{i} T}\right]}{8\pi^{3} T^{2}}\right\}=\\&&\nonumber\frac{-2\widetilde{\Gamma}_{\nu}\widetilde{E}_{\nu}}{\pi(\widetilde{E}_{\nu}^{2}+\widetilde{\Gamma}_{\nu}^{2})^{2}}
=\frac{-2\widetilde{\Gamma}_{\nu}\widetilde{E}_{\nu}}{\pi T^{4}_{K,\nu}}=\frac{-2\cos[\delta_{\nu}]\sin^{3}[\delta_{\nu}]}{\pi \widetilde{\Gamma}^{2}_{\nu}},
\end{eqnarray}
where $\Psi_{1(2)}(z)$ are the first and second derivative of the digamma function $\Psi_{0}$.
Equations (\ref{A13})-(\ref{A14}) determine correlators within SBMFA approach. Although quasiparticles in this approximation  are formally  treated as non-interacting, some of the effects of interaction are encoded in this approach by renormalization of parameters.\newline

For discussion of the case out of equilibrium ($V\neq0$) we extended MFA approach to slave bosons by including in self-consistent minimalization equations free energy supplemented by Fermi liquid interactions term $\Delta \widetilde{F}$ (\ref{20})
We integrated $\Delta \widetilde{F}$ putting $T_{K}=T^{\star}=1/(4\sqrt{\widetilde{\chi}_{\nu\nu}\widetilde{\chi}_{\nu'\nu'}})$ and taken the fluctuation of the occupation number $\delta n_{\nu}=(1/2)\sum_ {\alpha}f_{\alpha}-f_{\alpha}(E+\delta E_{\nu})\stackrel{T\mapsto0}=(1/2)\sum_{\alpha}\Theta(E\pm V/2)-\Theta(E+\delta E_{\nu}\pm V/2)$ ($f_{\alpha}$ is the Fermi-Dirac distribution and $\Theta$ is the Heaviside function).
In this generalized procedure appeared the new additional minimization parameter $\delta E_{\nu}$ (fluctuation of energy levels of FL quasiparticles). The corresponding minimalization equation is $\partial(\widetilde{F}_{f}+\Delta \widetilde{F})/\partial (\delta E_{\nu})=0$ and it supplements Equations (\ref{A4}) and (\ref{A5}). As it seen on Figure \ref{figA}, the SBMFA solutions do not exists for $V\gg T_{K}$. The proposed method interpolates slave boson approach for high voltages.
This approach extends the existence of auxiliary bosons in the voltage range $U\gg V>T_{K}$ (Figure \ref{figA}). This method is an alternative proposition to the problem of fluctuations in FL theory different from the earlier schemes presented e.g. in  \cite{Lavagna1990,Coleman2015,Fresard1997}.\newline
The condition $\partial^{2}|F|/\partial {V}^{2}=0$ determines the characteristic voltage $V^{\star}=\sqrt{\frac{G_{0}}{3c_{V,\nu}}}$, for which the picture of non-interacting Kondo particles breaks down and Landauer's description no longer applies.
\renewcommand{\thefigure}{A}
\begin{figure}
\includegraphics[width=0.8\linewidth]{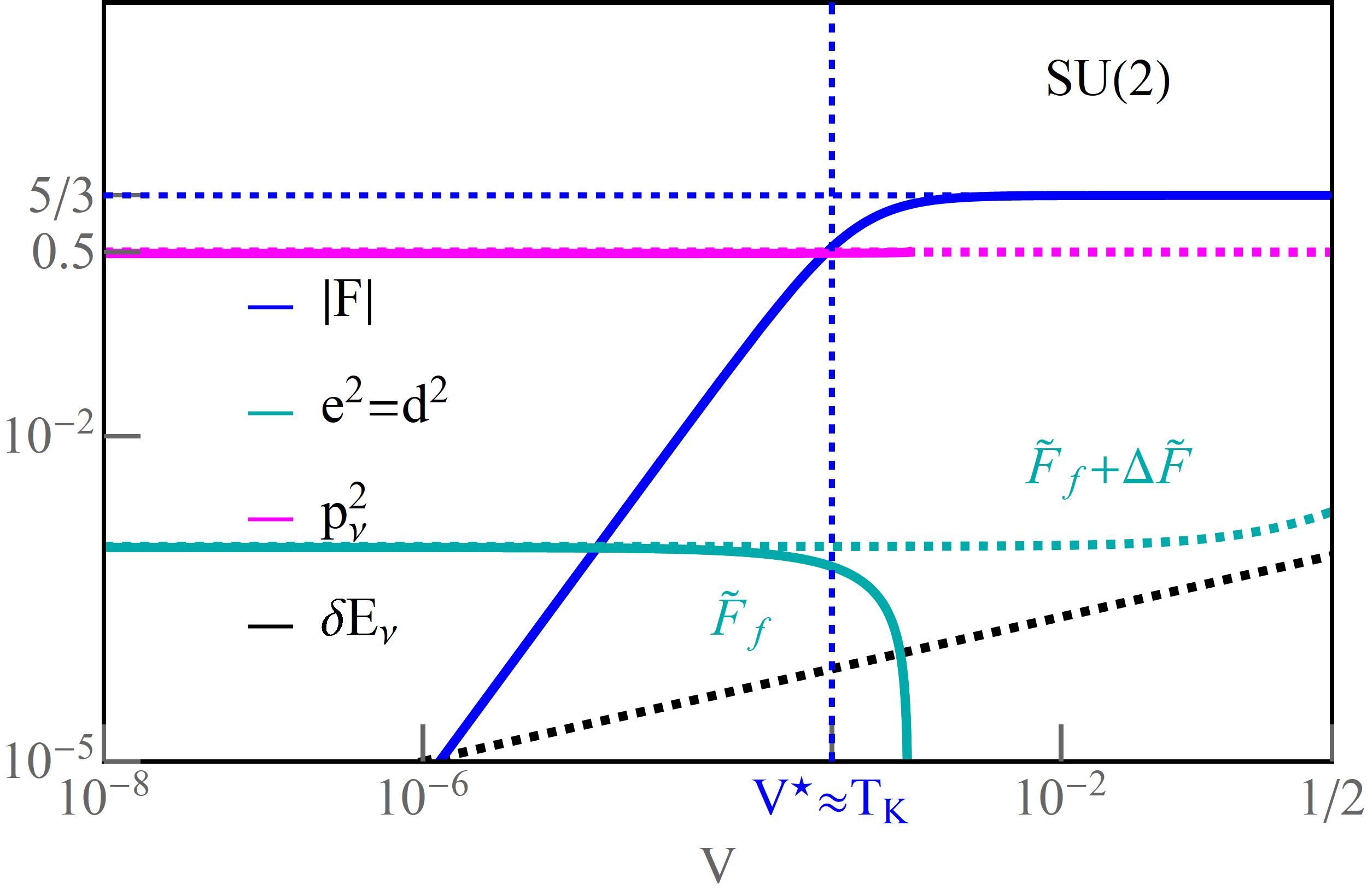}
\caption{\label{figA} Slave boson amplitudes $p_{\nu}^{2}$, $d^{2}$ and the fluctuation parameter $\delta E_{\nu}$ versus bias voltage (magenta dark cyan and black lines). Solid and dashed curves represent the self-consistent calculations without and with FL correction ($\Delta \widetilde{F}$) to free fermion energy $\widetilde{F}_{f}$. $T_{K}$ is shown on V axis. Blue line illustrates total $|F|$ versus V for SU(2) Kondo state ($U=3, T=0, E_{d}=-U/2, \Gamma=0.05$).}
\end{figure}

\section*{Appendix B. The analytical expression for the Kondo temperature in the SBMFA formalism}
\vskip 0.2in
\renewcommand{\theequation}{B\arabic{equation}}
\setcounter{equation}{0}
Here we only sketch the main points of the derivation,
a more detailed analysis, but only for the special case of SU(4) symmetry can be found in Appendix A of \cite{Krychowski20221}.
We restrict  to the fully symmetric SU(N) Kondo state. In this case the number of independent boson operators reduces from $2^{N}$ to $N+1$.
To derive an approximate analytical formula for $T_{K}$ applicable in  a given charge sector $\{Q\}$, for simplicity we take into account relationship of the minimization Equation 3 for a given charge sector $\{Q\}$ with analogous equations, but only from the nearest sectors $\{Q\pm1\}$. This simplification means that the dynamics within a given charge sector $Q=ne$ is influenced by fluctuations $(n-1)e\leftrightarrow ne$   and $ne\leftrightarrow(n+1)e$ alone. Let us denote the difference of two self-consistent equations for $n$ and $n-1$ and for given N by $\widetilde{\cal{K}}^{n}_{n-1}(N)$:
\begin{eqnarray}
&&\nonumber\widetilde{\cal{K}}^{n}_{n-1}=\frac{\partial \widetilde{\cal{H}}}{b_{n}\partial b_{n}^{\dagger}}
-\frac{\partial \widetilde{\cal{H}}}{b_{n-1}\partial b_{n-1}^{\dagger}}
=\frac{\Delta \widetilde{{\mathcal{H}}}_{{n}}}{b_{n}}-
\\&&\frac{\Delta \widetilde{{\mathcal{H}}}_{{n-1}}}{b_{n-1}}
+\Delta \widetilde{E}_{{n}}-\Delta \widetilde{E}_{{n-1}}\label{B1}
\\&&\nonumber=\sum_{k\alpha \nu}t\left(\frac{\partial z_{\nu}}{b_{n}\partial b_{n}^{\dagger}}
-\frac{\partial z_{\nu}}{b_{n-1}\partial b_{n-1}^{\dagger}}\right)c^{\dagger}_{k\alpha \nu}f_{\nu}+
\\&&\nonumber h.c.+\Delta \widetilde{E}_{{n}}-\Delta \widetilde{E}_{{n-1}}
\end{eqnarray}
For $V = 0$ and $T = 0$ the above formula simplifies:
\begin{eqnarray}
&& \left\langle\widetilde{\cal{K}}^{n}_{n-1}\right\rangle
=\\&&\nonumber\sum_{k\alpha \nu}t\left\langle\frac{\partial z_{\nu}}{b_{n}\partial b_{n}^{\dagger}}
-\frac{\partial z_{\nu}}{b_{n-1}\partial b_{n-1}^{\dagger}}\right\rangle \langle c^{\dagger}_{k\alpha \nu}f_{\nu}\rangle^{<}\label{B2}
\\&&\nonumber=\sum_{\nu}\left\langle\frac{z^{\dagger}_{\nu}\partial z_{\nu}}{b^{\dagger}_{n}\partial b_{n}}-\frac{z^{\dagger}_{\nu}\partial z_{\nu}}{b^{\dagger}_{n-1}\partial b_{n-1}}\right\rangle\frac{\Gamma}{\pi}\ln\left[\frac{T^{2}_{K}}{W^2}\right]
+\\&&\nonumber\Delta \widetilde{E}_{{n}}-\Delta \widetilde{E}_{{n-1}}
\end{eqnarray}
In the unitary limit $\lambda_{\nu}\approx E_{d}$ and then:
\begin{eqnarray}
&&\nonumber \left\langle\widetilde{\cal{K}}^{n}_{n-1}(N)\right\rangle
=\Lambda^{n}_{n-1}(N)\frac{2\Gamma}{\pi}\ln\left[\frac{|T_{K}|}{W}\right]+\\&&|E_{d}+(n-1)U|=0\label{B3}
\end{eqnarray}
and similarly
\begin{eqnarray}
&&\nonumber \left\langle\widetilde{\cal{K}}^{n+1}_{n}(N)\right\rangle
=-\Lambda^{n+1}_{n}(N)\frac{2\Gamma}{\pi}\ln\left[\frac{|T_{K}|}{W}\right]-\\&&|E_{d}+(n)U|=0\label{B4}
\end{eqnarray}
where the coefficients $\Lambda^{n}_{n-1}(N)$ and $\Lambda^{n+1}_{n}(N)$
are assigned to the virtual transitions between neighboring charge states. For $N=2$ $\{\Lambda^{n+1}_{n},\Lambda^{n}_{n-1}\}=\{4,4\}$, $N=3$ $\{\Lambda^{n+1}_{n},\Lambda^{n}_{n-1}\}=\{6,9/2\},\{9/2,6\}$, and for $N=4$ $\{8,16/3\},\{6,6\},\{16/3,8\}$.
For SU(5) and SU(6) Kondo symmetries $\{\Lambda^{n+1}_{n},\Lambda^{n}_{n-1}\}=\{10,25/4\},\{15/2,20/3\}$ and $\{12,36/5\},\{9,15/2\},\{8,8\}$ respectively.
For infinite U $\Lambda^{1}_{0}$ is different from zero and $\Lambda^{1}_{0}=N^{2}/(N-1)$, in particular for SU(4) symmetry $\Lambda^{1}_{0}=16/3$ \cite{Krychowski20221}. Comparing the left sides of the Equations A3 and A4 allows us to determine the SU(N) Kondo temperature  in selected charge sector $Q=ne$.
\begin{eqnarray}
&&\left\langle\widetilde{\cal{K}}^{n}_{n-1}\right\rangle|E_{d}+(n)U|=\\&&\nonumber\left\langle\widetilde{\cal{K}}^{n+1}_{n}\right\rangle|E_{d}+(n-1)U|\label{B5}
\end{eqnarray}
From (A5) it follows:
\begin{eqnarray}
&&T_{K}(N,n)=\\&&\nonumber We^{-\frac{|E_{d}+(n-1)U||E_{d}+nU|}{\frac{\Gamma}{\pi}(\Lambda^{n+1}_{n}|E_{d}+(n-1)U|+
\Lambda^{n}_{n-1}|E_{d}+nU|)}}\label{B6}
\end{eqnarray}
Kondo temperature is expressed in terms of the bare parameters of the Anderson model.
Formula (A6)  generalizes  K-R SBMFA expression for SU(2) onto the systems of higher symmetries SU(N), $N>2$.
For infinite U formula (A6)  simplifies:
\begin{eqnarray}
&&T_{K}(N)=We^{-\frac{|E_{d}|}{\frac{\Gamma}{\pi}\Lambda^{1}_{0}(N)}}\label{B7}.
\end{eqnarray}
For $N=2$ the above formula differs from Coleman's SBMFA expression \cite{Pavarini2015,Kim2016} and form commonly cite formula for $T_{K}$ derived from Schrieffer-Wolff  transformation \cite{Keller2014} by a factor two in the denominator of exponent. The difference is related to disparate representation of charge fluctuations in the K-R and Coleman approaches (different number of SB operator). For $N=4$ expression (B7)
has been derived earlier by us \cite{Krychowski20221}. The Kondo temperature increases with an increase of degeneracy. This is most clearly seen in the limit of infinite $U$, where the Kondo temperature scales by only one coefficient in the standard exponential dependence, i.e.,  $ \Lambda^{1}_{0} = \frac{N^2}{N-1}$ \cite{Hewson1997,Krychowski20221}. For high degeneracy $\Lambda^{1}_{0}\mapsto N$. The increase in $T_{K}$ along with the increase in degeneration opens a window to Kondo system applications. The increase of $T_{K}$ with N is observed in differential conductance spectroscopy \cite{Herrero2005,vanderWiel2000,Nygard2000}, where the width of the quasiparticle resonance increases with higher N \cite{Hewson1997}.

We can express two- and three-body correlation functions using Kondo temperature as follows:
\begin{eqnarray}
&&\widetilde{\chi}_{\nu\nu}=\frac{\widetilde{\Gamma}_{\nu}}{\pi T^{2}_{K}}\sim\frac{1}{T_{K}}\label{B8}\\
&&\widetilde{\chi}^{[3]}_{\nu\nu\nu}=\frac{-2\widetilde{E}_{\nu}\widetilde{\Gamma}_{\nu}}{\pi T^{4}_{K}}\sim\frac{-1}{T^{2}_{K}}\label{B9}.
\end{eqnarray}
Based on the above equations, we find $\widetilde{E}_{\nu}=\frac{-T^{2}_{K}\widetilde{\chi}^{[3]}_{\nu\nu\nu}}{2\widetilde{\chi}_{\nu\nu}}$ and we can write equivalently Kondo temperature (A6) as a function of not only two-body correlator, but also three-particle correlator:
\begin{eqnarray}
&& T^{2}_{K}=\frac{4}{(2\pi\widetilde{\chi}_{\nu\nu})^{2}
+\left(\frac{\widetilde{\chi}^{[3]}_{\nu\nu\nu}}{\widetilde{\chi}_{\nu\nu}}\right)^{2}}\label{B10}
\end{eqnarray}

\section*{Appendix C. Susceptibilities and Wilson ratio of isolated N-QD}
\vskip 0.2in
\renewcommand{\theequation}{C\arabic{equation}}
\setcounter{equation}{0}
Slave boson Hamiltonian of N-QD in the atomic limit ($t = 0$) reads:
\begin{eqnarray}
&& \nonumber \widetilde{{\mathcal{H}}}^{0}=\sum_{i\sigma}\widetilde{E}_{i\sigma}Q_{i \sigma}+U\sum_{i}d^{\dagger}_{i}d_{i}
+\\&& U\sum_{ij\sigma\sigma',i<j}d^{\dagger}_{ij\sigma\sigma'}d_{ij\sigma\sigma'}+\\&&\nonumber
3U\sum_{ij\sigma,i\neq j}t^{\dagger}_{i,j\sigma}t_{i,j\sigma}+ 3U\sum_{\sigma\sigma'\sigma''}t^{\dagger}_{\sigma\sigma'\sigma''}t_{\sigma\sigma'\sigma''}+\label{C1}
\\&&\nonumber6U\left(\sum_{i}f^{\dagger}_{i}f_{i}+
\sum_{ij\sigma\sigma',i<j}f^{\dagger}_{ij\sigma\sigma'}f_{ij\sigma\sigma'}\right)+\\&&\nonumber10U\sum_{i\sigma}q^{\dagger}_{i\sigma}q_{i\sigma}
+15Us^{\dagger}s+\lambda{\cal{I}},
\end{eqnarray}
The corresponding partition function is:
\begin{eqnarray}
&&{\mathcal{Z}}_{0}=\sum_{n}e^{-\beta \Delta \widetilde{E}_{n}'}\label{C2},
\end{eqnarray}
where $\beta=1/k_{B}T$, $\Delta \widetilde{E}_{n}'=\Delta \widetilde{E}_{n}(\{\lambda_{\nu}\}=\{\widetilde{E}_{\nu}\})$ and $\widetilde{E}_{\nu}$ is defined in the main text. The  probability distribution can be determined  by  the assignment:
\begin{eqnarray}
&& b^{2}_{n=1...64}=e^{-\beta \Delta \widetilde{E}_{n}'}/{\mathcal{Z}}_{0}\label{C3}.
\end{eqnarray}
The free energy of isolated N-QD is $F_{(0)}=-(1/\beta)\ln({\mathcal{Z}}_{0})$.
Now we present formulas for susceptibilities. For clarity and brevity of expressions, we illustrate the derivation of formulas  only  for the simplest case of SU(2) symmetry, which requires the use of four slave bosons. For higher symmetries considerations are completely analogous.
For SU(2) symmetry the set of SB operators is $\{b_{n}\}=\{e,p_{\nu},p_{\overline{\nu}},d\}$ and
$\Delta \widetilde{E}_{n}'=\{\lambda,\lambda+\widetilde{E}_{\nu},\lambda+\widetilde{E}_{\overline{\nu}},
\lambda+\widetilde{E}_{\nu}+\widetilde{E}_{\overline{\nu}}+U\}$. Susceptibilities are given by:
\begin{eqnarray}
&&\nonumber\chi_{\nu\nu}=-\frac{\partial^{2}F_{(0)}}{\partial \widetilde{E}^{2}_{\nu}}
\\&&\chi_{\nu\nu'}=-\frac{\partial^{2}F_{(0)}}{\partial \widetilde{E}_{\nu}\partial \widetilde{E}_{\nu'}}\label{C4}.
\end{eqnarray}
Using the inverted relation to Equation B3 we can write:
\begin{eqnarray}
&&\nonumber\chi_{\nu\nu}=\frac{(p^{2}_{\nu}+d^{2})(e^{2}+p^{2}_{\overline{\nu}})}{{\mathcal{Z}}_{0}T}=\frac{Q_{\nu}(I-Q_{\nu})}{{\mathcal{Z}}_{0}T}
\\&&\chi_{\nu\nu'}=\frac{(p_{\nu}p_{\overline{\nu}}+de)(-p_{\nu}p_{\overline{\nu}}+de)}{{\mathcal{Z}}_{0}T}
=\\&&\nonumber\frac{-Q_{\nu}Q_{\nu'}+d^{2}}{{\mathcal{Z}}_{0}T}=\frac{-Q_{\nu}Q_{\nu'}+Q_{\nu\nu'}}{{\mathcal{Z}}_{0}T}\label{C5},
\end{eqnarray}
where $Q_{\nu}=\langle \nu|\sum_{i\sigma}Q_{i\sigma}|\nu\rangle$ is the charge in state $\nu$ and
$Q_{\nu\nu'}=\langle \nu|\sum_{i\sigma}Q_{i\sigma}|\nu'\rangle$ is the inter-state charge correlator.

The Wilson ratio:
\begin{eqnarray}
&&W^{(0)}_{\nu\nu'}-1\equiv-\frac{\chi_{\nu\nu'}}{\sqrt{\chi_{\nu\nu}\chi_{\nu'\nu'}}}\label{C6}
\end{eqnarray}
can be expressed using Equation B5 as follows:
\begin{eqnarray}
&& W^{(0)}_{\nu\nu'}-1=\\&&\nonumber\frac{Q_{\nu}Q_{\nu'}-Q_{\nu\nu'}}{\sqrt{Q_{\nu}(I-Q_{\nu})Q_{\nu'}(I-Q_{\nu'})}}\label{C7}
\end{eqnarray}
It is important to note that while the corresponding bosonic expressions for the susceptibilities differ for individual symmetries, because the sets of auxiliary bosons are different, the formulas for  susceptibilities given by the charges and charge correlators have a general form, applying to any N.
According to the argument presented in the main text, we adopt the above  formula for $W_{\nu\nu'}$  (Equation B5) also for the case, when electrodes are weakly linked with the dot (see Equation 18).

Figure \ref{figC} presents dependence of generalized Wilson ratio on Coulomb interaction parameter $U$
exemplary for $Q = 1e$. For high values of $U$,  $W_{\nu\nu'}-1$ saturates and takes the values $1/(N-1)$ (inset in Figure C),
which agrees with the results of NRG and perturbation theory calculus  \cite{Hewson1993,Nishikawa2012,Oguri2018}.  As we have shown earlier, in the regime of strong coupling, the Wilson coefficient depends only on symmetry (N), and due to the smallness of quasiparticles amplitude $z^{2}_{\nu}\approx \frac{T_{K}}{\Gamma}$, it is well defined by charges $Q_{\nu}$ and inter-charge correlator $Q_{\nu\nu'}$ (Equation \ref{C7}). Lowering the coupling value with the electrodes $\Gamma$ effectively shifts the saturation point $(N-1)(W_{\nu\nu'}-1)$ toward lower $U$ values.
\renewcommand{\thefigure}{C}
\begin{figure}
\includegraphics[width=0.8\linewidth]{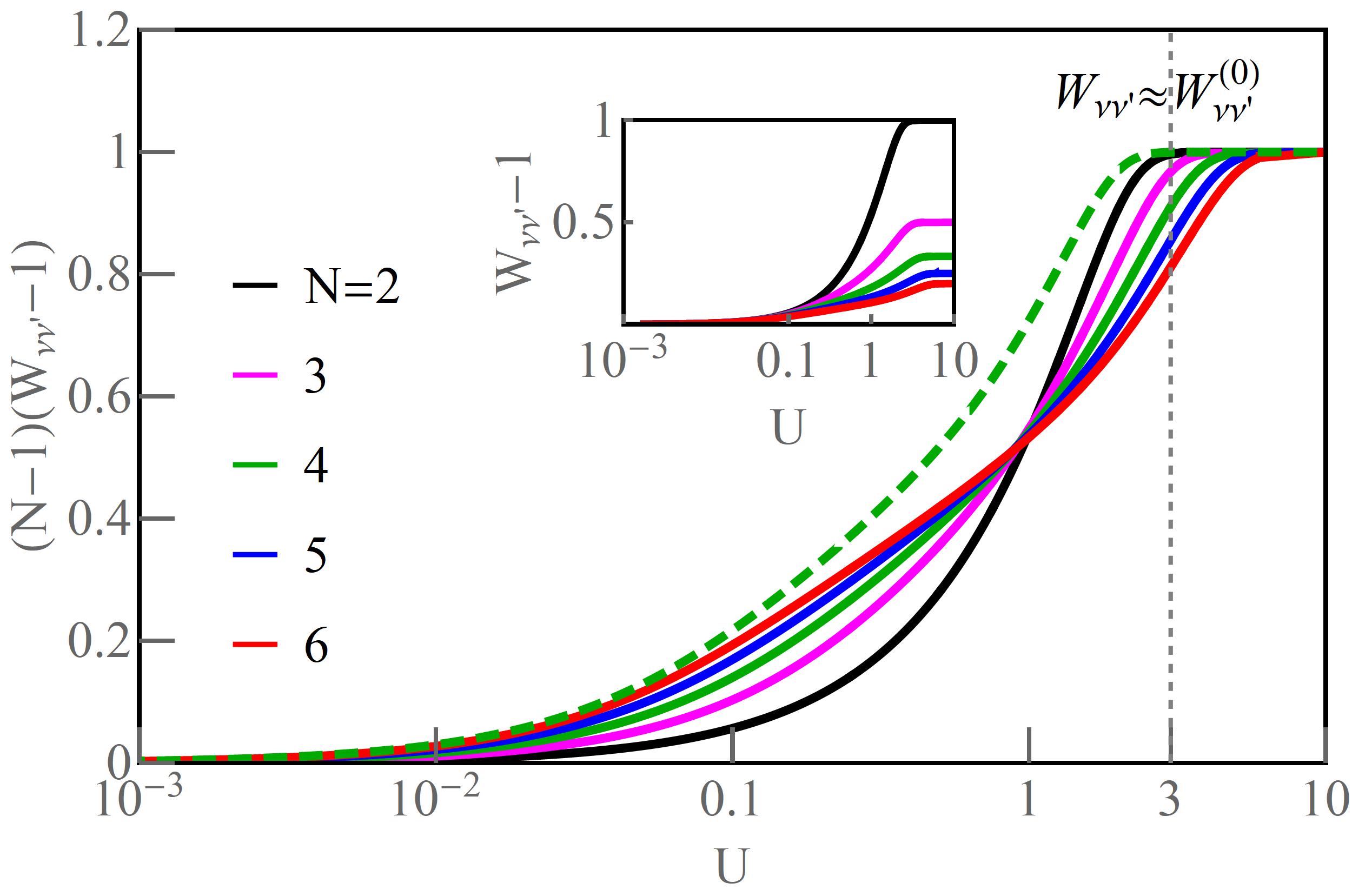}\caption{\label{figC} Generalized Wilson ratio $(N-1)(W_{\nu\nu'}-1)$ as a function of Coulomb interaction U in 1e charge region. Inset shows $W_{\nu\nu'}-1$ versus U ($\Gamma=0.05$). Green dashed line shows generalized Wilson ratio of SU(4) Kondo state for $\Gamma=0.025$. Dashed vertical gray line represents U used in calculations.}
\end{figure}

\section*{Appendix D. Nonlinear shot-noise of FL quasiparticles in SBMFA}
\vskip 0.2in
\renewcommand{\theequation}{D\arabic{equation}}
\setcounter{equation}{0}
As shown in \cite{Teratani2020,Oguri2022} the nonlinear coefficients $c_{V,\nu}$ and $c_{S,\nu}$ in the current and shot-noise can be expressed by the following high-order correlations and phase shifts as follows:
\begin{eqnarray}
&&\nonumber c_{V,\nu}=\frac{\pi^{2}}{12}(-\cos[2\delta_{\nu}](\widetilde{\chi}^{2}_{\nu\nu}+5\sum_{\nu'\neq\nu}\widetilde{\chi}^{2}_{\nu\nu'})+
\\&&(\widetilde{\chi}^{[3]}_{\nu\nu\nu}+3\sum_{\nu'\neq\nu}\widetilde{\chi}^{[3]}_{\nu\nu'\nu'})\frac{\sin[2\delta_{\nu}]}{2\pi})\label{D1}\\
&&\nonumber c_{S,\nu}=\\&&\nonumber\frac{\pi^{2}}{12}(\cos[4\delta_{\nu}]\widetilde{\chi}^{2}_{\nu\nu}+(2+3\cos[4\delta_{\nu}])\sum_{\nu'\neq\nu}\widetilde{\chi}^{2}_{\nu\nu'}
+\\&&4\sum_{\nu'\neq\nu}\cos[2\delta_{\nu}]\cos[2\delta_{\nu'}]\widetilde{\chi}^{2}_{\nu\nu'}
+\label{D2}v\\&&\nonumber
3\sum_{\nu'\neq\nu}\sum_{\nu''\neq\nu,\nu'}\sin[2\delta_{\nu}]\sin[2\delta_{\nu'}]\widetilde{\chi}_{\nu\nu''}\widetilde{\chi}_{\nu'\nu''}
-\\&&\nonumber(\widetilde{\chi}^{[3]}_{\nu\nu\nu}+3\sum_{\nu'\neq\nu}\widetilde{\chi}^{[3]}_{\nu\nu'\nu'})\frac{\sin[4\delta_{\nu}]}{4\pi})
\end{eqnarray}
The nonlinear Fano factor $F_{K}$ reads \cite{Oguri2022}:
\begin{eqnarray}
&&F_{K}=\frac{S-S_{0}}{2|e|(I-I_{0})}
=\\&&\nonumber\frac{S_{K}}{2|e|I_{K}}=\frac{\sum_{\nu}c_{S,\nu}}{\sum_{\nu}c_{V,\nu}}\label{D3}
\end{eqnarray}
Expressing two- and three-particle functions in (\ref{D1} and \ref{D2}) by slave-boson mean-field correlators we can rewrite the total shot noise formula for SU(N) Kondo quasiparticles up to $V^3$ as follows:
\begin{eqnarray}
&&\nonumber F=\frac{S}{2|e|I}=\frac{NA_{0}V+Nc_{S,\nu}V^{3}+0[V^{5}]}{NG_{0}V+Nc_{V,\nu}V^{3}+0[V^{5}]}=
\\&&(\frac{\widetilde{E}^{2}_{\nu}\widetilde{\Gamma}^{2}_{\nu}}{(\widetilde{E}^{2}_{\nu}+\widetilde{\Gamma}^{2}_{\nu})^{2}}V
+\\&&\nonumber \frac{[3+15(W_{\nu\nu'}-1)-6N(W_{\nu\nu'}-1)]\widetilde{E}^{4}_{\nu}\widetilde{\Gamma}^{2}_{\nu}V^{3}}{12(\widetilde{E}^{2}_{\nu}
+\widetilde{\Gamma}^{2}_{\nu})^{4}}
\\&&\nonumber+\frac{[-8-52(W_{\nu\nu'}-1)]\widetilde{E}^{2}_{\nu}\widetilde{\Gamma}^{4}_{\nu}V^{3}}{12(\widetilde{E}^{2}_{\nu}
+\widetilde{\Gamma}^{2}_{\nu})^{4}}
\\&&\nonumber+\frac{[18N(W_{\nu\nu'}-1)]\widetilde{E}^{2}_{\nu}\widetilde{\Gamma}^{4}_{\nu}V^{3}}{12(\widetilde{E}^{2}_{\nu}
+\widetilde{\Gamma}^{2}_{\nu})^{4}}
+\\&&\nonumber \frac{[1+9(W_{\nu\nu'}-1)]\widetilde{\Gamma}^{6}_{\nu}V^{3}}{12(\widetilde{E}^{2}_{\nu}
+\widetilde{\Gamma}^{2}_{\nu})^{4}})
/(\frac{\widetilde{\Gamma}^{2}_{\nu}V}
{\widetilde{E}^{2}_{\nu}+\widetilde{\Gamma}^{2}_{\nu}}
+\\&&\nonumber \frac{[-3-11(W_{\nu\nu'}-1)]V^{3}}{12(\widetilde{E}^{2}_{\nu}+\widetilde{\Gamma}^{2}_{\nu})^{3}}
+\\&&\nonumber \frac{6N(W_{\nu\nu'}-1)]\widetilde{E}^{2}_{\nu}\widetilde{\Gamma}^{2}_{\nu}
}{12(\widetilde{E}^{2}_{\nu}+\widetilde{\Gamma}^{2}_{\nu})^{3}}V^{3}+
\\&&\nonumber \frac{[1+5(W_{\nu\nu'}-1)]\widetilde{\Gamma}^{4}_{\nu}}{12(\widetilde{E}^{2}_{\nu}+\widetilde{\Gamma}^{2}_{\nu})^{3}}V^{3})\label{D4}
\end{eqnarray}
Formula (D4) explicitly show the dependence of $|F|$ on Kondo resonance characteristics ($\widetilde{E}_{\nu}$ and $\widetilde{\Gamma}_{\nu}$), degeneracy ($N$) and the Wilson ratio $W_{\nu\nu'}$. When $W_{\nu\nu'}=1$, the interaction corrections disappear, and the system behaves like a noninteracting system of Kondo particles.

\bibliographystyle{quantum}
\bibliography{DKSL}
\end{document}